\pdfoutput=1
\documentclass{aastex61}
\usepackage{subfloat}
\begin{document}
\title{The intrinsic characteristics of galaxies on the SFR-M$_{*}$ plane at $1.2<$\MakeLowercase{z} $<4$: I. the correlation between stellar age, central density and position relative to the main sequence.} 
\correspondingauthor{Bomee Lee}
\email{bomee@ipac.caltech.edu}

\author{Bomee Lee}
\affil{Department of Astronomy, University of Massachusetts, Amherst, MA 01003, USA}
\affil{Infrared Processing and Analysis Center, California Institute of Technology, Pasadena, CA 91125, USA}
\author{Mauro Giavalisco}
\affil{Department of Astronomy, University of Massachusetts, Amherst, MA 01003, USA}
\author{Katherine Whitaker}
\affil{Department of Astronomy, University of Massachusetts, Amherst, MA 01003, USA}
\affil{Department of Physics, University of Connecticut, Storrs, CT 06269, USA }
\affil{Hubble Fellow}
\author{Christina C. Williams}
\affil{Steward Observatory, University of Arizona, 933 N. Cherry Ave., Tucson, AZ 85721, USA}
\author{Henry C. Ferguson}
\affil{Space Telescope Science Institute, 3700 San Martin Drive, Baltimore, MD 21218, USA}
\author{Viviana Acquaviva}
\affil{Department of Physics, CUNY NYC College of Technology, 300 Jay Street, Brooklyn NY 11201, USA}
\author{Anton M. Koekemoer}
\affil{Space Telescope Science Institute, 3700 San Martin Drive, Baltimore, MD 21218, USA}
\author{Amber N. Straughn}
\affil{Goddard Space Flight Center, Code 665, Greenbelt, MD 20771, USA}
\author{Yicheng Guo}
\affil{UCO/Lick Observatory, Department of Astronomy and Astrophysics, University of California, Santa Cruz, CA 95064, USA }
\affil{Department of Physics and Astronomy, University of Missouri, Columbia, MO 65211, USA}
\author{Jeyhan S. Kartaltepe}
\affil{School of Physics and Astronomy, Rochester Institute of Technology, 84 Lomb Memorial Drive, Rochester, NY 14623, USA}
\author{Jennifer Lotz}
\affil{Space Telescope Science Institute, 3700 San Martin Drive, Baltimore, MD 21218, USA}
\author{Camilla Pacifici}
\affil{Goddard Space Flight Center, Code 665, Greenbelt, MD 20771, USA}
\affil{NASA Postdoctoral Program Fellow}
\author{Darren J. Croton}
\affil{Centre for Astrophysics \& Supercomputing, Swinburne University of Technology, PO Box 218, Hawthorn, Victoria 3122, Australia}
\author{Rachel S. Somerville}
\affil{Department of Physics and Astronomy, Rutgers, The State University of New Jersey, NJ 08854, USA}
\affil{Center for Computational Astrophysics, Flatiron Institute, New York, NY 10010}
\author{Yu Lu}
\affil{The Observatories, The Carnegie Institution for Science, 813 Santa Barbara Street, Pasadena, CA 91101, USA}

\begin{abstract}

We use the deep CANDELS observations in the GOODS North and South fields to
revisit the correlations between stellar mass ($M_*$), star--formation rate
(SFR) and morphology, and to introduce a fourth dimension, the
mass-weighted stellar age, in galaxies at $1.2<z<4$. We do this by
making new measures of $M_*$, $SFR$, and stellar age thanks to an improved SED fitting
procedure that allows various star formation history for each galaxy. 
Like others, we find that the slope of the Main Sequence (MS) of star formation in the $(M_*;SFR)$ 
plane bends at high mass. We observe clear morphological differences among 
galaxies across the MS, which also correlate with stellar age. At all redshifts, galaxies that 
are quenching or quenched, and thus old, 
have high $\Sigma_1$ (the projected 
density within the central 1 kpc), while younger, 
star--forming galaxies span a much broader range of $\Sigma_1$, which 
includes the high values observed for quenched galaxies, but also extends to much lower values. 
As galaxies age and quench, the stellar age and the dispersion of $\Sigma_1$ for
fixed values of $M_{*}$ shows two different regimes, one, at the low--mass end,
where quenching might be driven by causes external to the galaxies; the other, at the high--mass end, where
quenching is driven by internal causes, very likely the mass given the low
scatter of $\Sigma_1$ (mass quenching). We suggest
that the monotonic increase of central density as galaxies grow is one
manifestation of a more general phenomenon of structural transformation that
galaxies undergo as they evolve. 

\end{abstract}

\section{Introduction}

Deep multi-wavelength surveys like CANDELS (Cosmic Assembly Near-infrared
Extragalactic Legacy Survey: \cite{gro11, koe11}) have provided a tremendous
amount of new observational data for a large sample of galaxies over the last
few years. These surveys enable us to explore the Universe at redshifts
$z=1-3$, the peak epoch of star formation and active galactic nuclei (AGN)
activity \citep{mad14}, and very likely the epoch when the Hubble sequence
formed \citep{kri09, wuy11, szo11, lee13}. This peak in the star formation
rate density of the Universe signals a transition in the mode of galaxy
evolution, from the early universe galaxies, form stars rapidly, presumably
due to an abundance of cold gas, to an epoch of less star formation and more
passive evolution. Indeed, a number of studies have reported the emergence of
massive compact quiescent galaxies by $z\sim2-3$ \citep{cim04, tru06, tru07,
  van08, cas08, cas10}. Their number density increases rapidly, by a factor of
five, after $z\sim1$, and they are up to five times more compact in size than
local ones with similar stellar masses \citep{cas11, cas13}. However,
ellipsoidal, compact quiescent galaxies are morphologically very dissimilar
from their more extended disky star-forming counterparts \citep{kri09, cas11,
  guo12, che12, szo11, lee13}.

Quenching appears to be tied to morphological transitions, but this process 
is very poorly understood. Several quenching mechanisms have been proposed to 
explain the formation of quiescent galaxies. In general, very compact and 
massive galaxies are thought to be a result of a highly dissipative process, 
either (1) gas-rich mergers \citep{spr05, hop06, hop08, wuy10}, or (2) direct 
accretion of cold gas driven by violent disk instabilities in a compact disk 
(VDI; \cite{dek09,gen11}) or by gas traveling directly to 
the galaxy center and forming stars in--situ \citep{joh12}. The quenching of 
star formation subsequently takes place late when the gas supply is 
halted. Recently, it has been proposed that a significant fraction of quiescent
galaxies at $z>2$ are actually compact rotating disks \citep{vdw11, bru12}, as
opposed to systems stabilized by a significant, or dominant, fraction of
velocity dispersion. The presence of passive disks seems inconsistent with a
classical model, in which galaxy morphology is transformed from a disk into a
spheroid and star formation subsequently quenches. However,
recent hydrodynamic simulations predict that passive disks at $z>2$ will form when cold
gas inflows are halted, thus quenching star formation without transformation
of morphology \citep{ker05, dek08, wel15}.
Consistent with this scenario, \cite{wil14}
argued that compact galaxies simply assemble at very early times and evolve
through in--situ star formation by studying morphologies and volume densities
of massive early-type galaxies at $z\sim2$ and those of compact star-forming
galaxies at $z>3$. Moreover, using semi-analytic models, \cite{bre15} found
that the fractions of atypical galaxies (star-forming spheroids and quiescent
disks) are non-negligible and stay constant at $0<z<3$.  They showed that these
atypical galaxies follow different evolutionary paths compared to major
populations, star-forming disks and quiescent spheroids. The existence of such
atypical galaxies suggests that the physical mechanism responsible for
quenching star formation may, at least in some cases, be distinct from the process responsible for the
morphological transformation. Therefore, in principle, one should study the
morphologies of galaxies in different star formation regimes to constrain the
dominant quenching mechanism and solve the puzzle of galaxy evolution.

Classifying galaxies into different star formation regimes at high-redshift is
facilitated by the fact that star formation rate (SFR) and stellar mass
(M$_{*}$) of star-forming galaxies (SFGs) are strongly correlated out to at
least $z\sim4$ \citep{noe07, dad07, pan09, pan15, elb11, whi12a,whi14, lee15, sal15, sch15, tom16}. 
This correlation is commonly called the ``main
sequence of star formation" (MS). A common interpretation of the MS is that
the location of galaxies relative to the MS follows a different time evolution
of SFR \citep{ren09, dad10, rod11, sar12, ren15}. The tight MS with near unity
slope reflects that the majority of SFGs follow a steadily increasing star
formation history governed by a set of gradual physical processes like gas
exhaustion \citep{noe07}. A small fraction of galaxies exhibit
quasi-exponential mass and SFR growth, either through major mergers or through
strong bursts of star formation in the densest regions \citep{elb11, sar12}.
While typical galaxies therefore spend most of their time on the MS prior to
additional quenching processes, these starburst galaxies are located above the
MS and play a relatively minor role in the star formation history of the
Universe \citep{rod11}.  Galaxies located below the MS include quiescent
galaxies (QGs), with spheroidal-like structures and little star formation
activity, as well as fading SFGs with diminishing star formation activity. 
The transient galaxies, such as those in the green valley, can dominate the
lower region of the MS. At $z<1$, green valley galaxies are known to be off
the MS \citep{sch14} and they have intermediate morphologies combining 
disk-dominated and bulge-dominated systems \citep{sal09, men11, pan17}.

Star-formation activities of galaxies are strongly correlated with
morphologies \citep{wuy11, cam11, szo11, wan12, lee13}. \cite{wuy11}
investigated how the structures of galaxies depend on their location in the
log(SFR)--log(M$_{*}$) diagram since $z\sim2.5$, using large datasets from
four different fields (COSMOS, UDS, GOODS-South and North); They found strong
trends of specific star formation rate (sSFR=SFR/M$_{*}$) with galaxy
morphology, represented by S\'{e}rsic index (n) (see also \cite{whi15}). The
galaxies on and above the main sequence (MS) mostly show exponential light
profiles ($n\approx1$) and have blue rest-frame colors, while, relatively red
galaxies, reside below the MS, have ``de Vaucouleur'' (or similar) light
profiles.  However, most morphological studies on the MS have focused {\it
  only} on star-forming galaxies at $z>1$ \citep{wuy11, elb11, sal12, whi15},
and do not account for how galaxy morphology of the entire population
(including quiescent galaxies) differ throughout the log(SFR)-log(M$_{*}$)
plane.  Recently, \cite{bre17} studied structure of $z<2.5$ galaxies all the
way across and below the MS in CANDELS and a a semi-analytic model (SAM) of
galaxy formation.

It is important to constrain the intrinsic shape of the MS for all galaxies
across the log(SFR)-log(M$_{*}$) plane. \cite{spe14} reported that the sample
selection and other systematic uncertainties can significantly affect the
slope of the MS through an extensive compilation of 64 measurements of the MS
from 25 literature references using different SFRs, stellar masses and
different samples with different fitting methods out to $z\sim6$.  It is a
general conclusion at all explored redshifts and masses that the MS has a
constant dispersion of $\sim0.3$dex.  Furthermore, though it is well known
that the normalization of the MS increases with redshift, the derived single
power-law slope is sensitive to the SFR indicators, sample selections, and
initial mass function (IMF).  Recently, several studies have found that the MS
slope dramatically declines for massive galaxies at $z<2$ \citep{kar11,
  whi14, lee15, sch15, tom16}; therefore, a single power-law cannot explain
the MS slope, and a more complicated formula is necessary.  There is also
evidence that the stellar mass above which the MS flattens evolves with
redshift \citep{lee15, tom16, gav15}.  However, recently \cite{ren15}
re-defined the MS to be the main ridge line of the star formation peak in the
three dimensional SFR-mass-number space at $z\sim0$ (i.e., the mode and not
the mean/median).  Even when removing the pre-selection of star-forming
galaxies, they find that the best MS slope can be explained with a single
power law, without a bending of the MS at high masses. It is evident that the
robust characterization of the MS with careful sample selections, accurate
estimations of M$_{*}$ and SFR and better fitting methods are crucial.

A major complication in measuring the physical properties by fitting spectral
energy distribution (SED) of galaxies at $z>1$ is our inability to reliably
constrain the star formation histories (SFHs). In general, it would be ideal
to measure the real SFH of a galaxy, rather than assuming an analytic
function for it.  However, due to the complexity of the real SFH, the SED
fitting depends on SFH model with a simple functional form. There are
limitations in using SFH models: the degeneracy between SFHs and other
properties of galaxies such as dust extinction, metallicity, and redshift, as
well as the ``outshining effect'', where light from massive, young stars
dominates the observed SEDs with significantly higher luminosities than older
stellar populations. Despite these limitations, the most commonly used SFH
model is an exponentially decreasing star formation history, or a
$\tau$-model. The properties of remote spheroidal quiescent galaxies and low
redshift galaxies (including local spiral galaxies) are known to be reproduced
well using the $\tau$-model \citep{wuy09, bel00} because they clearly formed
stars at a higher rate in their past than at the epoch of
observation. However, using mock star-forming galaxies at $z>2$, \cite{lee09} and
\cite{wuy09} showed that the $\tau$-model cannot recover the intrinsic SFHs
for star-forming galaxies. Nevertheless, many studies still use the
$\tau$-model for all galaxy types in the high redshift Universe for
convenience, even though they do not necessarily show signs of star formation
decline. Recently, several studies have argued, either with observations
\citep{mar10, pap11, red12, pac13} or with simulations \citep{lee10, pfo12,
  beh13}, that the increasing SFH is a more appropriate model for star-forming
galaxies, especially at $z\gtrsim2$. The basic conclusion of these SFH
analyses is that various SFH models should be used to investigate diverse
galaxy populations \citep{con13}. In this regard, different SFH
parameterizations are suggested \citep{lee10, beh13, sim14, cas16}, but we
still do not know what the intrinsic SFH of a galaxy is nor how complicated
it is.

This paper is the first in a series of papers investigating the
characteristics of star-forming galaxies located on, above and below the MS
and quiescent galaxies at $1.2<z<4$ selected from the CANDELS (PI: S.Faber,
H. Ferguson). The primary goals of this study is to use an advanced SED fitting technique,
SpeedyMC, a Markov Chain Monte Carlo (MCMC) code based on Bayesian statistics
\citep{acq12}, to obtain new measures of the integrated parameters of the
galaxies' stellar populations, i.e. stellar mass, star--formation rate, median 
mass-weighted stellar age and dust obscuration, and study correlations with morphology. The
novelty of our approach is that the star--formation history of the galaxies is
not set equal to an assumed function but rather is treated as a
free ``parameter'', chosen from five simple models (constant, linearly
increasing, delayed, exponentially decreasing and increasing SFHs). We test
the robustness of our measures, especially the SFR and the mean stellar age,
against mock galaxies from SAM simulations, as well as with independent
measures (SFR only).  By adding the stellar age measures to the
  relation between morphology and the location of galaxies relative to the MS,
  we also investigate the evolution of galaxies and their quenching
  process. We also use various morphological diagnostics measured from the
  CANDELS/WFC3 H-band images, including parametric (S\'{e}rsic index,
  half-light radius), non-parametric (G and $M_{20}$) measures, and projected
  mass densities ($\Sigma_{50}$ and $\Sigma_{1}$: see Section 5.2 for
  definitions). The second paper in this series will focus on the
  non-parametric measures and $\Sigma_{1}$ to understand the morphological
  transformation of galaxies as they evolve.

The structure of this paper is as follows. The optical and NIR data from
CANDELS and the infrared data from \emph{Herschel} and the sample selection
used in this study are introduced in Section 2. In Section 3, the galaxy
properties (stellar mass, age, and SFR) obtained using various SFH models with
SpeedyMC are explained in detail, and the validation of the SED fitting
results through comparisons with different SFR indicators and simulations are
also found. With the robust estimation of M$_{*}$ and SFR using the best-fit
SFH, we investigate the MS in the log(SFR)--log(M$_{*}$) diagram in Section 4
and classify galaxies according to their positions relative to the MS. We
then present an extensive morphological analysis associated with the location
in the log(SFR)-log(M$_{*}$) plane using parametric (S\'{e}rsic index and
half-light radius), non-parametric (G and $M_{20}$) measures, and projected
mass densities in Section 5. The synthesis of morphological trends relative 
to the MS is presented in Section 6. We discuss our results in the
context of quenching processes of galaxies in Section 7 and conclude with 
summary in Section 8.

\section{DATA}

All data used in this study are based on the WFC3/F160W (H-band) selected
multi-wavelength catalogs from the CANDELS \citep{gro11, koe11}. We use
CANDELS/Deep fields (GOODS-S and -N) having deeper and fully panchromatic
images relative to CANDELS/Wide fields (COSMOS, EGS, and UDS). The Deep fields
cover about 130 square arc minutes and have a $5\sigma$ point source limiting
depth of $H_{AB} = 27.5$.  The multi--wavelength photometry has been obtained
using a software package with an object template-fitting method (TFIT,
\cite{lai07}). This catalog includes photometry from the {\it HST}/ ACS images
in the BViz and F814W; from WFC3/IR images in the F098M (only in the GOODS-S),
F105W, F125W, F160W; from VLT/VIMOS U and VLT/ISAAC Ks images; and from the
\emph{Spitzer}/IRAC images at 3.6, 4.5, 5.8 and 8.0 $\mu$m (GOODS-S from \cite{guo13} 
and GOODS-N from Barro et al., in preparation). We use CANDELS
best estimated photometric redshifts (photo-z) measured for all galaxies by \cite{dah13}
, unless spectroscopic redshifts (spec-z) are available (about 6\%
of our sample).

In order to investigate the
characteristics of IR detected galaxies among our sample, we use the
public GOODS-{\it Herschel} DR1 catalogs in the GOODS-S and -N
\citep{elb11} and identify IR detected galaxies by matching the positions in both 
CANDELS and GOODS-{\it Herschel} catalogs. The source catalogs containing 
{\it Spitzer}/MIPS 24$\mu$m, 70$\mu$m (only GOODS-S) and {\it Herschel}/PACS 
100$\mu$m, 160$\mu$m bands are used to compute total infrared luminosities 
in Section 3.1. {\it Herschel}/PACS flux densities and uncertainties are extracted 
from the PSF fitting using the {\it Spitzer}/MIPS 24$\mu$m prior positions. 

\subsection{Sample Selection}

\begin{figure}
\centering
\includegraphics[width=4in]{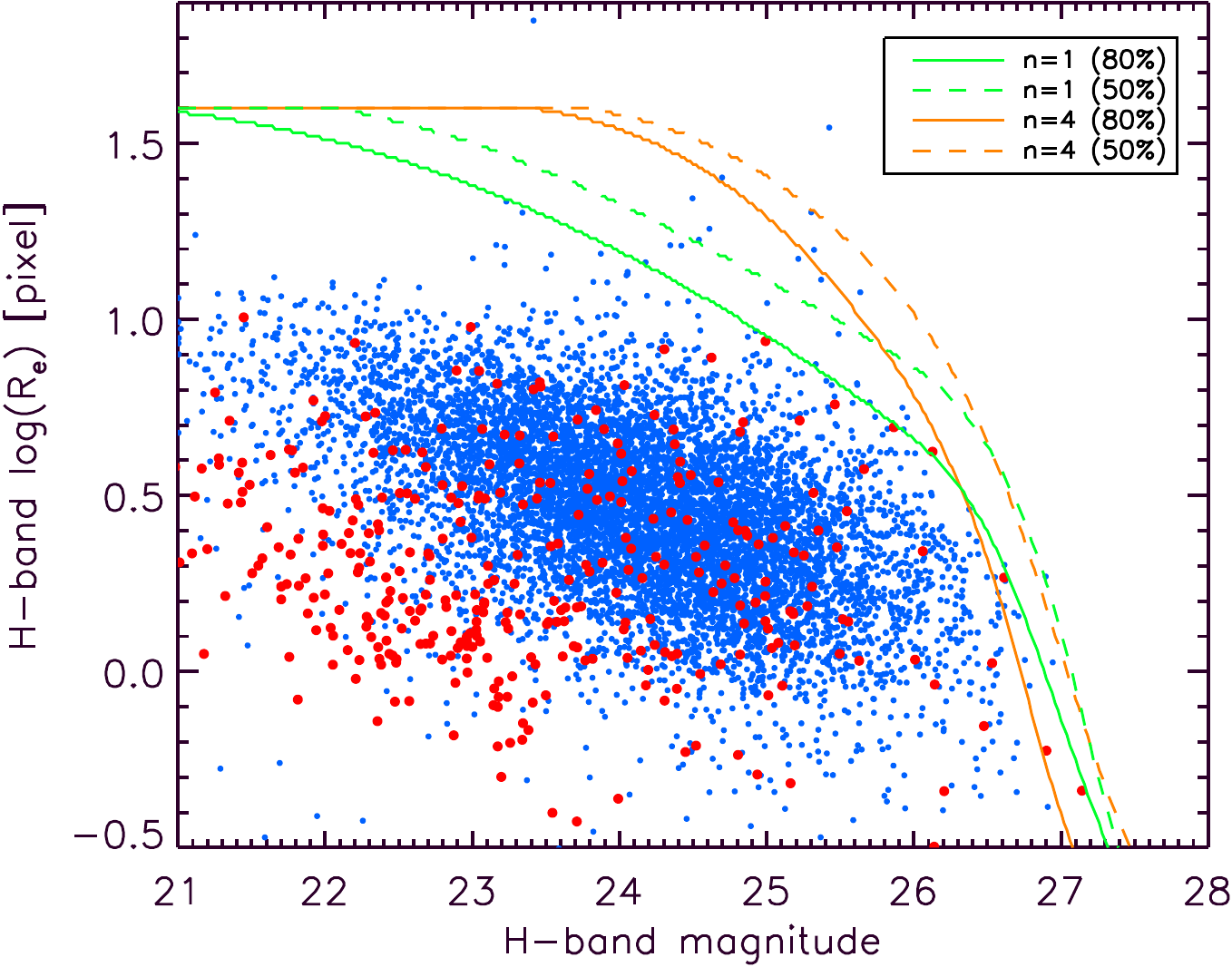}
\caption{Completeness curves for the CANDELS GOODS H-band images in the
  GOODS-S fields from the Monte Carlo simulations by \cite{guo13} plotted against 
  our sample in the half-light radius ($R_e$) vs. H-band mag plane. Blue and red 
  points represent UVJ selected star-forming and quiescent galaxies, respectively. 
  Both 50\% (dashed) and 80\% (solid) completeness curves are shown for two 
 morphological types, the exponential disk (S\'{e}rsic $n=1$) and the De Vaucouleurs spheroid
 ($n=4$) with green and orange lines, respectively. About two dozen 
 galaxies are outside the 80\% curve of disks and about
 one dozen outside the same curve for spheroids, suggesting that our sample is about
 80\% complete at all redshift and stellar mass that we have considered.}
\label{fig:complete}
\end{figure}

The goal of this paper is to study galaxies at $1\lesssim z \lesssim 4$, the
peak epoch of the star formation \citep{mad14}. We identify 23,580 galaxies at
$1 \le z<4$ in the CANDELS portion of the GOODS-N and GOODS-S fields after
limiting our sample to sources with H-band signal-to-noise ratio 
$SNR(H)>10$. We also eliminated from the sample those galaxies for which the 
SExtractor PhotFlag and SpeedyMC quality flags revealed problems with the
fitting procedures. SExtractor PhotFlag is used to designate
suspicious sources that fall in contaminated regions \citep{guo13}. We only
use galaxies having PhotFlag $=0$, indicating non-contaminated sources without
detections of star spikes, halos, and bright stars, as well as excluding
sources that are either artifacts or falling at the edge of the image. SpeedyMC 
quality flag is a warning sign, which 
represents the quality of the convergence of the MCMC chains. The quality flag is 
assigned using the length of the chain and the Gelman and Rubin ``R" test (Gelman \& Rubin 1992), 
which compares the variance of the mean within and between chains. 
If several long chains are present and $R-1 < 1$, then flag $=1$. 
If only one (long) chain is used, or if several chains are used, but there is 
a convergence problem ($R-1 > 1$, common if there are multiple peaks), then
 flag $=2$. If there are no long chains, then flag= 3. Following the definition of the quality flag, 
 galaxies with flag $=3$ are excluded in our final sample.

In this study, we separate star-forming galaxies from quiescent galaxies using the UVJ 
color-color selection. Previous studies have
shown that the galaxy selection in sSFR can effectively discriminate between
quiescent and star-forming galaxies \citep{kar11, lee13}.  Because it might
cause an artificially clean correlation between SFR and $M_{*}$, it is
preferable to use a quiescent selection independent of SFR and/or
$M_{*}$. Color-color selection has been widely used to classify quiescent
galaxies out to $z\sim 3$ during the last decade (e.g., BzK by \cite{dad07,
  lee13}; e.g., UVJ by \cite{wil09, bra11, whi11}). Quiescent galaxies are
characterized by red U-V colors and bluer V-J colors relative to star-forming
galaxies having same U-V colors. In this study, we classify quiescent galaxies
at all explored redshifts as,

\begin{equation}
U-V > 0.88 \times (V-J)+0.49,~U-V > 1.3,~V-J < 1.6. 
\label{eq:uvj}
\end{equation}

Using slightly different definitions of quiescent galaxies is common in
different surveys: while \cite{whi12a} changed to a redshift independent
rest-frame UVJ selection, \cite{muz13} adopt one that has weak redshift
evolution, dating back to the original work of \citet{wil09}. We use the
definition of \cite{sch15} for all CANDELS fields up to $z=4$. 
The rest-frame colors are measured using the EAZY
software \citep{bra08}, which performs a template-based interpolation
of the observed photometry. We use the template set of \citet{muz13}, using
the Bessel filters in the optical and Palomar filters in the NIR. From
Equation 1, we find that about 3$\%$ (29\% for
M$_{*}>3\times10^{10}$M$_{\odot}$) galaxies are classified as quiescent
galaxies. 

We have studied the completeness of our sample as a function of the H-band
limiting magnitude and surface brightness using the simulations by \cite{guo13}, 
in which artificial galaxies with an exponential profile (S\'{e}rsic parameter $n=1$) 
or De Vaucouleurs profile ($n=4$) have been inserted into the H-band images and
retrieved and analyzed with the same procedures as real galaxies. This 
analysis only tests for incompleteness in the detection process; our
additional cut in H-band SNR ensures that the subsequent morphological analysis
is likely to succeed for each detected galaxy. Figure~\ref{fig:complete} summarizes 
the results of the simulations and shows the 50\%-- and 80\%--completeness curves in 
the half-light radius ($R_{e}$) vs. H-band plane together with our sample, namely 
all galaxies with stellar mass $M_*>10^9$ M$_{\odot}$ (see later). Star--forming 
galaxies are shown in blue and passive ones in red. Except for about two dozen 
galaxies who are outside the disk 80\% curve and for one dozen galaxies which 
are outside the spheroid 80\% curve, the rest sample is located within the 80\% disk. 
This leads us to conclude that the sample is at least 80\% complete at all redshift 
and stellar mass considered here, and that incompleteness is not likely to affect 
any of our conclusions. 

Lastly, $\sim 1.5\%$ of our sample are known X-ray, IR and, radio
AGN candidates among our final sample (GOODS-S: \cite{xue11} (X-ray),
\cite{don12, pad11} (IR); GOODS-N: \cite{ale03} (X-ray)). Most of 
these AGN candidates (87\%) are classified as star-forming galaxies from the 
UVJ diagram, and about half of them are IR detected galaxies. Although our FIR 
measurements can be polluted by the light of dust-obscured AGN, we do not 
exclude AGN candidates from the further study
because inclusion of them does not significantly change our results.

\section{SED fitting allowing various star formation histories}

We have investigated how the choice of star formation history affects the
behavior of the spectral energy distribution (SED) fitting by comparing galaxy
properties obtained using five commonly used star formation history (SFH)
models: linearly increasing, constant, delayed (linearly increasing at early
time, then exponentially decreasing after $t=\tau$), exponentially decreasing
($\tau$-model), and exponentially increasing (inverted $\tau$-model) SFHs (see
Table 1 for definitions of SFH models). SEDs are fit with SpeedyMC \citep{acq12}
 to measure the properties of stellar populations including
stellar mass, age, dust reddening, and star formation rate. 
SpeedyMC is an updated version of GalMC, which is a publicly available
Markov Chain Monte Carlo (MCMC) code for the SED fitting based on Bayesian
statistics \citep{acq11}. In SpeedyMC, the model spectra are computed once at
a grid of locations exploring the entire parameter space. Then, the MCMC
exploration of the parameter space is carried out, but a multi-linear
interpolation between the pre-computed spectra is used to compute the model
SED at each step.

\begin{table}[h]
\centering
\label{table1}
\begin{tabular}{|c|c|}
\hline
SFH & Definition \\ \hline
Linearly increasing SFH (LinInc) & $SFR(t) \propto t$ \\ \hline
Constant SFH (CSF) & $SFR(t)$ = constant $\propto$ galaxy mass/age \\ \hline
Delayed SFH (Delay) & $SFR(t) \propto \frac{t}{\tau^{2}}e^{(-t/\tau)}$ \\ \hline
Exponentially decreasing SFH (Tau) & $SFR(t) \propto \frac{1}{\tau}e^{(-t/\tau)}~(\tau>0)$ \\ \hline
Exponentially increasing SFH (InvTau) & $SFR(t) \propto \frac{1}{\tau}e^{(-t/\tau)}~(\tau<0)$ \\ \hline
\end{tabular}
\caption{Definitions of five SFHs. SFR(t) is the instantaneous star formation
  rate and the star formation timescale, $\tau$, is the free parameter during
  the SED fitting, ranging from 0.1Gyr to 5 Gyr. }
\end{table}

\begin{figure}
\centering
\includegraphics[width=4in]{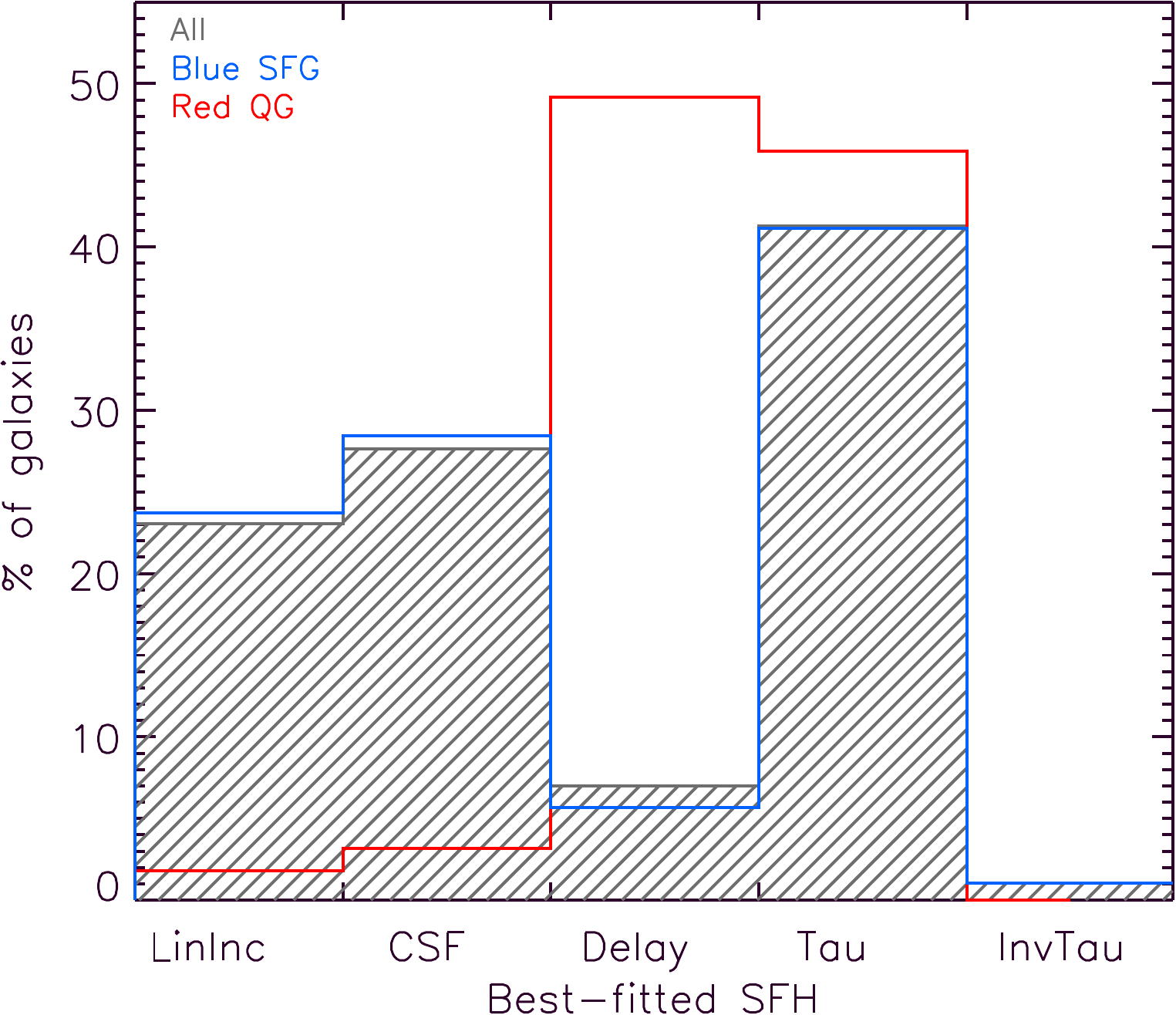}
\caption{The histogram of the percentage of galaxies having the Best Fit SFH
  as linearly increasing SFH (LinInc), constant SFH (CSF), delayed model
  (Delay), $\tau$-model (Tau) and inverted-$\tau$ model (exponentially
  increasing SFH; InvTau). The grey shaded histogram represents all galaxies,
  whereas blue and red are for rest-frame UVJ selected SFGs and QGs. The SEDs
  of most of the QGs are best-fit using decreasing model such as delayed and
  $\tau$-models, while SFGs have various SFHs.}
\label{fig:bf_sfh}
\end{figure}

One of the advantages of MCMC technique is that it provides posterior
distributions of the galaxy properties estimated from the code, allowing an
accurate computation of expectation values and confidence intervals even in
the case of non-Gaussian probability distribution functions. As a final
output, we adopt the mean values computed from the posterior distribution. In
most cases, the mean and the best-fit values are close, especially when the
posterior distribution is approximately Gaussian. However, the best-fit values
are not meaningful when the probability distribution is skewed or broad. In a
case of bimodal distributions, the mode of the probability distribution would
be a better estimate than the mean. But we find that less than 3\% have
bimodal distributions in any SFHs and that the differences between mode and
mean values of the bimodal distributions are not significant.

We fit the CANDELS multi--band photometry to the \cite{bru03} (BC03) spectral
population synthesis library, with a Chabrier initial mass function (IMF) and
metallicity fixed to the solar value. The Calzetti law \citep{cal00} is used
for the dust obscuration model, together with the Madau prescription for the
opacity of intergalactic medium (IGM) \citep{mad95}. Redshifts are fixed to
the CANDELS photo-z (or spec-z if available) during the fitting. We include
the flux from the nebular continuum and line emission by tracking the number
of Lyman-continuum photons and by assuming case B recombination. We then model
the empirical line intensities relative to H$\beta$ for H, He, C, N, O, and S
lines as a function of metallicity according to the prescription in
\cite{and03, sch09}. SpeedyMC performs the SED fitting on the three (or four)
dimensional parameter space defined by stellar mass, age, dust extinction and
e-folding time, $\tau$ (in the case of $\tau$-related SFHs). We define the stellar 
age as the median stellar mass weighted age of the galaxy, i.e., the lookback
time in which 50\% of the stellar mass has been built. 
Dust extinction (reddening) is parameterized by the color excess, E(B-V),
assuming the Calzetti dust absorption law. Overall, our procedure is
conceptually similar to that of \cite{pfo12} and \cite{mar10} 
to study galaxies at $1.4<z<2.9$, the key differences being that they used the
stellar population models by \cite{mar05} and did not include the nebular
emission. Our work also considers a wider choice of SFH and takes advantage of
the broader wavelength coverage and deeper sensitivity of the CANDELS/Deep
survey.

We present the best-fit star formation history (Best Fit
SFH) of individual galaxies determined by maximum likelihood obtained from the
SpeedyMC. Figure~\ref{fig:bf_sfh} depicts the fraction of galaxies having the
Best Fit SFH as one of five SFHs. Most of the quiescent galaxies classified
from the rest-frame UVJ diagram are best-fit with decreasing SFHs, either 
Delay or Tau SFHs. 47\% of SFGs are also best-fit with decreasing
SFHs, and the rest of them with increasing SFHs (LinInc, CSF, and InvTau).  
For SFGs at high redshift, Delay and InvTau SFHs have been suggested as more 
suitable functional forms than the Tau (Delay: \cite{lee10,spe14}; InvTau: 
\cite{mar10,pfo12}).  However, we show here that they are
not the preferred SFH in the SED fitting of SFGs at $1<z<4$. Only 5.7\% and
1.1\% of SFGs have the Delay and InvTau as the Best Fit SFH,
respectively. As we will further demonstrate in Section 3.1 and 3.2, our results
suggest that using only one SFH for all galaxies is not necessarily the
correct approach. 

\subsection{Comparison of SFRs with various SFR indicators}

Without dust extinction, SFR estimated from rest-frame UV light is more
sensitive than SFR estimated from IR or radio data by orders of magnitude
because it originates mainly from young, massive stars \citep{mad14}. However,
most energy radiated by young stars is heavily obscured by dust at least out
to $z=2.5$; \cite{whi17} showed that $>$50\% of star formation is
obscured at log(M$_{*}$/M$_{\odot}$)$>$9.4 (see also e.g., \cite{mag09, mur11, bou16}). 
Since IR emission
represents re-emitted UV emission from completely dust obscured stars,
correctly estimated total IR luminosity (L$_{IR}^{tot}$) is a key to measuring
the actual SFR of distant galaxies.  Currently, {\it Herschel} observations
enable us to {\it directly} measure re-processed star light, and hence the
total bolometric IR luminosity of individual massive
galaxies. But, a significant fraction of SFGs are missed at $z>1$ and
far-infrared (FIR) luminosity is limited to the brightest galaxies because of
the sensitivity limit of {\it Herschel} \citep{elb11}. In the absence of FIR
data, various SFR indicators have been used to {\it indirectly} estimate the
dust attenuation of individual galaxies \citep{dad07, kar11, sob12, lee11,
  arn13}.

\begin{figure}
\centering
\epsscale{0.8}
\plotone{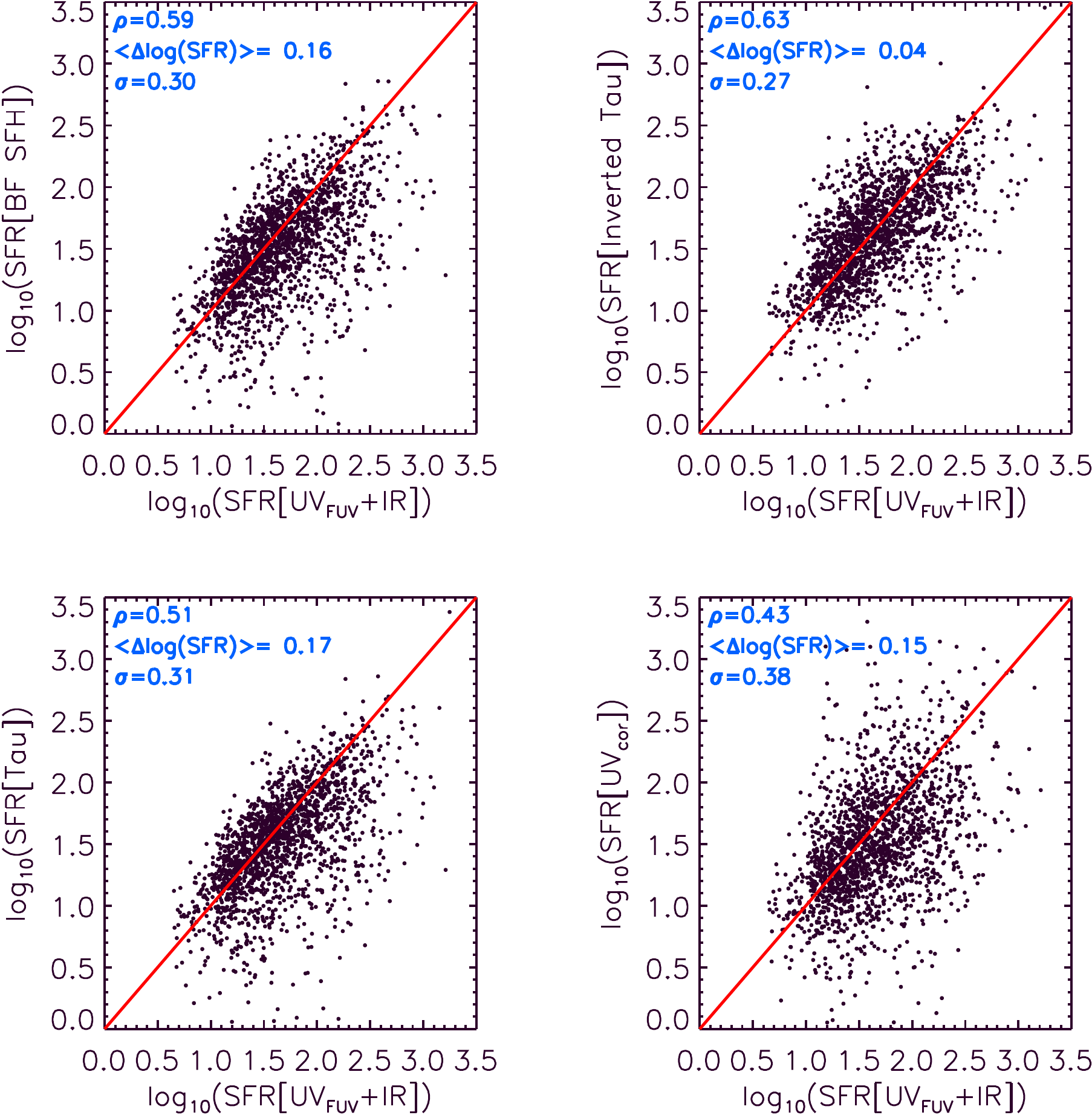}
\caption{Comparison of total SFRs for IR detected SFGs with SFRs obtained
  using the Best Fit SFH (BF SFH), $\tau$, inverted-$\tau$ models and dust-corrected UV
  SFR. The red line is for the one-to-one correlation. $\rho$, $<\Delta SFR>$,
  $\sigma$ are the Pearson correlation coefficient, average and mean absolute 
  deviation (MAD) of
  differences of log(SFR), respectively. When $\rho$ is close to unity, SFRs
  in x and y-axes are linearly correlated, and as $<\Delta SFR>$ and $\sigma$
  are close to zero, SFRs in x and y-axes are identical. Even though statistics 
  indicate that the SFR obtained using the inverted-$\tau$ model shows the best
correlation with SFR(UV$_{FUV}$+IR), the relatively good correlation between 
SFR obtained using the Best Fit SFH and SFR(UV$_{FUV}$+IR) are also shown in 
this comparison. }
\label{fig:sfr}
\end{figure}

The total SFR, adding SFR from IR and UV emission, might be the best indicator
of SFR for SFGs because it combines the light from the unobscured stars (UV)
and that which is re-processed by dust (IR). In Figure~\ref{fig:sfr}, we
compare four different SFR measurements to total SFR, $SFR(UV_{FUV}+IR)$,
for 2,006 SFGs that have 24$\mu$m and/or {\it Herschel} detections at $1<z<3$ and 
log(M$_{*}$/M$_{\odot}$)$>$9. We
only use galaxies at $z<3$ due to the sensitivity limits of the GOODS-{\it
  Herschel} observations \citep{elb11}. Here are the definitions of each SFR
we use in this analysis:

\begin{itemize}

\item \textbf{Instantaneous SFR}: The instantaneous SFR obtained from the SED
  fitting, including prescriptions of dust obscuration, is commonly used for
  galaxies lacking spectroscopic star formation tracers.  However, it is
  strongly dependent on the choice of SFH \citep{mar10}. In this study, the SFR is obtained
  from the best-fit SED assuming the Best Fit SFH for each galaxy. For
  comparison, we also show the instantaneous SFR obtained using the $\tau$ and
  inverted-$\tau$ models.

\item \textbf{SFR from Dust-corrected UV emission}: The dust-corrected UV SFR,
  SFR(UV$_{\mathrm{cor}}$), is computed using the conversion factor by \cite{con09} 
  with an assumption of the Chabrier IMF,
\begin{equation}
\mathrm{SFR (UV_{cor})} [\mathrm{M}_{\odot}/\mathrm{yr}]=0.82\times10^{-28}L_{1500}(erg/s/Hz). 
\label{eq:uv}
\end{equation}
We adopt the empirical correlation between dust obscuration and the slope of
the rest-frame UV of starburst galaxies \citep{meu19, cal00} to derive the 
dust-corrected UV luminosity at the rest-frame
$1500\AA$ ($L_{1500}$), and subsequently SFR(UV$_{\mathrm{cor}}$).

\item \textbf{Total SFR}. This SFR is a combination of light from both the UV
  and IR, SFR$_{tot}$= SFR(UV$_{FUV}$)+ SFR(IR) \citep{mad14}. SFR(IR) is defined as
\begin{equation}
\mathrm{SFR_{IR}}[\mathrm{M}_{\odot}/\mathrm{yr}]=1.09\times10^{-10}L_{\mathrm{IR}}(L_{\odot}), 
\label{eq:ir}
\end{equation}
where $L_{\mathrm{IR}}$ is the total infrared luminosity, $L(8-1000\mu
\mathrm{m})$.  We measure $L_{\mathrm{IR}}$ by fitting the mid or far-infrared
SED with IR templates for high redshift SFGs ($0.5<z<3$) introduced by
\cite{kir15}. We use a maximum of 3 bands 70$\mu$m from {\it Spitzer}/MIPS, 100$\mu$m, and
160$\mu$m from {\it Herschel}/PACS, if available. For galaxies only having 
24$\mu$m {\it Spitzer}/MIPS observation, we use 24$\mu$m to compute 
$L_{\mathrm{IR}}$. The
far-UV (FUV) components of star formation is derived from the observed FUV
luminosity at 1500\AA~with no correction for extinction, and the
SFR(UV$_{FUV}$) is subsequently estimated using the conversion factor by 
\cite{con09}. The overall conversion factor of the FUV and IR contribution
assumes the Chabrier IMF.
\end{itemize}

In Figure~\ref{fig:sfr}, we use the Pearson correlation coefficient ($\rho$),
the average ($<\Delta SFR>$), and the mean absolute deviation ($\sigma$) of
the differentiation of SFRs to test correlations between the different SFR
estimates. The SFRs obtained using the inverted-$\tau$ model show the best
correlation with SFR(UV$_{FUV}$+IR); these models have the largest
$\rho$ (0.63), smallest $<\Delta SFR>$ (0.04), and lowest $\sigma$
(0.36). However, despite these statistics, we note that only 1.1\% of SFGs are
formally best-fit with an inverted-$\tau$ model.  We also find that SFRs
obtained using the Best Fit SFH have a similarly large $\rho$ (0.59), only a
modest offset in $<\Delta SFR>$ (0.16), and relatively small $\sigma$ (0.30)
when compared to results obtained using the $\tau$-model or dust-corrected UV
SFR. This implies that the SFRs obtained using the Best Fit SFH are better 
correlated with total SFRs relative to the individual SFHs. Furthermore, we propose 
that the Best Fit SFH of each galaxy is a robust
approach that yields more accurate SED-derived SFRs for the overall galaxy
population.

\subsection{Validation with Simulated Galaxies}

\begin{figure}
\epsscale{0.65}
\plotone{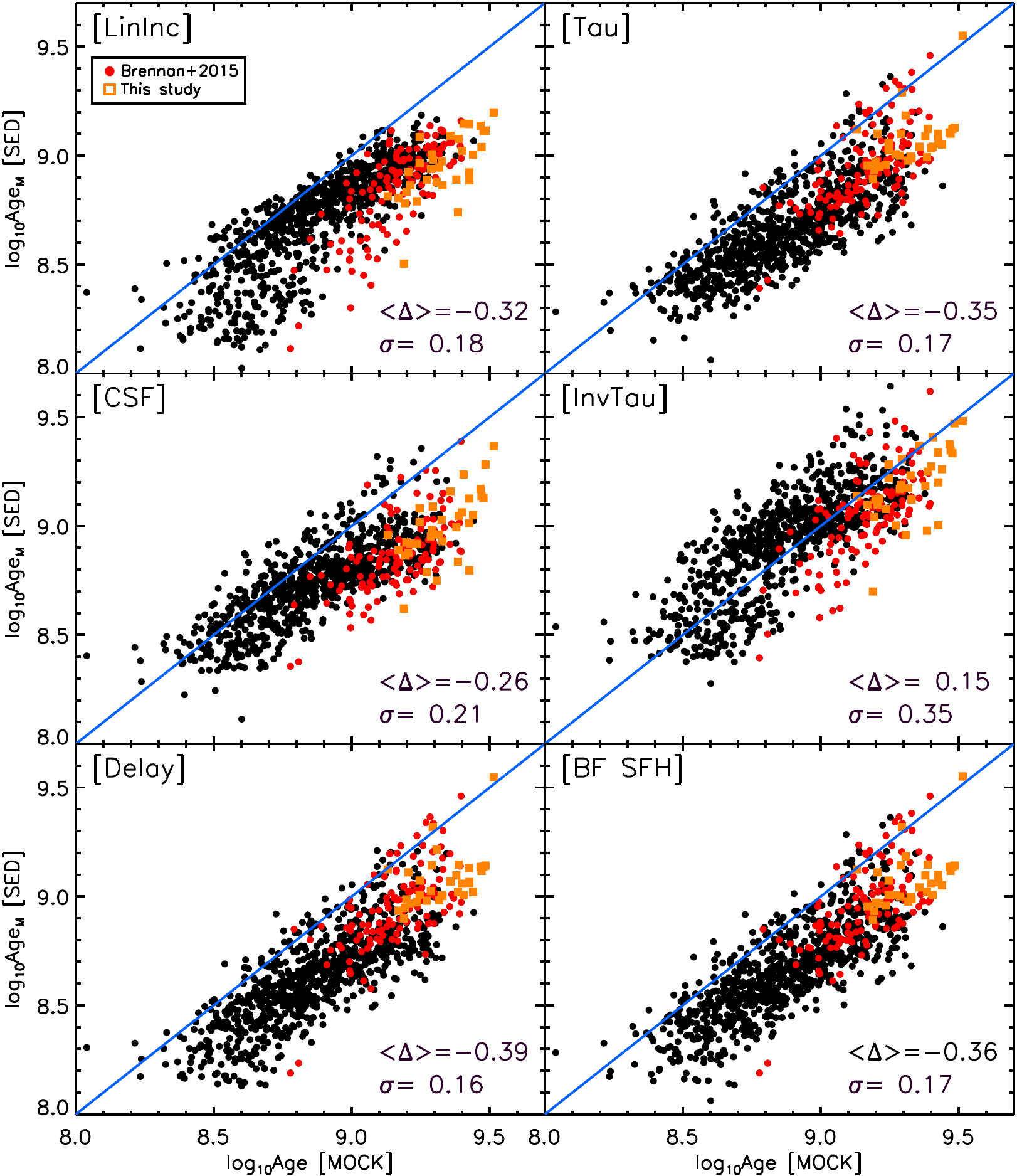}
\caption{Comparison of median stellar--mass weighted age (Age$_{M}$ [SED]) obtained 
from SpeedyMC to the intrinsic age from the simulation (Age[MOCK]) for 
mock galaxies. Blue lines indicate a linear correlation. $<\Delta>$ and $\sigma$ are 
the mean and mean absolute deviation of 
(Age$_{\mathrm{derived}}$-Age$_{\mathrm{intrinsic}}$)/Age$_{\mathrm{intrinsic}}$, 
respectively. Quiescent galaxies are classified as galaxies having $R_{SB}  < 0.25$ 
(red circles, 150 galaxies) following the definition from \cite{bre15} and 
$R_{SB} < 1/30$ (orange squares, 33 galaxies) from this study (see Section 4.2). 
Based on $<\Delta>$ , ages obtained using InvTau are the closest to the 
intrinsic absolute ages, while the deviations ($\sigma$) are the largest. 
Though the relative ages are generally 
robust, all models underestimate the absolute ages.}
\label{fig:mock_age}
\end{figure}

Next, we test how well we can recover the various best--fit galaxy properties
from SpeedyMC, in particular the SFH, stellar age, SFR and stellar mass, by
using a sample of artificial galaxies created using simulations of 
CANDELS light cones (Somerville et al., in preparation), based on a
combination of the ROCKSTAR halo catalogs by \cite{beh13} extracted from the
Bolshoi dark matter N--body simulations of \cite{kly11}, together with the
semi--analytic models (SAMs) of \cite{som12}, dubbed the ``Santa Cruz
models''. The SAMs implement gas cooling, star formation, and the growth of
supermassive black holes, as well as including the effects of merging,
stellar--driven winds, and black hole feedback. The synthetic magnitudes of
the galaxies are computed using BC03 models convolved with the star formation
history and chemical enrichment evolution of each galaxy as predicted by the
SAM, assuming a Chabrier IMF. The Santa Cruz models implement the effects of
dust based on a two-component model of the extinction \citep{cha00} including
diffuse cirrus in the disk and the dense birth clouds surrounding new
stars. To derive the actual extinction, a `slab' model is used to compute the
inclination dependent extinction (see \cite{som12} for further details). The
total amount of energy absorbed by the dust is assumed to be re-emitted in the
IR using the templates of \cite{cha01} to determine the SEDs of the dust
emission.

\begin{figure}
\epsscale{0.65}
\plotone{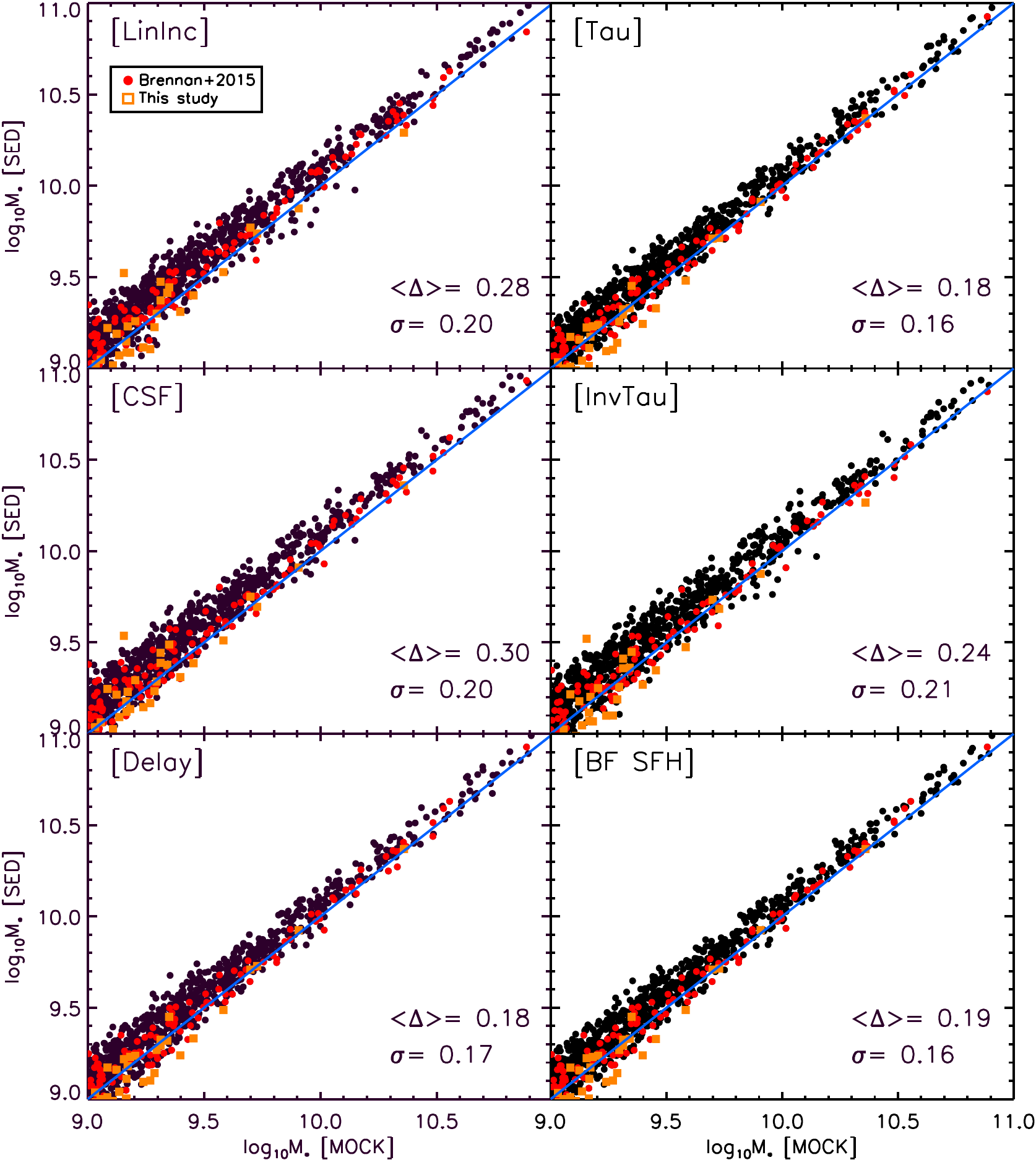}
\caption{Comparison of stellar masses (M$_{*}$ [M$_{\odot}$]) obtained from
  SpeedyMC relative to simulation, with red and orange points
    representing QGs classified using different definitions (\cite{bre15} and
    this study, respectively) and black points SFGs. The M$_{*}$ obtained
    using five SFHs strongly correlates with the intrinsic M$_{*}$, indicating
    that the stellar mass estimation is insensitive to the assumed SFH and
    galaxy types, i.e., SFGs and QGs. Generally, the derived stellar mass
    overestimates the intrinsic stellar mass about $\sim$0.2 dex (or less).}
\label{fig:mock_mass}
\end{figure}

\begin{figure}
\epsscale{0.65}
\plotone{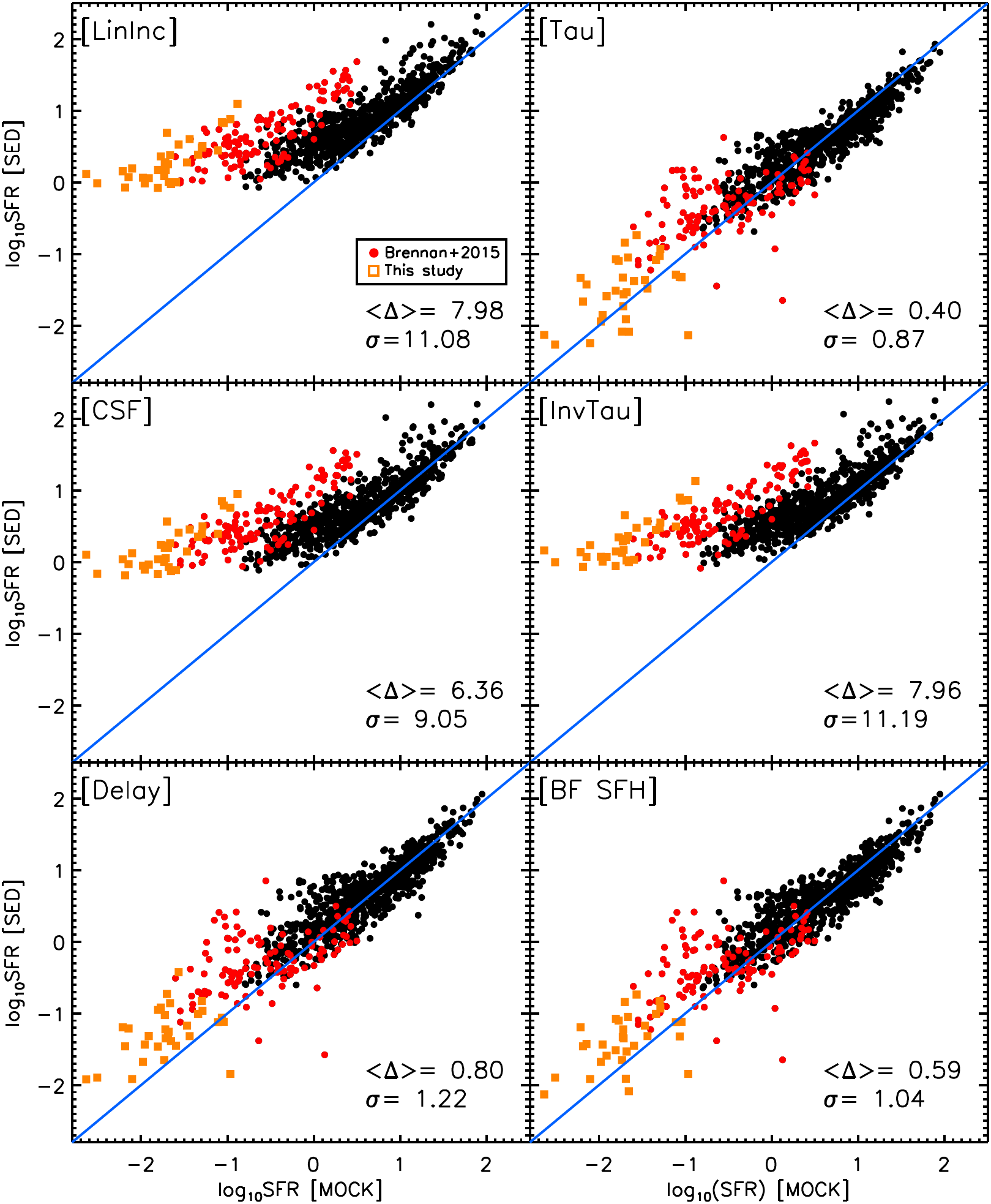}
\caption{Comparison of SFRs obtained from SpeedyMC relative to simulations,
  with red and orange points representing QGs classified using different
  definitions (\cite{bre15} and this study, respectively) and black points
  SFGs. SFRs obtained using Tau show the best correlation with intrinsic SFRs
  for both the mock SFGs and QGs.}
\label{fig:mock_sfr}
\end{figure}

\begin{figure}
\epsscale{1.0}
\plotone{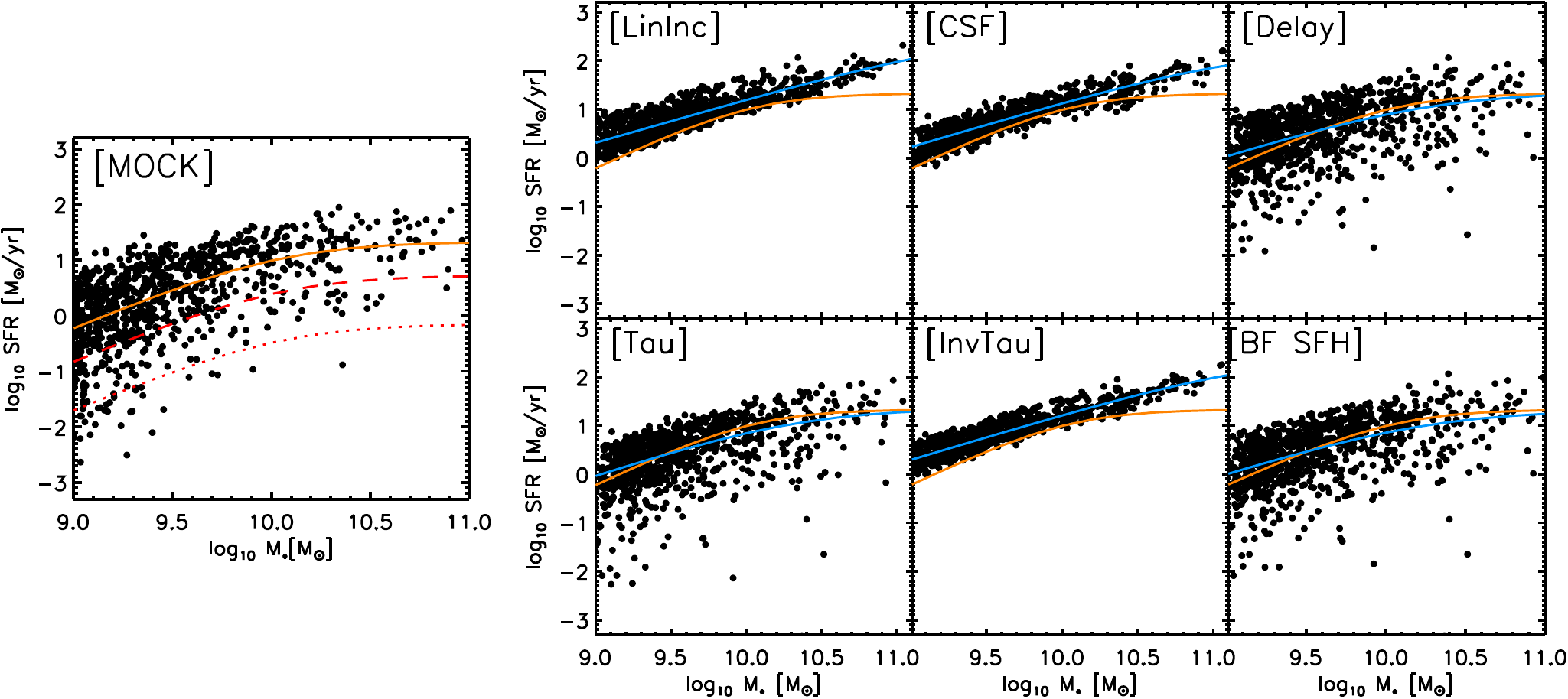}
\caption{log(SFR) vs. log(M$_{*}$) of simulated galaxies from the CANDELS/GOODS-S mock
  catalog. Left: SFR and M$_{*}$ are intrinsic values from the
  mock catalog.  The fit of the MS is shown as orange line and the slope is about 1.1 below 
  the turnover mass ($\sim10^{10}M_{\odot}$) and 0.32 above the turnover mass. The red dashed line 
  represents $R_{SB}=0.25$ and the red dotted line is for $R_{SB}=1/30$. Galaxies below these lines 
  are considered as quiescent galaxies in \cite{bre15} and this study (see Section 4), respectively.
  Right: SFR and M$_{*}$ are obtained using five SFHs and the Best Fit 
  SFH (BF SFH) and the blue lines are the derived MS slopes. The intrinsic MS slope
  (orange) is over-plotted for a comparison. The MS slope using Delay and Tau, and Best Fit SFH 
  are close to the intrinsic slope, indicating that intrinsic properties are best recovered when assuming 
  those SFH models.}
\label{fig:mock}
\end{figure}

Specifically, for our test of SpeedyMC we use the CANDELS/GOODS--S mock
catalog, from which we have randomly extracted 1000 galaxies with the same
selection criteria we have used for the observed galaxies in Section 2, 
redshift in the range $1<z<4$, stellar mass
  $>10^{9}M_{\odot}$. To estimate the photometric errors in each band for a
  simulated galaxy, we extract a random gaussian variable using median errors
  and scatters of each observed band in the GOODS-S field. We then use them as
  the photometric errors of simulated galaxies in this MOCK catalog and run
  SpeedyMC for those 1000 galaxies. True synthetic fluxes and the randomly 
  extracted photometric errors are being used as input into SpeedyMC. Note that, in principle, 
  it is suggested to perturb the synthetic fluxes by some photometric errors to properly 
  simulate observed fluxes of galaxies. However, this only slightly increase the observed scatters in 
  Figure~\ref{fig:mock_age}-\ref{fig:mock} and leave the qualitative results unchanged.
    
  After excluding 32 galaxies that have SpeedyMC quality flag $=3$,
  we compare the ``observed'' output parameters
from the Best Fit SFH and five SFHs (age, stellar mass, and SFR) with the
input values from the mock catalog. We do not perform a detailed comparison of
the Best Fit SFH with the intrinsic one because, in general, the latter is
rather different from our simple five analytic functions and the focus of this
paper is not on the reconstruction of the SFH of the galaxies. Rather, we use
the comparison of the derived stellar age, M$_{*}$, and SFR with the intrinsic
values to quantify the effectiveness of the fit, including the determination
of an approximate simple SFH that is capable of returning robust measures of
SFR and stellar age.

We first consider the stellar age, as this is the parameter that most directly
relates to the SFH. We initially considered two estimators of the stellar age,
the age since the onset of star formation(Age$_O$), and the median stellar
mass--weighted age (Age$_M$). After verifying that Age$_M$ correlates
  with intrinsic age from the MOCK significantly better than Age$_O$ (smaller
  scatters in all cases of SFHs), however, we have used the median stellar
  mass--weighted age for subsequent analysis. In Figure~\ref{fig:mock_age},
we show the comparison of the input Age$_M$ in the case of the Best Fit SFH
(bottom right panel) relative to when the SFH is forced to be only one of the
five analytical models. Red circles and orange squares represent
  quiescent galaxies (QGs) selected from different definitions. Normally, the
  dividing cut which made for observed galaxies does not work well for model
  galaxies because the distribution of specific star formation rates (and the
  distribution of galaxies on the color-color diagram) in the model is not
  bimodal. We, therefore, cannot apply the same UVJ color-color selection to
  the model galaxies. \cite{bre15} defined the quiescent galaxies as having
  less than 25 per cent of the sSFR of the main-sequence (MS) line for both
  observations (CANDELS) and model galaxies (SAMs). Adopting their definition,
  we separate quiescent galaxies from star-forming ones based on their
  distances from the MS defined as, $R_{SB} =$ sSFR/sSFR of the MS \citep{elb11}
  (see Figure~\ref{fig:mock} for the MS slope estimation of 
  mock galaxies). Two QG selection cuts used here are 1) $R_{SB}<0.25$ 
  from \cite{bre15} (red circles) and 2) $R_{SB} <1/30$ from Section 4.2 in 
  this study (orange squares).  As a result, we classify 150 and 33 
  quiescent galaxies among 968 mock sample, respectively.

In Figure~\ref{fig:mock_age}, there is a clear correlation between the input
and output age, irrespective of the adopted SFH. The robustness of the
measure, however, varies with the SFH and with galaxy type, namely SFGs (black
points) or QGs (red and orange points). In all cases, the ages of QGs are 
underestimated by $\sim$0.1--0.4 dex. Forcing the SFH to the Lininc,
Constant or InvTau results in larger scatters to both SFGs and QGs, and even
worse for QGs. This is not surprising, as these adopted SFHs are not typical
of QGs. While the InvTau model produces stellar mass-weighted ages for SFGs
that are closest to reality (i.e., closest in absolute age), there is a
significantly larger scatter. 
Apparently, the ages derived with the Tau, 
Delay and Best Fit SFHs perform
better, preserving the relative age of both QGs and SFGs. We cannot
differentiate between the overall quality of Age$_M$ from these three SFHs, as
they all have similar order offsets in terms of absolute age and a small
intrinsic scatter. 
The range of recovered age versus input age is located
parallel to, but downshifted by $\sim$0.5 dex relative to the diagonal line
(where recovered age equals input age). In other words, while we can robustly
recover the relative ages, we underestimate the absolute age by a constant
amount (in log--log scale). It may be that this systematic offset relative to
the absolute age is due to our simplifying assumption that the metallicity is
fixed to a constant value during the fit. However, one could use the results
of these comparisons with simulated galaxies, where we know ``truth'', to
correct to the absolute ages.  In this work, we are not so much interested in
the absolute age dating of the galaxies as we are in the relative ones. We
therefore proceed estimating age differences adopting the mass--weighted age
estimated by the Best Fit SFH for the analysis of the real data.

Next, we consider how well we can recover the stellar mass of galaxies using
the various SFHs. In Figure~\ref{fig:mock_mass}, we compare the measured
stellar mass of the simulated galaxies for each of the five SFHs and
the Best Fit SFH relative to the input values. This comparison confirms what
has been found in many previous studies: the stellar mass is the most robust
recovered parameter by SED fitting procedures (e.g. \cite{muz09, lee10,san15}) 
and it is largely degenerate to the assumptions of
SFH. We do observe that stellar masses of both SFGs and QGs are nicely 
recovered by the SED fit, while there is a systematic bias toward large 
value of $\sim$0.2 dex or less; this result is largely independent of the adopted 
SFH. The independence of the measured stellar mass from the assumed SFH for the
simulations is also clearly observed in the case of real galaxies, for which
the stellar masses for each assumed SFH are compared against each other 
(full discussion forthcoming in Lee et al. in preparation).

Figure~\ref{fig:mock_sfr} shows that the star--formation rate, on the other
hand, does indeed depend on the assumed SFH. We also see this 
result evidently from our comparison of the measurements of real galaxies for 
each the five SFHs considered relative to each other. For the simulations, 
the recovered SFR assuming LinInc, CSF, and InvTau tends to overestimate 
the intrinsic values, with the largest deviations occurring at the low end of the 
SFR distribution for both SFGs and QGs. 
Similar to the age measurements, the best measures are obtained
by either assuming the Tau, Delay or Best Fit SFH, with Tau being the best of
the three. It is interesting to note that the SFRs of QGs are generally overestimated 
for all SFHs, with increasing SFHs showing the largest deviations.  
These are the same galaxies that are forced to rely on SED SFRs, as they are far below
the detection limits of the deepest existing IR surveys.  While the FUV+IR SFRs
are generally considered robust for SFGs, they are likely upper limits for
QGs.  For example, the \emph{Spitzer}/MIPS 24$\mu$m calibrations tends to
overestimate the SFRs for galaxies with log sSFR$<$-10 yr$^{-1}$ (e.g., 
\cite{hay14, uto14, fum14}).  The results
here suggest that SED SFRs also tend to overestimate the SFRs for this same
population of galaxies.

As a final test, we evaluate the ability of our SED fitting procedure to
reconstruct the distribution of the mock galaxies in the log(SFR)
vs. log(M$_{*}$) plane.  In Figure~\ref{fig:mock}, we plot the
log(SFR)-log(M$_{*}$) diagram using intrinsic values from the mock catalog
(left panel) and derived values assuming different SFHs (right panel). The 
red lines in the left panel represent different QG selection cuts, 
dashed : $R_{SB}<0.25$ \citep{bre15} and dotted: $R_{SB}<1/30$ (this study), 
respectively. We then measure the main sequence (MS) using an 
equation 4, where the MS slope of the 
intrinsic values (orange) is about unity below the turnover 
mass ($\sim10^{10}M_{\odot}$) and becomes flatter (about 0.32) at larger mass, and 
the blue lines are the best-fit slopes for each respective SFH.
Apparently, the MS slopes using Tau and Delay SFHs are close to 
the intrinsic slope (orange) with a bended MS even though the turnover 
mass is slightly larger ($\sim 10^{10.3}M_{\odot}$). When
factoring in the dispersion in addition to the marginal differences in slope
and normalization, it becomes clear that the intrinsic properties of the MS
are best recovered when assuming the Tau and Delay SFHs. 
In cases of LinInc, CSF, and InvTau, the MS slope is rather explained 
by a single power-law (linear fit) and the galaxy distribution in the 
log(SFR) and log(M$_{*}$) plane is totally different from the intrinsic one.
As expected, the SFRs of QGs are drastically
overestimated when using increasing SFHs (LinInc, CSF, InvTau); the properties
of QGs cannot be recovered when adopting these SFHs. It appears that the different 
SFH clearly affect the scatter in the log(SFR) and log(M$_{*}$) relation.

The correlation between log(SFR) and log(M$_{*}$) using the Best 
Fit SFH (last plot in the right panel) are similar to ones using Tau and Delay SFHs because 
most of galaxies have these decreasing SFHs as their Best
Fit SFH. While adopting the Tau 
SFH in particular is not a poor choice overall, the comparisons with the
simulation further support our results in Section 3.1 that the Best Fit SFH
better constrains the intrinsic galaxy properties relative to those obtained
using one simple SFH. In particular, we find significant systematic biases in 
the overall population when adopting either the CSF, LinInc or InvTau SFHs.

\section{log(SFR)--log($M_{*}$) diagram at $1.2\le${\MakeLowercase{z}}$<4$ in CANDELS}
\label{sec:sfr}

\begin{figure}
\epsscale{0.6}
\plotone{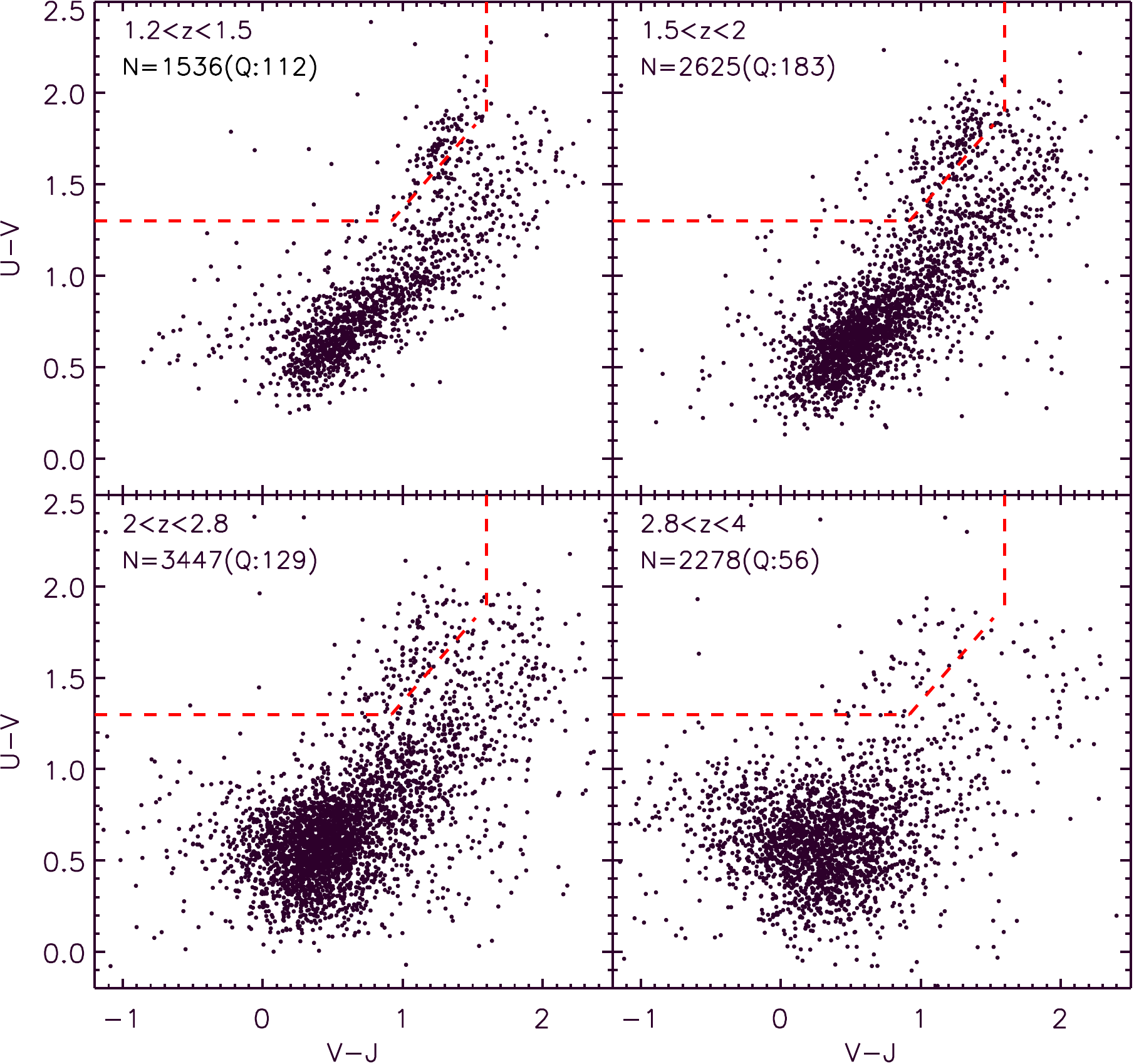}
\caption{Rest-frame U--V vs. V--J color diagrams at 4 redshift bins ($\Delta t \sim 1Gyr$), $1.2<z<1.5$,
  $1.5<z<2$, $2<z<2.8$, $2.8<z<4$. Red quiescent and bluer star--forming
  galaxies are classified by their rest-frame (U-V) and (V-J) colors with
  Equation~\ref{eq:uvj} (red dashed line).  The number indicates the total
  number of galaxies included in each redshift bin (Q stands for the number of
  quiescent galaxies). }
\label{fig:uvj}
\end{figure}

We now explore the main sequence of star-formation (MS) at $1.2 \le z<4$ using the 
robust estimate of stellar masses and SFRs of galaxies in CANDELS. 
For our analysis of galaxies in the log(SFR) and log(M$_{*}$) plane, we use 9,888 galaxies 
at $1.2 \le z<4$ after limiting our sample with stellar mass, 
$M_{*}>10^{9}M_{\odot}$ (over 80\% completeness limit as shown in Figure~\ref{fig:complete}).
Among them, about 17\% 
have IR detections. The rest-frame UVJ diagrams in four redshift bins with $\Delta t\sim1$ 
Gyr are shown in Figure~\ref{fig:uvj}. Using Equation 1 in Section 2.1, we distinguish 
about 6\% quiescent galaxies from our galaxy sample at $1.2 \le z <4$. The fraction of 
quiescent galaxies is mass dependent. Above a stellar mass limit of $10^{9}$M$_{\odot}$ (
$3\times10^{10}$M$_{\odot}$), the quiescent fraction varies from 9$\%$ (27\%) 
at $1.2<z<1.5$ to 2.6$\%$ (30\%) at $2.8<z<4$.

\subsection{The Main Sequence of Star Formation}

\begin{figure}
\epsscale{1.0}
\plotone{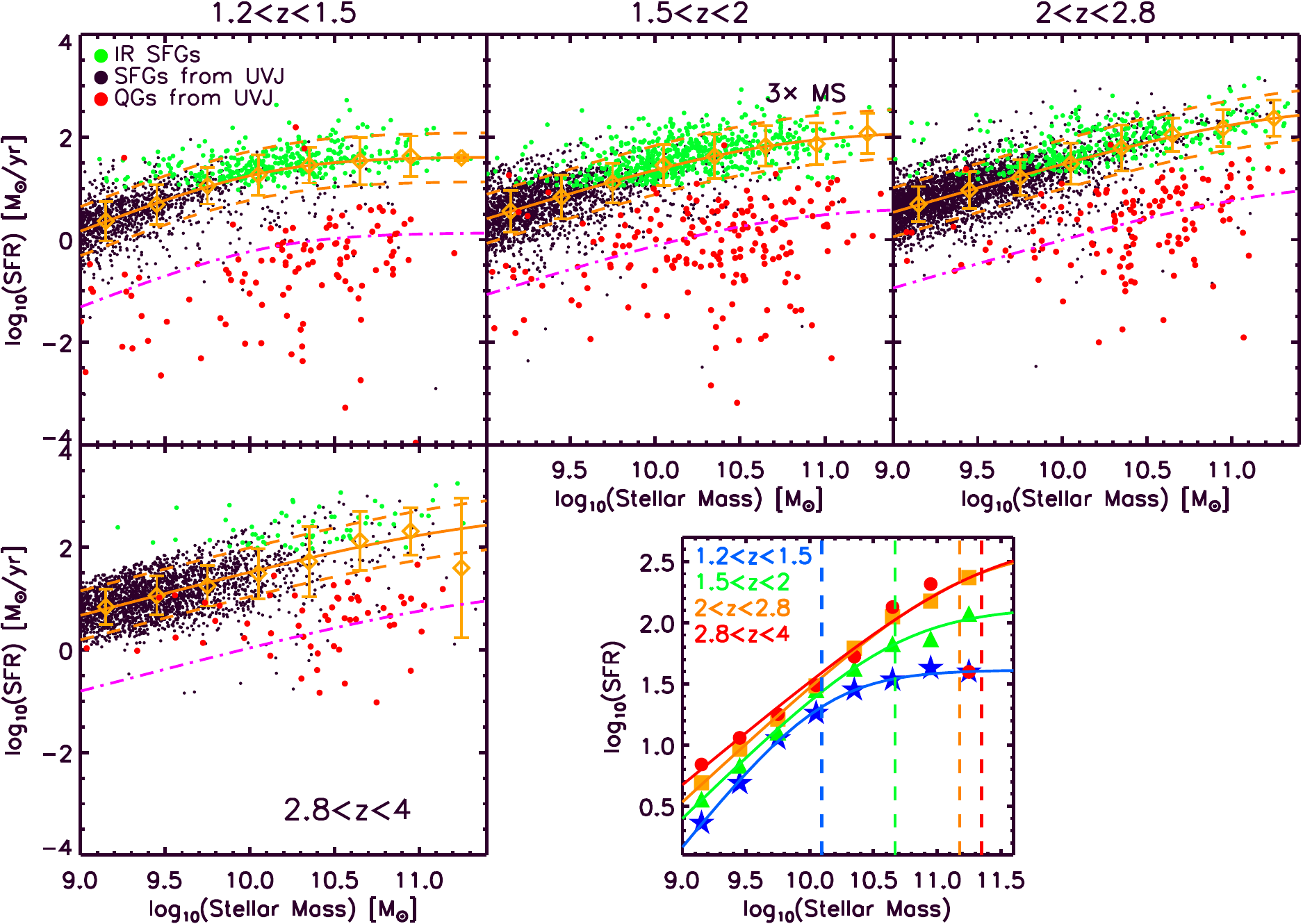}
\caption{log(SFR) vs. log(M$_{*}$) at four redshift intervals corresponding to
  $\Delta t\sim1$ Gyr. We fit the MS {\it only} using SFGs with SFR and
  M$_{*}$ estimated from the SED fitting using the best-fit SFHs (black).
  SFR(IR+UV$_{FUV}$) is used in the case of IR detected galaxies at $z<3$
  (green). The red points are QGs classified from the UVJ colors. The
  orange line and the error bar represent the MS slope estimated from Equation
  4 and the standard deviation of log(SFR) in each stellar mass bin ($\Delta
  log(M_{*})=0.3$), respectively. The orange dashed lines are the factor of three above and
  below the MS slope. The magenta line is the boundary adopted to identify
  quiescent galaxies relative to the MS. The final panel shows the fitting results obtained from
  Equation 4 for the four redshift bins, $1.2<z<1.5$ (blue), $1.5<z<2$
  (green), $2<z<2.8$ (orange), $2.8<z<4$ (red). The vertical dashed line is
  the turnover mass, $M_{0}$, measured for each redshift bin. More details on
  the MS can be seen in Figure~\ref{fig:comp_ms}. }
\label{fig:ms}
\end{figure}

Figure~\ref{fig:ms} illustrates log(M$_{*}$) as a function of log(SFR), 
sliced into four redshift bins (where each bin has $\Delta
t\sim1$ Gyr). Red and black points represent QGs and SFGs classified via
rest-frame UVJ colors (Figure ~\ref{fig:uvj}), respectively. We use the
$SFR(UV_{FUV}+IR)$ for IR detected galaxies (green) at $1.2<z<3$ and the SFR
obtained using the Best Fit SFH for the rest of non-IR detected
galaxies. Stellar mass is measured using the Best Fit SFH as well. Orange
points and error bars represent the mean and dispersion of SFRs of SFGs distributed in
a stellar mass bin, $\Delta=$0.3dex, computed using bisquare
weighting. A tight main sequence of SFGs exists at all explored redshifts.  To
parameterize the main sequence, we fit the mean log(M$_{*}$) and log(SFR) at
each redshift bin with a polynomial model provided by \cite{lee15},

\begin{equation}
S=S_{0}-log\Big[1+\Big(\frac{10^{M}}{10^{M_{0}}}\Big)^{(-\gamma)}\Big]
\end{equation}

where $S=log(SFR)$ and $M=log(M_{*}/M_{\odot})$. $\gamma$ is the power-law
slope at lower stellar masses and $S_{0}$ is the maximum value of $log(SFR)$
that the function is asymptotically approached at higher stellar masses.  In
particular, we use this model to quantify the turnover mass, $M_{0}$, which is
a break of the power-law slope (for a detailed explanation of the model, see
section 4.1 of \cite{lee15}). The mean SFR in each mass bin is plotted with a
MS fit and $M_{0}$ in the last panel of Figure~\ref{fig:ms}. In agreement with
earlier studies, we show that there is a break of the MS at $M_{0}$
\citep{whi14,lee15} and observe an evolution of $M_{0}$ to larger values with
increasing redshift \citep{sch15,tom16}.  A broken power-law is used to
measure the best-fit MS slope below and above the $M_{0}$. The MS slope
measurements are listed in Table 1. While the slope is nearly unity
($\beta_{1} \sim 0.85-1.0$) below the turnover mass, it becomes flatter ($\sim 0.2-0.3$) 
above $M_{0}$ at $z<2$. This bending of the MS has been explored from the local to $z\sim4$
universe \citep{whi14, lee15, gav15, tom16}.  $\beta_{1}$ is slightly
shallower than the finding of \cite{whi14} at $1.5<z<2.5$, but agrees well at
$1.2<z<1.5$. This may be because they fixed $M_{0}$ to
log(M$_{*}$/M$_{\odot}$)=10.2 at all redshifts. At $z>2$, the slope can be rather explained 
by a single power-law with $\beta_{1} \sim 0.8$. 

\begin{figure}
\epsscale{0.8}
\plotone{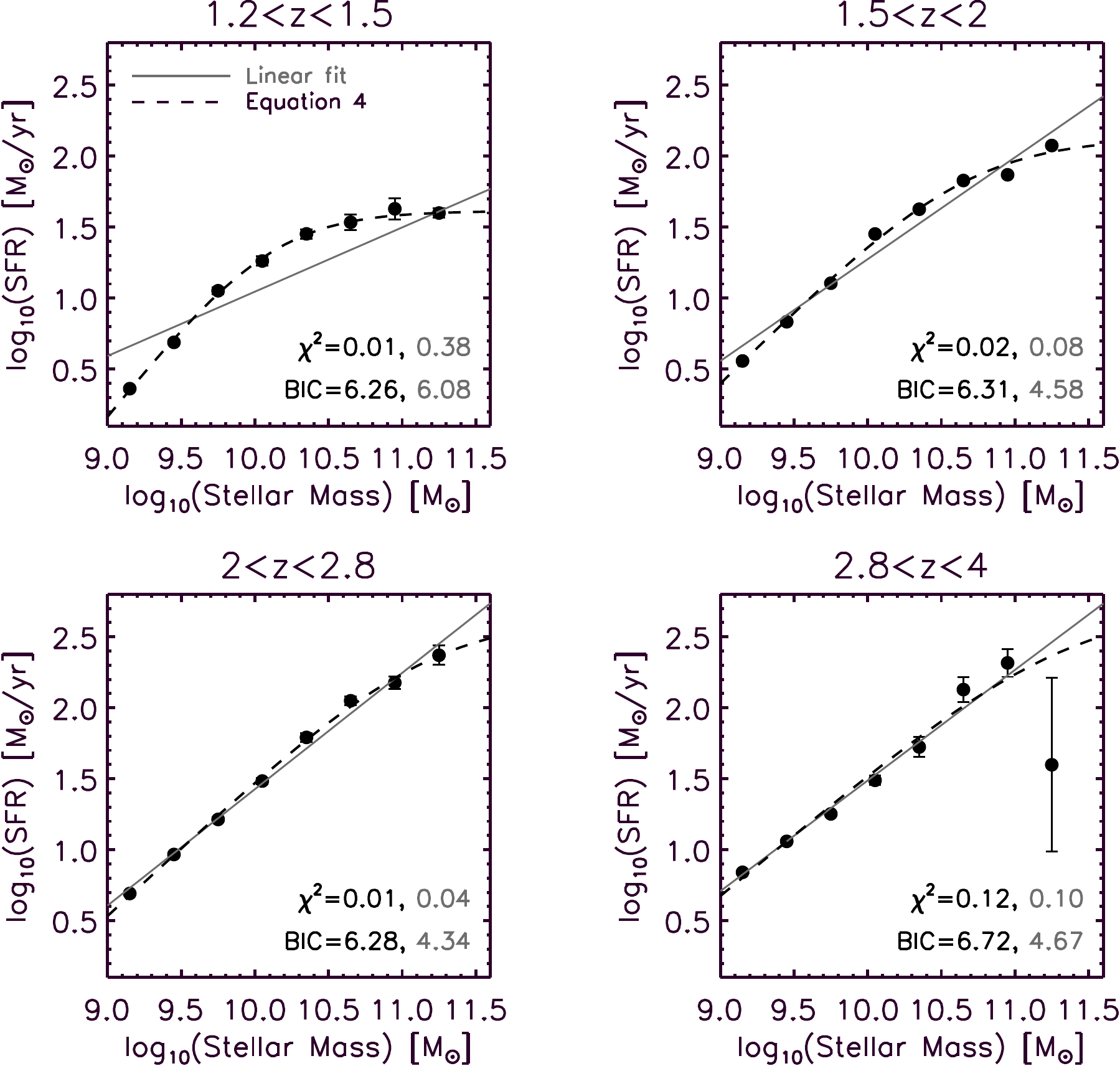}
\caption{Comparison of the linear fit and polynomial fit (Equation 4) to the
  MS in the log(SFR)--log(M$_{*}$). The points and error bars represent
  the mean and standard error ($\sigma / \sqrt{N}$) of SFRs in each stellar
  mass bin. We fit the data (black points) to the single power-law (linear
  fit, gray) and Equation 4 (dashed black), and compute a reduced $\chi^{2}$ 
  and Bayesian Information Criteria (BIC) between the fit and data using 
  $\sigma$ of the MS from Figure~\ref{fig:ms}. There is no significant
  difference between two fittings.}
\label{fig:comp_ms}
\end{figure}

A recent paper by \cite{ren15} suggested that the MS slope is better explained
by a single power-law at $z\sim0$, in contrast to Figure 2 in \citep{gav15}.
In Figure~\ref{fig:comp_ms}, we compare the linear (gray) and analytic
  function fit (black dashed line: Equation 4) to investigate how the MS fit
differs from the single power-law. Although the analytic function fit 
is almost identical to the linear fit at lower stellar masses, we find an 
analytic function to be a better fit to the decreasing SFRs at
highest masses. Because it is hard to discriminate visually, particularly in
the case of the Equation 4 having an extra parameter ($M_{0}$), we test which
model is a better fit using the reduced $\chi^{2}$ and the Bayesian
  information criterion (BIC). The BIC is defined as $\chi^{2}+kln(n)$, where
  k is the number of model parameters and n is the number of data points.
Figure~\ref{fig:comp_ms} shows that the linear fit has lower BIC than the
polynomial fit over all redshifts.  However, the differences between the BIC
values are less than two, indicating that the two models are
similar. Also, the difference of reduced $\chi^2$ values between
  linear and Equation 4 is insignificant except in the redshift bin
  $1.2<z<1.5$, which has a
  smaller reduced $\chi^2$ values with an analytic function fit. Note the 
  small SFR dispersion in the highest mass bin at $1.2<z<1.5$ due to three 
  galaxies that accidentally have similar
  SFRs, which might be responsible for the apparent downward bending of the
  MS. We believe that the bending is real, though, since we still observe the
  decline of the MS slope in the high mass bin when we increase
  the bin size to have better statistics.
Based on this result, we suggest that a single power-law can generally explain
MS with and without the bending of the MS, but the analytic function would be
suitable to investigate the curved MS at high masses.

\begin{table}[h]
\label{table2}
\centering
\begin{tabular}{|cccccc|}
\hline
& {\footnotesize \# of SFGs (IR) /QGs} & {\footnotesize $M_{0}$} & {\begin{tabular}[c]{@{}c@{}} \footnotesize $\beta_{1}$\\ \footnotesize $(M<M_{0})$\end{tabular}} & { \begin{tabular}[c]{@{}c@{}} \footnotesize $\beta_{2}$\\ \footnotesize $(M>M_{0})$\end{tabular}} & {\footnotesize $\sigma$ of MS/IQR of MS} \\ \hline
\footnotesize$1.2 \leq z<1.5$ & \footnotesize 1424 (367) /112 & \footnotesize10.09 & \footnotesize 1.08$\pm$0.06 & \footnotesize 
0.18$\pm$0.03 & \footnotesize 0.35 / 0.37\\ \hline
\footnotesize $1.5\leq z<2$ & \footnotesize 2442 (637) /183 & \footnotesize 10.67 & \footnotesize 0.84$\pm$0.04 & \footnotesize 0.28$\pm$0.11 & \footnotesize 0.40 /0.49 \\ \hline
\footnotesize $2\leq z<2.8$ & \footnotesize 3318 (537) /129 & \footnotesize 11.18 & \footnotesize 0.86$\pm$0.03 & & \footnotesize 0.37 /0.40 \\ \hline
\footnotesize $2.8\leq z<4$ & \footnotesize 2222 (68**) /56 & \footnotesize 11.35 & \footnotesize 0.78$\pm$0.07 & & \footnotesize 0.58(0.47)* /0.77(0.64)* \\ \hline
\end{tabular}
\caption{First column: the number of SFGs (IR detections) and QGs. 
  Second column: turnover mass, $M_{0}$.  Third and fourth column:
  $\beta_{1}$ and $\beta_{2}$ are the MS slope below and above the $M_{0}$,
  respectively.  Fifth column: $\sigma$ is the mean value of the
  $log_{10}(SFR)$ dispersions (IQR: inter-quartile range).  *The number in the
  parenthesis indicates a $\sigma$(or IQR) of the MS excluding the highest
  mass bin having the largest SFR dispersion with only five galaxies. **Note
  again that we do not include IR detected galaxies at $z>3$. }
\end{table}

At $z<2.8$, the observed dispersion of of the log(SFR)-log(M$_{*}$) relation
(MS dispersion) is consistently measured to be $\sigma \sim 0.35-0.4$ dex at
all masses. The MS dispersions obtained in this work are analogous to ones
reported by \cite{lee15}. At $z>2.8$, the MS dispersion is 0.47 dex. It is
possible that increased uncertainties in the photometric redshifts and
consequently all derived physical parameters (i.e., SFR) at high redshifts
lead to this larger dispersion. However, it is also the case that the
adoption of different SFHs in the SED fitting can induce a larger MS
dispersion \citep{sal15, cas16}.  This may be why our measured MS dispersion
is larger than results based on UV+IR SFRs alone \citep[e.g.,][]{whi12a,sch15}.
The overall normalization of the MS increases with redshift (see the last
panel of Figure~\ref{fig:ms}). It has been suggested that this higher sSFR of
distant galaxies is related to their larger gas fractions at high redshifts
\citep{sar14}.

Overall, our results show a good agreement with other studies which explored the
same epoch with different samples and SFR indicators \citep{kar11, whi14, sch15, tom16}; 
all of these studies report that the MS slope
below a certain stellar masses ($>2\times10^{10}M_{\odot}$ at $z>1.2$) is
close to unity, 0.8--1.0, while the slope dramatically declines for massive
galaxies starting around $z=2$.

\subsection{The Properties of Galaxies in the log(SFR)--log($M_{*}$) Diagram }

\begin{figure}
\epsscale{0.8}
\plotone{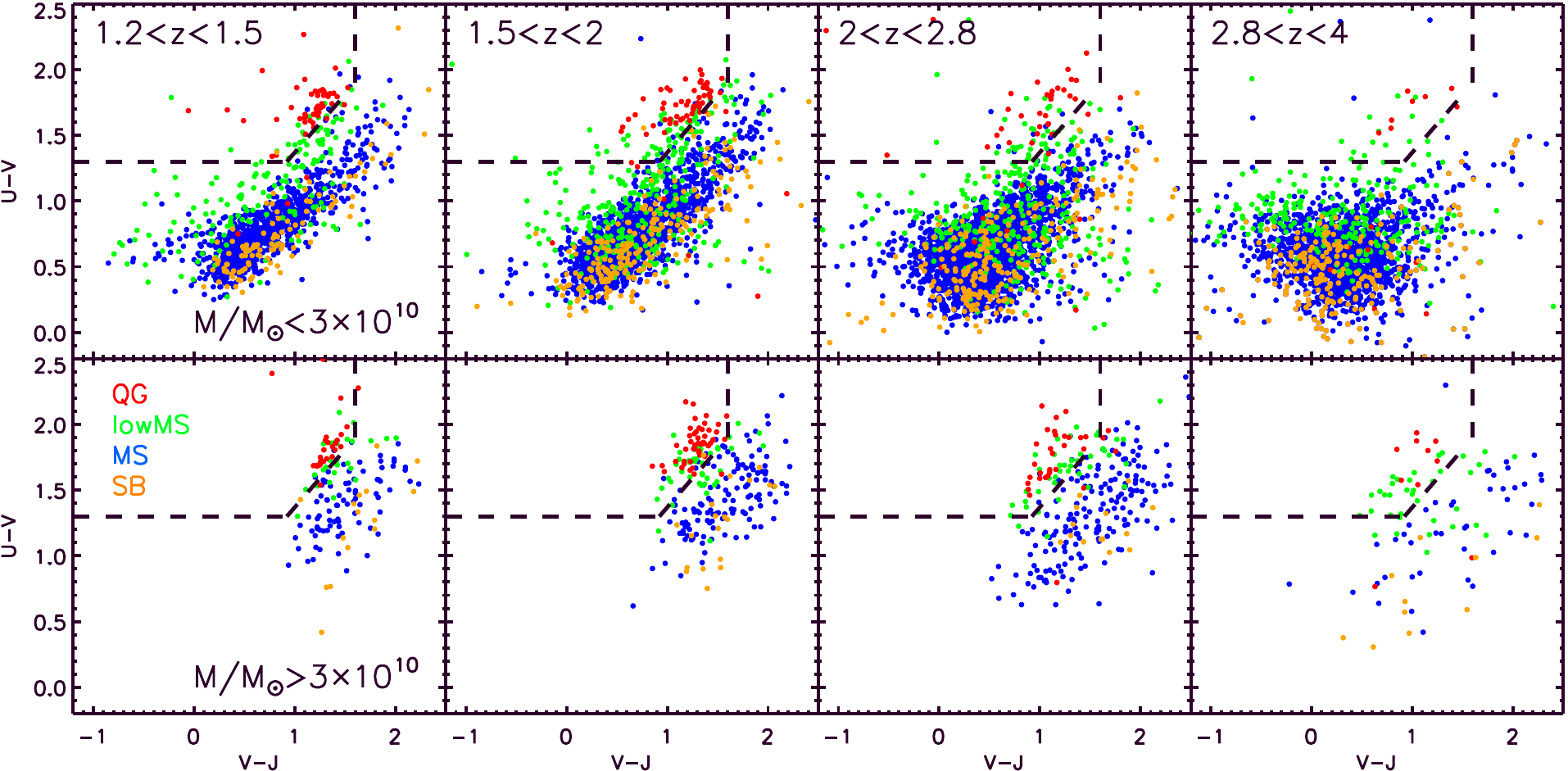}
\caption{Rest-frame UVJ diagram with four different galaxy populations:
  starbursts (SB: orange), normal SFGs on the MS (MS: blue), galaxies located
  below the MS (sub-MS: green) and quiescent galaxies (QG: red). We divide the
  sample into two stellar mass bins, $1\times10^{9} < M_{*}/M_{\odot} < 3\times10^{10}$ (top) 
  and $M_{*}/M_{\odot} > 3\times10^{10}$ (bottom). Rest-frame colors of the sub-MS galaxies are
  intermediate between QG galaxies and normal SFGs on the MS.}
\label{fig:ms_uvj}
\end{figure}

In the log(SFR)-log(M$_{*}$) diagram, there are non-negligible populations of
galaxies that do not lie on the MS.  Using our best estimates of galaxy
properties, we make a robust characterization of galaxies into four classes
based on their positions in the log(SFR)-log(M$_{*}$) plane by measuring the
excess in $sSFR$, or ``starburstiness'' ($R_{SB}$).  Starburstiness (\citep{elb11}) 
is defined as $R_{SB}=sSFR/sSFR_{MS}$, which is the distance from the MS 
measured using Equation 4 in Section 4.1 at a given stellar mass. 
Here, galaxies are classified as starburst
galaxies (SB), normal SFGs on the MS (MS), galaxies below the MS with little
star formation activity (sub-MS), and quiescent galaxies (QG) based on
$R_{SB}$.  We use $R_{SB}=3$ to identify galaxies outside 1-$\sigma$ of the MS at all
redshifts. Consequently, the SB galaxies located above the Main Sequence 
are defined as galaxies having $R_{SB}>3$. The MS galaxies are
defined as those whose $R_{SB}$ is between $1/3<R_{SB}<3$ (within 3 times
above and below the MS, or 1$\sigma$). The sub-MS galaxies are defined as
$1/3<R_{SB}<1/30$ (10 times below the lower cut of the MS: magenta line in
Figure ~\ref{fig:ms}). Lastly, we call any galaxies with $R_{SB}<1/30$
quiescent (QG). This classification results in 895 SB (9.0\%), 7510 MS
(76.0\%), 1181 sub-MS (11.9\%) and 302 QG (3.1\%) galaxies. As expected, the
MS galaxies are dominant, but the number of other populations, particularly SB
and sub-MS galaxies, are not negligible.

Rest-frame (U-V) and (V-J) colors of the four populations are illustrated in
Figure~\ref{fig:ms_uvj}.  The galaxies are divided into two stellar mass bins,
above and below $M_{*}/M_{\odot}=3\times10^{10}\sim log(M_{*})=10.5$, which 
roughly corresponds to the turnover mass at $z\sim2$. SB galaxies tend to be located in
the bulk of SFG color-color region.  About 83\% of the QG galaxies would also
be UVJ-selected quiescent galaxies. Interestingly, the rest-frame colors of
the sub-MS galaxies are almost exclusively intermediate between the QG and MS
galaxies for the low stellar mass bin.  Whereas the same is true for more
massive galaxies, there also exist sub-MS massive galaxies with a range of
rest-frame colors consistent with normal star-forming galaxies. We find out
15.6\% of the sub-MS galaxies are rest-frame UVJ classified as quiescent
galaxies and there is no sub-MS galaxy with $sSFR<0.01$ Gyr$^{-1}$.

\section{The Morphology of Galaxies and their Position in the log(SFR)-log(M$_{*}$) plane}
\label{sec:morphology}

The relationship between galaxy morphology and star formation history has been
studied at $z\sim2$ \citep{wuy11, cam11, szo11, wan12, lee13} using various
diagnostics beyond visual inspections, including non-parametric measures (Gini
(G): \cite{abr03}; M$_{20}$: \cite{lot04}; multiplicity
($\Psi$): \cite{law07}; Concentration (C), Asymmetry (A), and Clumpiness
(S): \cite{con03}), parametric measures (S\'{e}rsic index,
half-light radius: \cite{vdw12,whi15}), and
projected mass density \citep{fra08, cas11, cas13, fan13, bar13, whi17}. In
this section, we explore how morphologies of galaxies correlate with their
positions on the log(SFR)-log(M$_{*}$) diagram.  We use ``starburstiness''
($R_{SB}$), defined in Section 4.2, as a metric quantifying the position of
galaxies in log(SFR)-log(M$_{*}$) space relative to the main sequence (MS).

We consider how R$_{SB}$ correlates with S\'{e}rsic index (n) and half-light
radius ($R_{e}$) in Section 5.1, stellar mass surface density ($\Sigma_{50}$
and $\Sigma_{1}$) in Section 5.2, and G and $M_{20}$ in Section 5.3.  Each
morphological parameter is introduced in greater detail in the subsequent
sections, where we perform an extensive analysis of galaxy morphologies at
$1.2<z<4$.

Throughout this section, Spearman's rank correlation coefficient, $r_{s}$, is
used to compute the correlation between morphologies and $R_{SB}$
statistically.  This coefficient describes how two variables are monotonically
related. Thus, $+1$ ($-1$) means that two values have a perfect monotonic
increasing (decreasing) relationship, while zero indicates that there is no
correlation between two values.  We interpret the results of $r_{s}$ adopting
the following guidelines. :
\begin{enumerate}
\item $\vert r_{s}\vert<0.19$ is considered to show no
significant correlation.
\item $0.20<\vert r_{s}\vert<0.39$ is a weak correlation.
\item $0.40<\vert r_{s}\vert <0.59$ is a moderate correlation. 
\item $0.60<\vert r_{s}\vert <0.79$ is a strong correlation.
\item $0.80<\vert r_{s}\vert <1.0$ corresponds to a very strong correlation.
\end{enumerate}

\subsection{Fitting the S\'{e}rsic Light Profile}

\begin{figure*}
\epsscale{0.75}
\plotone{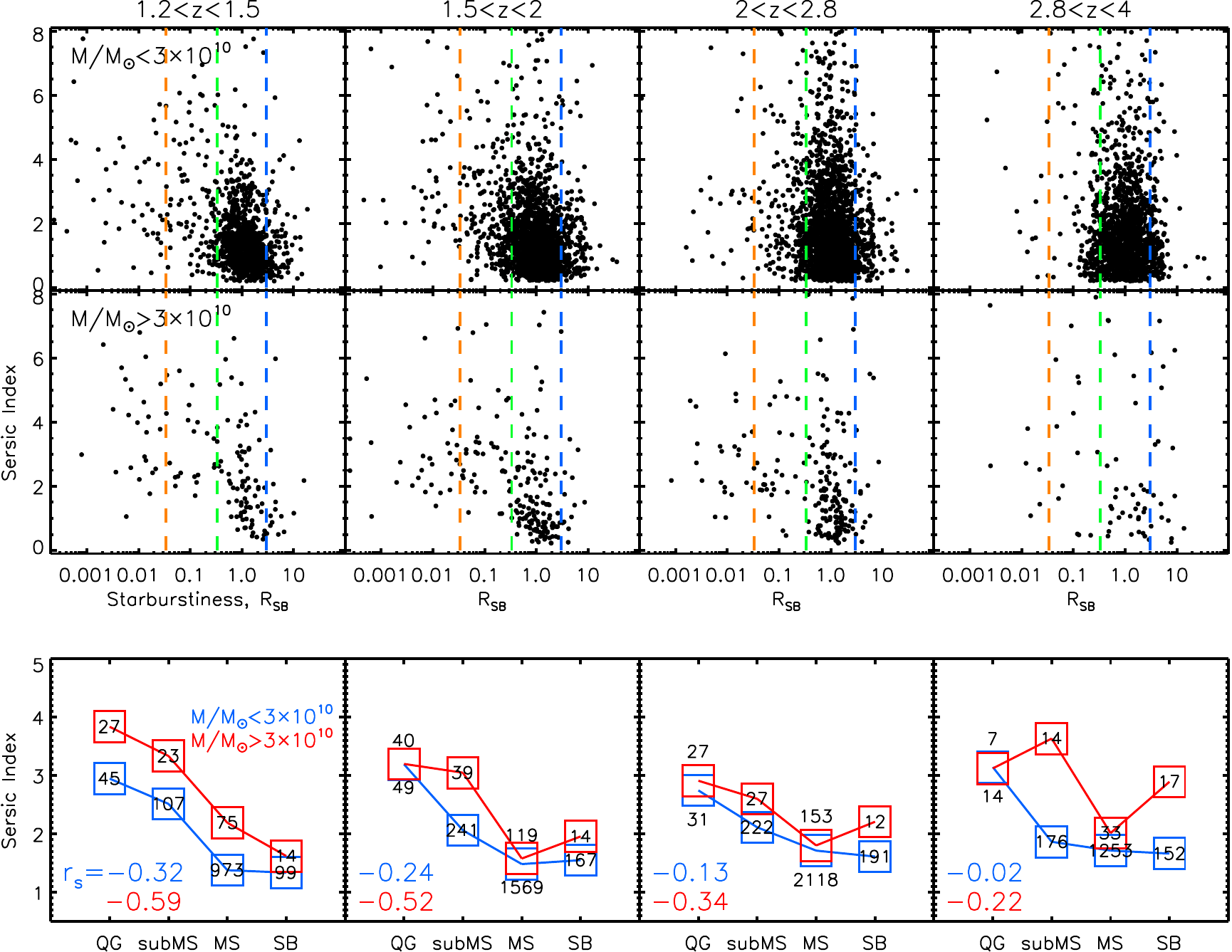}
\plotone{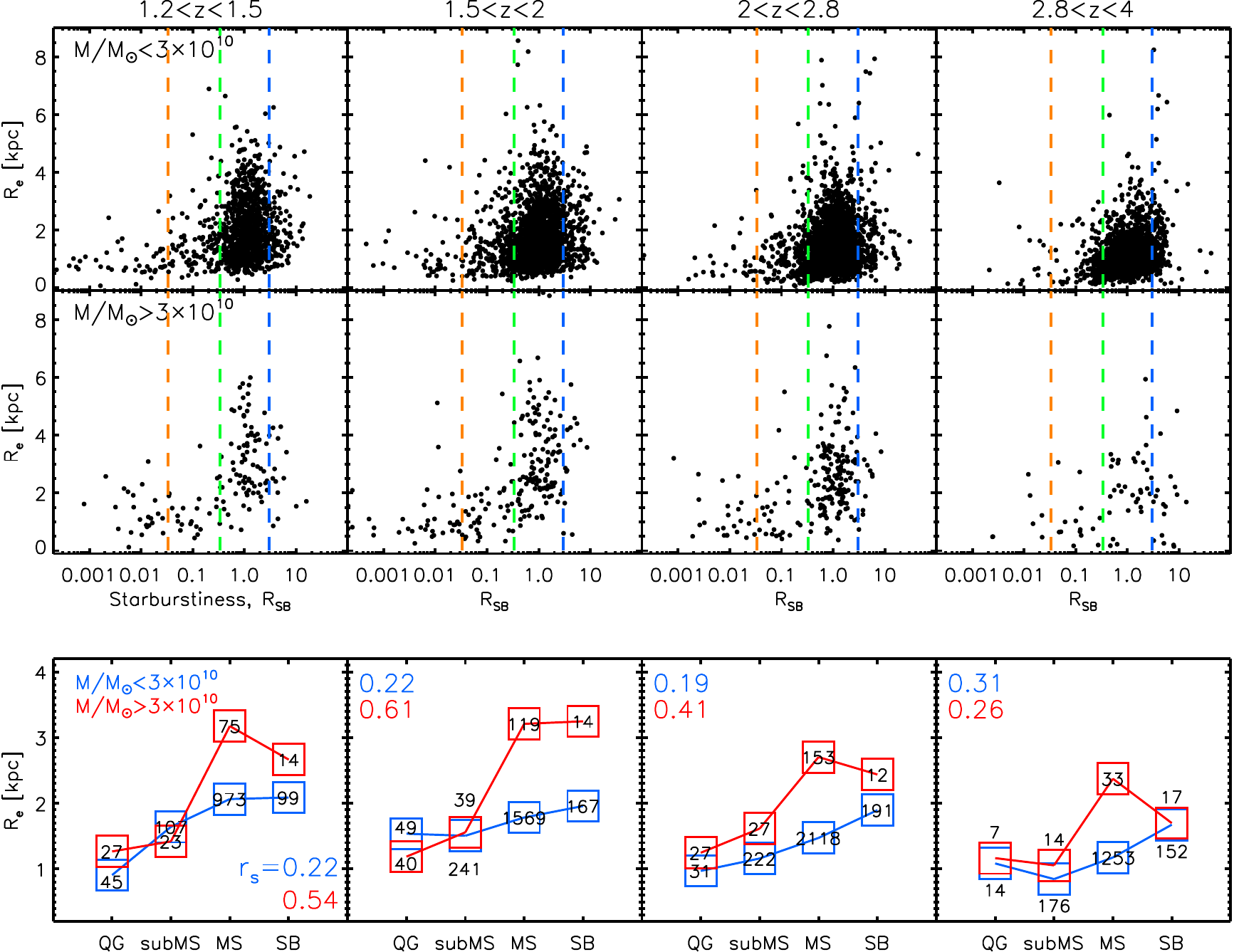}
\caption{We study the correlation between $R_{SB}$ and S\'{e}rsic index(n)/Half-light radius($R_{e}$) for
  four different galaxy populations on the log(SFR)-log(M$_{*}$) relation
  across four redshift bins. In the top two panels, blue, green and orange vertical lines indicate $R_{SB}=3,
  1/3, 1/30$ for the classification of galaxies as SB, MS, and sub-MS,
  respectively. Galaxies to the left of the orange line are classified as the QG
  galaxies, irrespective of their rest-frame UVJ colors.  The top 
  panels show distributions of n/$R_{e}$ as a function of $R_{SB}$ for galaxies having
  stellar masses below (above) $3\times10^{10}M_{\odot}$. The bottom panel
  shows the average S\'{e}rsic indices/$R_{e}$ of the four galaxy populations for the
  high mass (red) and low mass (blue) samples. The number inside the square
  symbol indicates the number of galaxies in each galaxy population. $r_{s}$
  is the Spearman's rank correlation coefficient between n/$R_{e}$ and $R_{SB}$ for
  each redshift and mass bin. Top figure: $R_{SB}$ vs. S\'{e}rsic index(n). On average, galaxies tend to have higher n as
  $R_{SB}$ decreases. The overall correlation between n and $R_{SB}$ is weak,
  especially at higher redshifts and lower masses.  We find a moderate correlation for massive galaxies at $z<2$.
  Bottom figure: $R_{SB}$ vs. Half-light radius($R_{e}$). On average, as $R_{SB}$ increases, the sizes
  of the galaxies increase.  The correlation between $R_{SB}$ and $R_{e}$ is
  weak in all redshift bins for lower mass galaxies and the highest redshift
  massive galaxies, whereas it is a moderate to strong trend for massive
  galaxies at $z<2.8$. }
\label{fig:mean_sersic}
\end{figure*}

The most commonly used parametric diagnostics of galaxy morphology include the
S\'{e}rsic function, describing the light profile, and the effective radius
enclosing half of the light. We fit the S\'{e}rsic function to the HST/WFC3
F160W (H-band) images using the GALFIT package \citep{vdw12} (hereafter,
VDW12), which returns the S\'{e}rsic index (n), semi-major axis (SMA), axis
ratio, position angle, and various flags by fitting a single S\'{e}rsic
profile to a galaxy. To avoid large systematic
and random uncertainties, we exclude galaxies having GALFIT FLAG$\neq 0$
($\sim18\%$). VDW12 also suggested that faint galaxies are expected to produce
biased results. Although we use deeper images( 10-epoch) than what was 
used in VDW12 and most of our sample is relatively bright ($H<26$), we 
repeat our morphological analysis in Appendix A using only
galaxies having $H<24.4$, which is a suggested magnitude limit of VDW12 for
S\'{e}rsic index. We find that restricting the sample to bright galaxies does
not change the overall results.  Therefore, we do not limit the sample based
on the galaxy's magnitude in this study. 

The distributions of S\'{e}rsic index and half-light radius as a function of
$R_{SB}$ are shown in Figure~\ref{fig:mean_sersic}. The circularized half-light radius,
$R_{e}$, is computed in units of kpc using the formula,
$R_{e}=SMA\sqrt{axis~ratio}$. On average, n increases (see the top figure in Figure~\ref{fig:mean_sersic}) 
and $R_{e}$ decreases (the bottom figure) as $R_{SB}$ decreases. Most galaxies located below
the MS (QG and sub-MS) have $n>2.5$ and $R_{e} < 2kpc$ over all stellar mass
and redshift ranges.  Based on $r_{s}$ (see numbers in the bottom panel), we
find that massive galaxies show moderate to strong correlations
(anti-correlations) between $R_{e}$ and $R_{SB}$ at $z<2.8$, while n 
of massive galaxies at $z<2$ is moderately correlated with $R_{SB}$. All
correlations at $z>2.8$ are either weak or insignificant.  These results are
in broad agreement with those presented in \cite{wuy11, bre17, pan17}, who present the
correlation between specific star formation rate (similar to $R_{SB}$) and
S\'{e}rsic index across the log(SFR)-log(M$_{*}$) diagram since
$z\sim2.5$. For lower mass galaxies, $R_{SB}$ is only weakly correlated with
$R_{e}$.  Similarly, $R_{SB}$ is also weakly correlated with n for low mass
galaxies at $z<2$, with very weak or insignificant correlations at $z>2$.

\subsection{The Effective Surface Density and Projected Central Mass Density}

\begin{figure*}
\epsscale{0.75}
\plotone{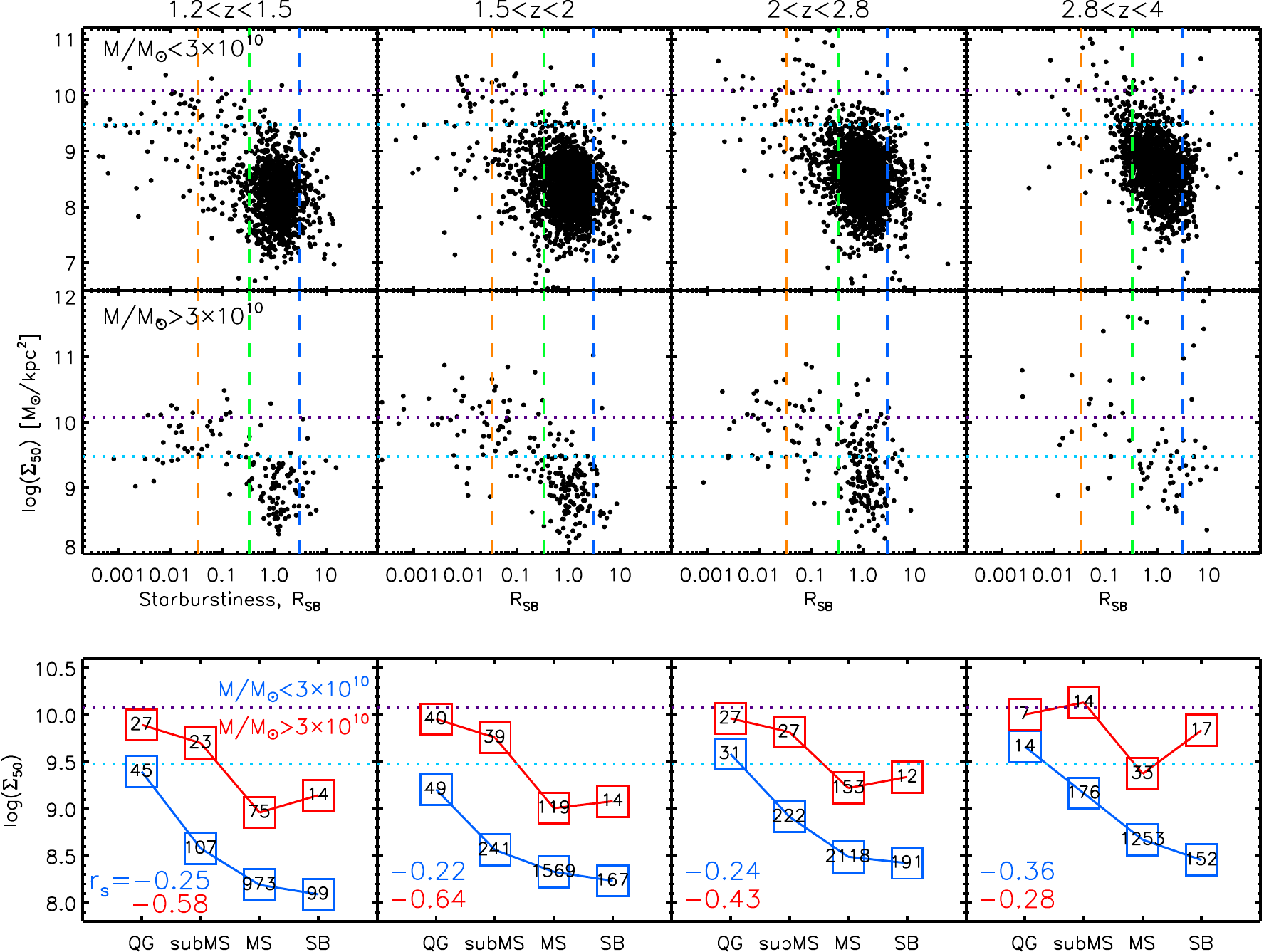}
\plotone{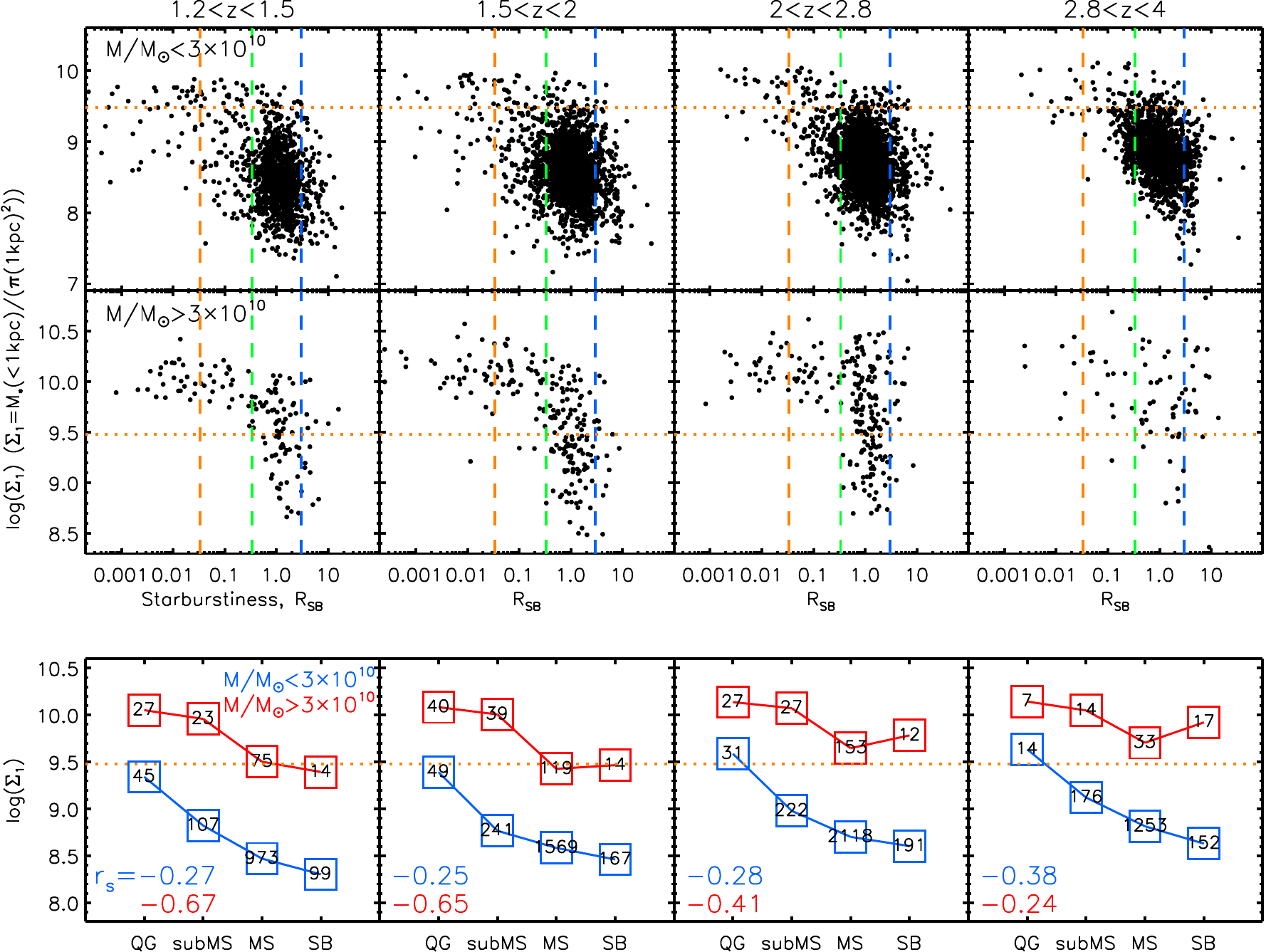}
\caption{Top figure: The surface mass density within the effective radius, $\Sigma_{50}$,
  vs. $R_{SB}$ (two top panels) and the average $\Sigma_{50}$ for the four
  galaxy populations classified based on $R_{SB}$ (bottom panel). The
  horizontal lines represent the classification of high density and ultra-high
  density galaxies defined from \cite{cas11} (sky blue:
  $\Sigma_{50}=3\times10^{9}$, violet: $1.2\times10^{10}$). On average,
  massive galaxies at $z<2.8$ show moderate to strong anti-correlations,
  becoming denser as $R_{SB}$ decreases. The trend is present but weaker in
  the highest redshift bin.  Similarly, we find a weak trend between
  $\Sigma_{50}$ and $R_{SB}$ for low mass galaxies.
  Bottom figure: The extrapolated projected central density, $\Sigma_{1}$,
  vs. $R_{SB}$ (two top panels) and the average of $\Sigma_{1}$ for the four
  galaxy populations (bottom panel). The horizontal line is for the
  classification of high central density galaxy, $\log(\Sigma_{1})=9.5$, from \cite{bar17}.  
  The $r_{s}$ values for $\Sigma_{1}$ are similar to
  those for $\Sigma_{50}$, but with a less scatter in the correlation (see
  top panel). }
\label{fig:mean_den1}
\end{figure*}

It has been suggested that compactness of a galaxy is closely connected to
quiescence \citep{bel12, lan14, mci14}. S\'{e}rsic index, bulge-to-total ratio
(B/T), and the stellar density are commonly used to identify compact
structures \citep{cas11, cas13, che12, fan13, lan14, bar13, bar17, sch16}.  At
$z<0.8$, \cite{che12} and \cite{fan13} suggested that the projected central
mass density is a stronger predictor for quenching of star formation (SF) than
B/T and S\'{e}rsic index.  In this study, we compute two mass densities, the
effective surface mass density ($\Sigma_{50}$) and the projected mass density
within the central 1kpc ($\Sigma_{1}$).  We investigate trends between
position of galaxies in the log(SFR)-log(M$_{*}$) plane and mass density,
which is often used as a proxy of compactness. The first, $\Sigma_{50}$, is
defined as a half of the stellar mass divided by a surface surrounded by a
half-light radius, $\bf{\Sigma_{50}=M_{*}/ 2\pi R_{e}^{2}~[M_{\odot}/kpc^{2}]}$. In
\cite{cas11}, compact QGs at $1.2<z<2.5$ are defined as the galaxies located
1-$\sigma$ below the distribution of local QGs on the mass-size relation,
while QGs 0.4 dex smaller than the local QGs are called ultra-compact
galaxies.  Adopting their definition, we identify galaxies having $\Sigma_{50}
> 3\times 10^{9} M_{\odot}/kpc^{2}$ ($\log\Sigma_{50}\sim9.5$) and
$\Sigma_{50} > 1.2\times 10^{10}M_{\odot}/kpc^{2}$ as high mass density and
ultra-high mass density galaxies, respectively (sky blue, violet vertical
lines in the top of Figure~\ref{fig:mean_den1}).

The projected central mass density, $\Sigma_{1}$, is the extrapolated
projected central density, defined as $M_{*}(<1kpc)/\pi(1kpc)^{2}$.  The
stellar mass within 1kpc is calculated numerically with the following
equation:
\begin{equation}
M_{*}(<1kpc) =\frac{\int_{0}^{1 kpc}I(r)2\pi r dr}{\int_{0}^{\infty}I(r)2\pi r dr} \frac{L_{galfit}}{L_{phot}}M_{phot},
\end{equation}
where $I(r)$ is the S\'{e}rsic profile and $M_{phot}$ is the stellar
mass. $L_{galfit}$ is the total luminosity as computed by integrating a
S\'{e}rsic profile obtained from GALFIT.  $L_{phot}$ is the total luminosity
of a galaxy from the CANDELS H-band catalog.  \cite{bar17} selected compact
galaxies with $\log(\Sigma_{1})>9.5$, defined from the tight correlation
between the $\Sigma_{1}$ and stellar masses of the quiescent galaxies. We use
this same definition to classify galaxies having the high central density
(orange horizontal line in the bottom of Figure~\ref{fig:mean_den1}).

As illustrated in Figure~\ref{fig:mean_den1}, the
correlation coefficient $r_{s}$ of the relationships $\Sigma_{50}$
vs. $R_{SB}$ and $\Sigma_{1}$ vs. $R_{SB}$ is similar to that of $R_{e}$
vs. $R_{SB}$ and they are higher than that for the S\'{e}rsic index. In other
words, the effective surface mass density, projected central mass densities
and effective radius all show moderate to strong correlations with $R_{SB}$ at
$z<2.8$, while the correlation is weaker for the S\'{e}rsic index $n$. 

While qualitatively similar, the relationship between log($\Sigma_{1}$) and
$R_{SB}$ is characterized by significantly less scatter than that of
log($\Sigma_{50}$), suggesting that the latter is a noisier statistic than
the former. As discussed in Appendix B, this is supported by the direct 
comparison of log($\Sigma_{1}$) and log($\Sigma_{50}$) as well as the
comparison with the non--parametric statistics, Gini and $M_{20}$, which
provide an alternative description of the degree of ``compactness'' and
``nucleation'' of galaxies. We suspect that the higher statistical noise of
$\Sigma_{50}$, which is solely derived from the parameter $r_e$, is due to the
comparatively large covariance between $n$ and $r_e$ in the fit of the light
profile to the S\'{e}rsic function. The value of $\Sigma_1$, which depends on both
$r_e$ and $n$, evidently turns out to be much better constrained than the two
parameters individually. Thus, we reach the same conclusions as \cite{whi17}
that the extrapolated projected central stellar density is a better parameter
to investigate the connection between the galaxies' central morphology and
their star formation activities. 

The relationship between $\Sigma_1$ and $R_{SB}$ in the bottom of 
Figure~\ref{fig:mean_den1} illustrates what appears to be a general
property of galaxies, namely that as long as galaxies are on or above the Main
Sequence, i.e. actively forming stars, their central density spans a broad
range of values, to first order homogeneously distributed. The central density
of galaxies below the Main Sequence or quenched, on the other hand, is
systematically restricted to the approximately top quartile of the
distribution of $\Sigma_1$, whose value is constant. This result is in very
good qualitative and quantitative agreement with similar measures made by 
\cite{bar17, bre17}. Adopting the value $log(\Sigma_{1})=9.5$ to define 
galaxies having the high central
density, the fraction of high--central density galaxies in our
sample is 70 (98)\%, 23 (95)\%, 6 (57)\% and 6 (56)\% for QG, sub-MS, MS and
SB galaxies, respectively, where the value in parentheses represents massive
galaxies only (i.e., $M_{*}>3\times 10^{10}$ M$_{\odot}$). 

\begin{figure}
\epsscale{1.0}
\plotone{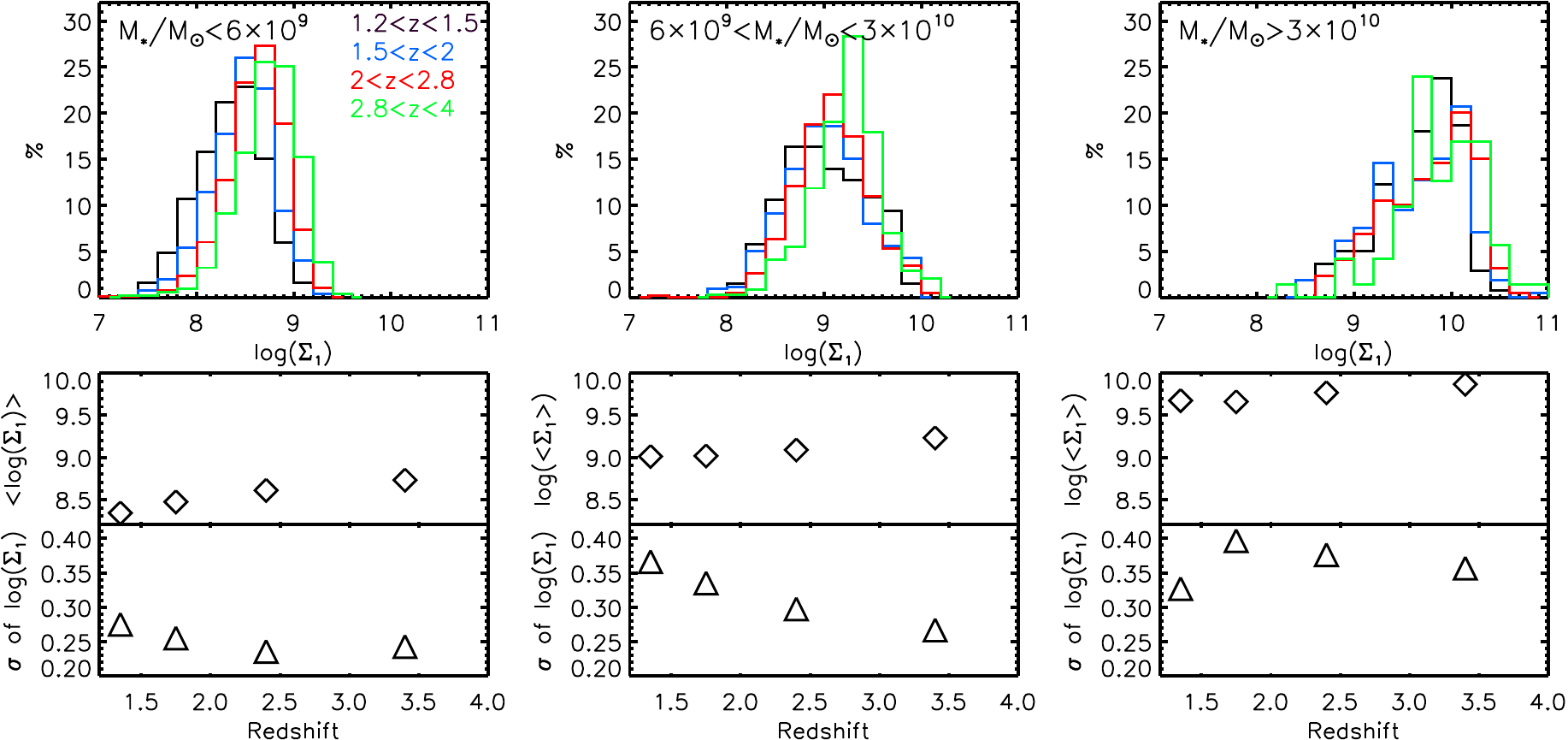}
\caption{The dependence of the distribution of log$\Sigma_1$ on stellar mass
  and redshift. Here, we split our low--mass bin
 ($10^9<M_{*}/M_{\odot}<3\times10^{10}$) into two sub-bins to study the 
 trends in more careful detail. The upper panels show the evolution of the distribution, 
 where we have color-coded the histograms in each of our four redshift bins. 
 The lower panels show the redshift evolution of the mean and the variance of 
 log$\Sigma_{1}$. The range of values of log($\Sigma_1$) in each mass bin 
 decreases with decreasing mass, reflecting the well-known correlation between 
 stellar mass and central density (e.g., Woo et al. 2015). But overall, at fixed stellar 
 mass, there are only subtle changes in the dependence of log($\Sigma_1$) with 
 redshift. The range of values covered by the distribution, as well as its peak 
 remains fairly constant, as reflected by the very mild redshift dependence of the mean. 
The distribution tends to become more extended towards lower log($\Sigma_1$) 
values as the redshift decreases, which causes the variance to increase correspondingly.}
\label{fig:hist_sigma1}
\end{figure}

This trend between central density and star formation activity is observed in
all redshift bins that we have considered here (the larger scatter in the
highest redshift bin is consistent with large random errors in the measures),
and this independence from redshift is further observed in
Figure~\ref{fig:hist_sigma1}, which plots the redshift evolution of the
distribution of $log(\Sigma_1)$ as a function of stellar mass (top panels),
and the mean and variance of log($\Sigma_1$) for the overall population as a
function of redshift (bottom panels). While the range of values of
log($\Sigma_1$) depend on the stellar mass, monotonically drifting from low 
to high values as the mass increases, in any given mass bin the range remains
fairly constant with redshift.  This is also evident in the redshift evolution
of the mean (bottom panels), which shows little to no change with time.  The
shape of the distribution also undergoes only a subtle change, becoming slightly
broader towards lower log($\Sigma_1$) values, while keeping the peak
substantially unchanged and spanning the same global range of values. This is
reflected in the variance that increases towards low redshift (the only
exception is the abrupt decrease of the variance in the lowest redshift bin of
the massive galaxies).

\subsection{Non-Parametric morphology: Gini and $M_{20}$}

Although the value of $\Sigma_1$ is obtained from the S\'{e}rsic function
parameters fit to the full light profile of each galaxy, it only charaterizes
the morphology of a galaxy's central regions, i.e. it is a local metric of
morphology. A galaxy, however, can have a high value of $\Sigma_1$ either if
it consists of a predominantly compact nuclear source embedded in a diffuse
and fainter component or if it consists of only the compact source (this case
is commonly referred to as a ``nugget''). Since $\Sigma_1$ cannot
discriminate between these two cases, which actually might represent the
outcomes of different evolutionary and quenching mechanisms 
\citep{bar13,woo15,zol15}, we have also characterized
the morphology of the galaxies in terms of the Gini coefficient, $G$, and the
second-order moment of the brightest 20\% of the galaxy pixels,
$M_{20}$. These diagnostics have been used to describe galaxy morphologies both
in the local universe and at high redshift universe \citep{abr03, lot04,
  lot08, wan12, lee13, pet16}.
Since not all galaxies are described by smooth and symmetric light
profiles, non-parametric measures are known to better characterize the
morphologies of irregular galaxies, which are more common at high redshifts
\citep{lot04}. For example, \cite{lee13} found that the combination of
non-parametric and parametric measures of the CANDELS galaxies at $z\sim 2$
provides a more complete description of the morphological properties of
high--redshift galaxies than using {\it only} S\'{e}rsic index or $R_e$.

We have measured $G$ and $M_{20}$ from the WFC3/HST F160W(H) images of our
samples using the definitions by \cite{lot04}. The Gini coefficient provides a
powerful description of how nucleated the light of a galaxy is, reagrdless of
its size and mass: light
distributed over only a few pixels has $G\sim 1$, whereas uniformly
distributed light corresponds to $G\sim 0$. 
The $M_{20}$ parameter quantifies
the tendency of the light distribution to be in structures (bars, spiral arms,
clumps): galaxies with high $M_{20}$ values are clumpy objects, whereas those
with low (negative) values of $M_{20}$ are relatively compact objects with one
bright core. Thus, for example, a ``nugget" would have a large value of
$\Sigma_1$, a large value of $G$ and a highly negative value of $M_{20}$.  On
the other hand, a galaxy with a massive, compact nuclear component embedded in
an extended disk also has a large value of $\Sigma_1$ but a lower value of $G$
compared to the nugget. Among massive quiescent galaxies, mean values of $G$
and $M_{20}$ of visually classified nuggets (and non-nuggets) are $0.61$
(0.58) and $-1.87$ (-1.76), respectively.

A complete discussion of the non--parametric morphological measures and their
evolution with redshift will be the subject of the second paper in this
series. Here we will mostly use the Gini and $M_{20}$ coefficients for a
descriptive characterization of the morphology of galaxies on and around the
Main Sequence of star--formation.

\begin{figure*}
\epsscale{0.75}
\plotone{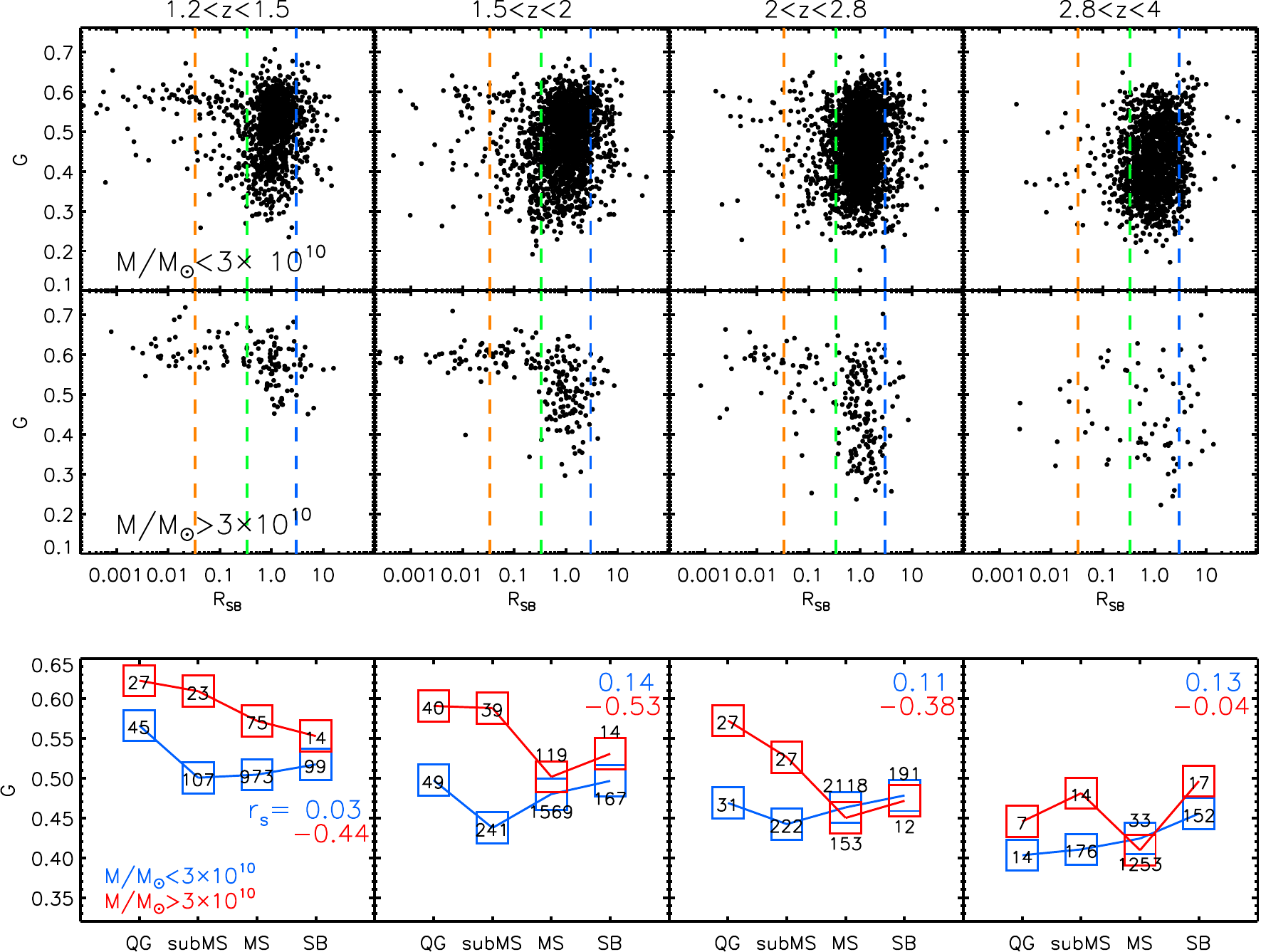}
\plotone{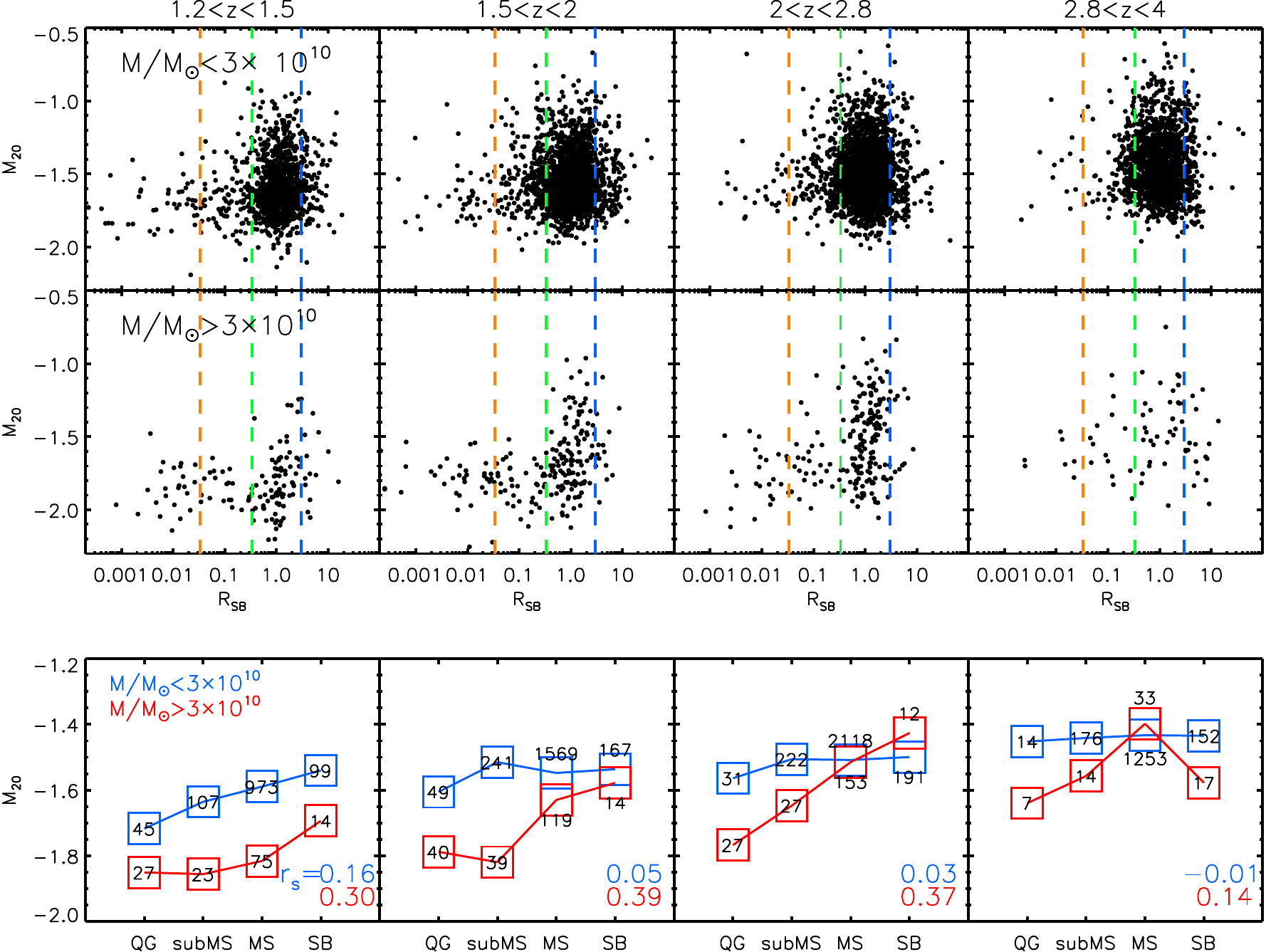}
\caption{Gini coefficients ($G$)/$M_{20}$  vs. $R_{SB}$ for individual galaxies at four 
  redshift epochs (top) and the average $G$/$M_{20}$  
  across the log(SFR)-log($M_{*}$) plane, as parameterized by $R_{SB}$. All
  lines, colors and symbols correspond to Figure~\ref{fig:mean_sersic}. Top figure: G vs. $R_{SB}$. QG 
  galaxies have the highest $G$ on average, which is an indication of compact
  structure. $G$ of massive galaxies at $z<2.8$ shows moderate/weak 
  correlation with $R_{SB}$. The low
  mass sample and galaxies at the highest redshift has no significant 
  correlation between $R_{SB}$ and $G$. Bottom figure: $M_{20}$ vs. $R_{SB}$. SB galaxies have the highest $M_{20}$, which
  is indicative of a clumpy sub-structure.  For low-mass galaxies at all
  redshifts, as well as massive galaxies at $z>2.8$, there is no significant
  correlation between $M_{20}$ and $R_{SB}$. $M_{20}$ shows a weak to moderate
  correlation amongst massive galaxies at $z<2.8$. }
\label{fig:mean_gini}
\end{figure*}

Figure~\ref{fig:mean_gini} depicts the distributions of
$G$ and $M_{20}$ as a function of $R_{SB}$, respectively. On average, the
galaxies have higher $M_{20}$ (more clumpy) and lower $G$ (more extended) as
$R_{SB}$ increases.  In other words, the galaxies with more active star
formation appear to have more diffuse structures, but the overall correlation
between G/$M_{20}$ and $R_{SB}$ for massive galaxies at $z<2.8$ is weaker than
$\Sigma_{1}$ and $R_{e}$.  $r_{s}$ of $G$ and $M_{20}$ for less massive
galaxies is about $<0.15$ at all redshifts, an insignificant
correlation. Similarly, we find that the correlation of $G$ and $M_{20}$ with
$R_{SB}$ barely exists at $z>2.8$.  Such weak or no correlation between
non-parametric measures and $R_{SB}$ are observed even though we only consider
very bright galaxies ($H<24.4$) as shown in Figure~\ref{fig:m20_ex} (Appendix A).

\section{Synthesis of Morphological Trends}
\label{sec:synthesis}

A general result of the analysis presented in Section~\ref{sec:morphology} is
that Main Sequence and Starburst galaxies cover a broad spectrum of
morphologies in all redshift and stellar mass bins considered here, as
quantified by our indicators (primarily $\Sigma_1$).  All
the while, quiescent galaxies systematically occupy a more restricted range of
the same indicators, located towards the extreme of the distributions where
high--central density, compact and nucleated galaxies are found. The
morphology of the sub-MS galaxies is on average intermediate between MS and QG
galaxies. The overall trend is that as $R_{SB}$ decreases, galaxies have
larger projected (central) stellar density ($\Sigma_1$ and
$\Sigma_{50}$), smaller half-light radius $r_e$ and larger S\'{e}rsic index
$n$, larger $G$ and lower (more negative) $M_{20}$, namely they are more
compact and nucleated. This trend is more pronounced in more massive galaxies.
Note that there is in fact either a weak or no significant correlation between
$R_{SB}$ and $n$/non-parametric measures considered for the highest redshift
bin at $z>2.8$ regardless of their magnitudes (see Appendix A).

The trend is particularly evident in $\Sigma_1$ versus $R_{SB}$. 
For example, Figure ~\ref{fig:mean_den1} (and also 
Figure~\ref{fig:den1_age_hm} in Section 7.1) shows that the range of $\Sigma_1$ values of 
quiescent galaxies is approximately $1/3$ (in log space) that of star--forming
ones. It is less evident when plotting $R_e$ and $n$ versus $R_{SB}$, very likely
because both are noisier metrics, less representative of a galaxy's overall
morphology. In particular, we note that we do not seem to reproduce the tight
correlation between $n$ and sSFR observed by \cite{wuy11}. Approximately the
same compression of the dynamic range of morphological indicators of quenched 
galaxies is observed in every redshift bin. Looking at the panel at $2<z<2.8$
of Figure~\ref{fig:den1_age_hm}, for example, this means that the non--compact
star--forming galaxies in this redshift range have either (1) not quenched by
the time they are observed in the two lower redshift panels (the panels are
spaced in cosmic time by $\Delta t\approx 1$ Gyr), or (2) if they have
quenched, they have developed a high--central density central region, as
quantified by $\Sigma_1$. 

In fact, the distribution of galaxies in the $\Sigma_1$ vs. $R_{SB}$ diagram
has been interpreted as evidence of morphological transformation during the
quenching process itself and possibly of a causal relationship between
development of high stellar density and quenching \citep{che12, fan13, bar13, 
bar17, zol15, whi17, bre17}. \cite{lil16} suggested that ``progenitor bias'' (i.e. the fact that
galaxies were smaller and denser in the past) is at least in part responsible
for the observed differences between quenched and star-forming galaxies at any
epoch. Spectroscopic measures of stellar age at redshift $z<1.5$ 
\citep{fag16,wil17,bel14,bel15} found that compact
galaxies are indeed $\approx 1$ Gyr older than normally--sized ones, adding
support to the idea that progenitor bias plays at least some role.

The extent to which progenitor bias contributes, in part or all, to the
apparent morphological transformation of galaxies as they quench illustrated
in Figure~\ref{fig:mean_den1} (and Figure~\ref{fig:den1_age_hm}) 
remains to be quantified. However, it is possible that by only using
the projected central density $\Sigma_1$ (or its noisier sibling
$\Sigma_{50}$) to quantify the process of ``compactification'' of galaxies as
they complete the transition from the star--formation phase to quenching we
may miss the overall features of the phenomenon, since this parameter only
describes the transformation at the center of galaxies and does not capture
the complexity of morphological evolution and transformation in the whole
structure of a galaxy. In the second paper of this series, we will compare
$\Sigma_1$ as a probe the growth of the core with the non--parametric
morphology indicators, $G$ and $M_{20}$, which inform us about the relative
proportions of light (mass) in the core, or in one or more compact sources,
and in more diffuse structures in the galaxies.

\subsection{Characteristics of the galaxies located below the MS}

Based on our galaxy classification relative to the MS, we identify a large
number of the sub-MS galaxies located below the galaxies on the MS, 
defined as $1/3<R_{SB}<1/30$. The sub-MS 
galaxies have little star formation (SF) activity
but are not yet quenched. We show that the rest-frame colors of the low-mass
sub-MS galaxies are intermediate between QG and MS galaxies in
Figure~\ref{fig:ms_uvj}, while massive galaxies are also intermediate but
exhibit a broader range of rest-frame colors consistent with the overall SFG
population. The sub-MS galaxies are more spheroidal-like than the MS
galaxies as shown in Section 5. These
galaxies are seen to be similar to green valley galaxies observed at $z<1$ in
that they have intermediate colors between the red sequence and blue
cloud. \cite{sch14} found that green valley galaxies at $0.02<z<0.05$ are
located below the MS regardless of morphology and are in the process of
quenching. At $z\sim1$, \cite{men11} showed that green valley galaxies are
typically disk galaxies with high concentrations. Thus, the sub-MS galaxies 
may be under transition from the star-forming to quiescent populations. 
The existence of galaxies having intermediate colors and/or morphologies 
at $z\sim2$ have been reported;
\cite{whi12b} found that galaxies below the MS are redder SFGs having lower
sSFR and low-dust attenuation. \cite{pan17} defined the transition galaxies at
$z<3$ as galaxies located between 0.6 dex and 1.4 dex below the main
sequence. In a similar vein, Fang et al. 2017 classified ``fading
galaxies" located in the SFG region of the UVJ diagram, but below the MS
($\Delta logsSFR < -0.45dex$), which is a similar definition with our
study. These ``fading galaxies" have intermediate colors and smaller radii as
well as lower dust attenuation than normal SFGs, indicative of a transition
from SFGs to QGs as stopping SF due to the loss of their ISM. We also find
that the sub-MS galaxies in our sample have the lower E(B-V) than MS galaxies
on average, which further supports that a sub-MS galaxy is transforming to a
quiescent galaxy as quenching its SF.

\subsection{Morphologies of starbursts}

In this study, starbursts are classified as the galaxies located three times
above the MS (SB galaxies). Their very high SFR can be explained by either
larger gas reservoirs or a higher star formation efficiency
(SFE). \cite{sar14} suggested that the most extreme SFRs observed in
high-redshift starbursts would be caused by the SFE boost induced by major
mergers, since internal gas reservoirs are depleted quickly due to short-lived
SFR boosts. With {\it Herschel} data, \cite{elb11} investigated the
  sizes of starbursts relative to MS galaxies using stacks of rest-frame UV
  images and found that starbursts at $z\sim2$ have more compact 
  star formation core than MS galaxies on average. However, we find
that the average rest-frame optical morphologies of the SB galaxies are
larger, clumpier (higher M$_{20}$ and lower G), and less dense than ones of the MS galaxies.  
Most of the SB galaxies in our sample (94\%) have log($\Sigma_{1}$) $<9.5$ and shallow light profiles
($<n>\sim1.5$ and $2.3$ for less massive and massive SB galaxies). The massive
SB galaxies with $M_{*}>3\times10^{10}$ have $<R_{e}>\sim 2.74$ kpc at
$1.2<z<4$, which is more than a factor of two smaller than QGs, and have
$<R_{e}>\sim 4.0$ kpc at $2<z<2.5$, which is about a factor of four smaller
than QGs at the same redshift and stellar mass ranges. Morphologically, they
are rather different from quiescent galaxies, suggesting that it is unlikely
that during the post-starburst phase they can shrink their size to match that
of compact quiescent galaxies on a short timescale. This is inconsistent with
the previous finding \citep{wuy11}, suggesting a rapid build-up of the central
mass concentration in starbursts that have a higher $n$ than the galaxies on
the main sequence at $z\sim2$. One has to be very careful when looking at 
morphologies of starburst galaxies only using the S\'{e}rsic fits, as morphologically 
disturbed objects would tend to have poor fits \citep{bre17}.

\section{Discussion}

In this paper, we use the deepest CANDELS data, i.e. the GOODS fields, which 
have amongst the broadest and most dense coverage of photometric bands, 
to investigate general correlations between the position of galaxies at $1.2<z<4$
on and around the Main Sequence and their morphology. We find
that the morphology of galaxies clearly correlates with their position in the
log(SFR)-log(M$_{*}$) plane. On average, we reproduce previous results that galaxies
below the MS tend to have smaller sizes, larger S\'{e}rsic indices, a compact
central structure with high projected light (mass) densities. By adding measures
of stellar age from our new MCMC SED fitting,
however, we also find
additional more subtle trends that we are now going to discuss.

\subsection{Possible formation scenarios of massive, compact quiescent galaxies}

Quenching mechanisms in galaxies can be broadly classified by the timescales
on which they operate; namely, fast track and slow track quenching \citep{fan12, fan13, 
bar13, dek14, sch14, sch16}. Slow track quenching appears to dominate at 
low redshifts or later
stages of the galaxy evolution because it happens when star formation
gradually vanishes over several Gyrs.  This can be explained by slow gas exhaustion at a
critical halo mass,$\sim 10^{12}M{_\odot}$, without a disruptive external
trigger, such as major merging \citep{fan13, sch14}. As a
galaxy on the slow track becomes redder, the inner part of a galaxy becomes
denser while the outer disk fades away, and it subsequently moves off to the
MS as SF decreases. On the other hand, the fast track quenching is a rapid
process occurring with timescales less than 1 Gyr. The gas reservoir of a
gas-rich star-forming disk is destroyed by an intensive triggering event such
as a merger-induced starburst or a VDI associated with stellar feedback. Then,
the gas disk rapidly transforms its morphologies into an early-type and
immediately quenches SF and moves to below the MS. The fast track quenching is
probably the dominant mechanism at early times and has been proposed to
explain the formation of high-redshift massive compact quiescent galaxies 
\citep{van08, bar13, muz13}. Here, we
investigate any evidence for slow and fast track quenching based on our
morphological analyses of galaxies relative to the MS.

\subsubsection{Central density, Quenching and Compactification}

We find that at any redshift, QGs always occupy a restricted range of values of the 
morphological parameters $\Sigma_1$, $G$ and $M_{20}$ relative to the spread covered by
SFGs. It has been suggested that morphological transformation must take place
during the quenching phase \citep{bar13, bar14, pat13, lan14, zol15, tac16a, tac16b}.
In this scenario, SFGs are required to shrink
their size and develop a compact core within a relatively short time 
scale, and quench. Highly dissipative gas accretion into the center, or compaction,
and adiabatic contraction have been identified in simulations and are
considered as viable mechanisms for morphological transformation 
\citep{dek13, dek14, zol15, tac16a, tac16b}.

\begin{figure}
\epsscale{1.1}
\plottwo{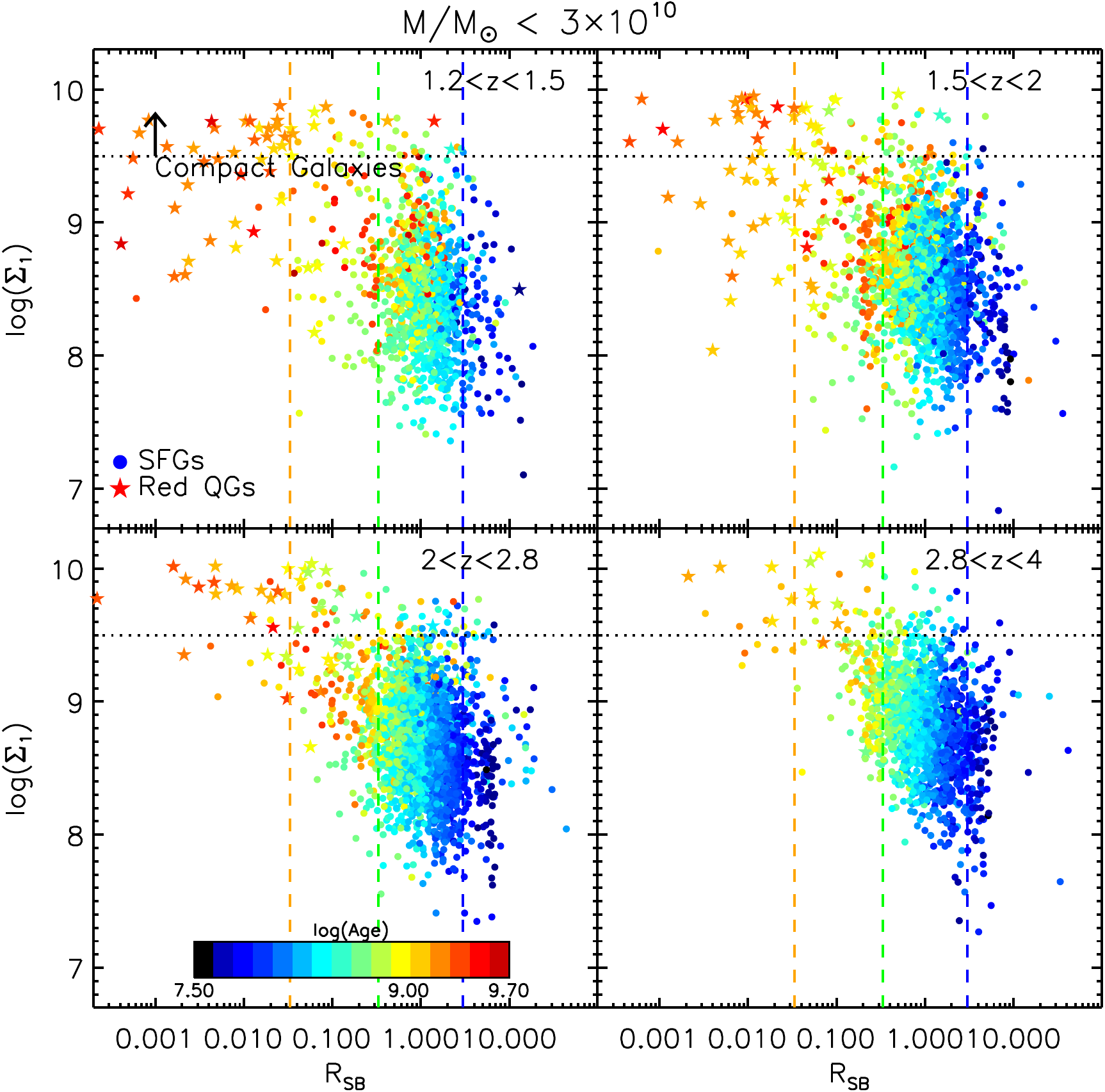}{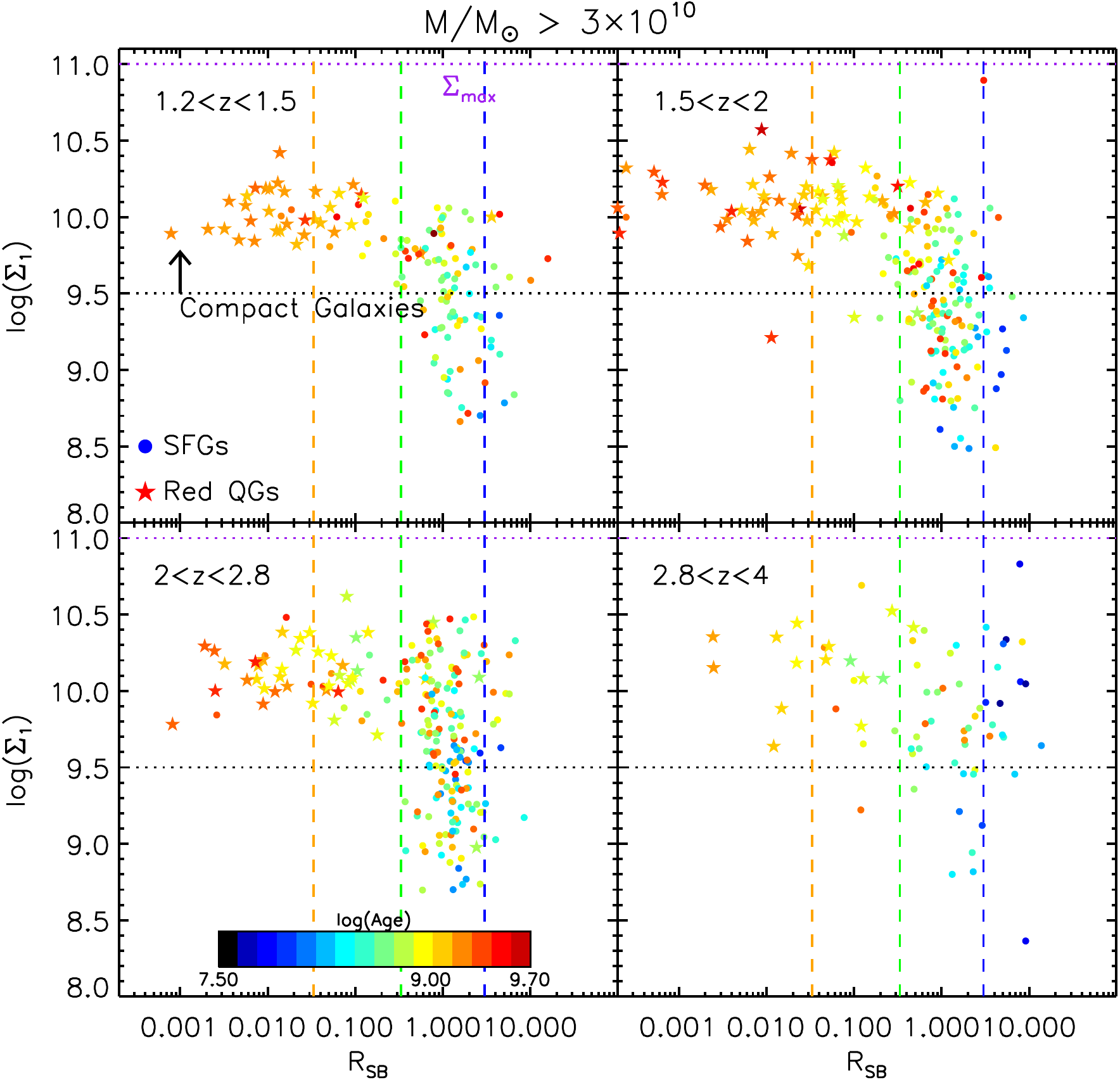}
\caption{$\Sigma_{1}$ vs. $R_{SB}$ with color-coded median-mass weighted ages for two stellar mass bins, 
$M_{*}/M_{\odot}<3\times10^{10}$ (left) and $M_{*}/M_{\odot}>3\times10^{10}$ (right). Circle and 
star symbols represent SFGs and red QGs, respectively. 
Orange, green and blue lines are to divide galaxies into QG, sub-MS, MS and SB. Black dotted 
horizontal line is $log(\Sigma_{1})=9.5$. Galaxies are older as $R_{SB}$ decreases, but there 
is no correlation between the projected central density and galaxy age. We also note that 
massive SFGs are relatively older than less massive SFGs.}
\label{fig:den1_age_hm}
\end{figure}

In the simulations, compaction creates dense massive central cores but does
not shrink the existing non-dissipative component, namely the
stars. Adiabatic compression can compactify the stars, but this should largely take 
place on the same spatial scales as the gaseous compaction. Thus,
it is not clear that the current families of compaction simulations are
actually capable of reproducing the observed compact QGs, where the majority
do not exhibit extended light profiles on top of the compact central core
\citep{szo10, szo13}.

In any case, if galaxies keep their morphology as they quench, then the
progenitors of compact QGs should be the compact star--forming ones.
\cite{wil14, wil15, van15, nel14, zol15} showed that the number density, SFR,
mass distribution and estimated quenching time of compact SFGs are consistent
with them being the progenitors of QGs. Note that in the log(SFR)-log(M$_{*}$) diagram,
the bulk of compact SFGs are located on the MS or slightly above the MS 
\citep{wil14, bar17}, suggesting that the progenitors should be
looked for among normal star--forming galaxies and not starburst ones. This 
is also supported by the discovery of the short depletion time scales of 
$\lesssim 100$Myr for compact SFGs \citep{bar13, bar14, spi16}.

\begin{figure}
\epsscale{0.8}
\plotone{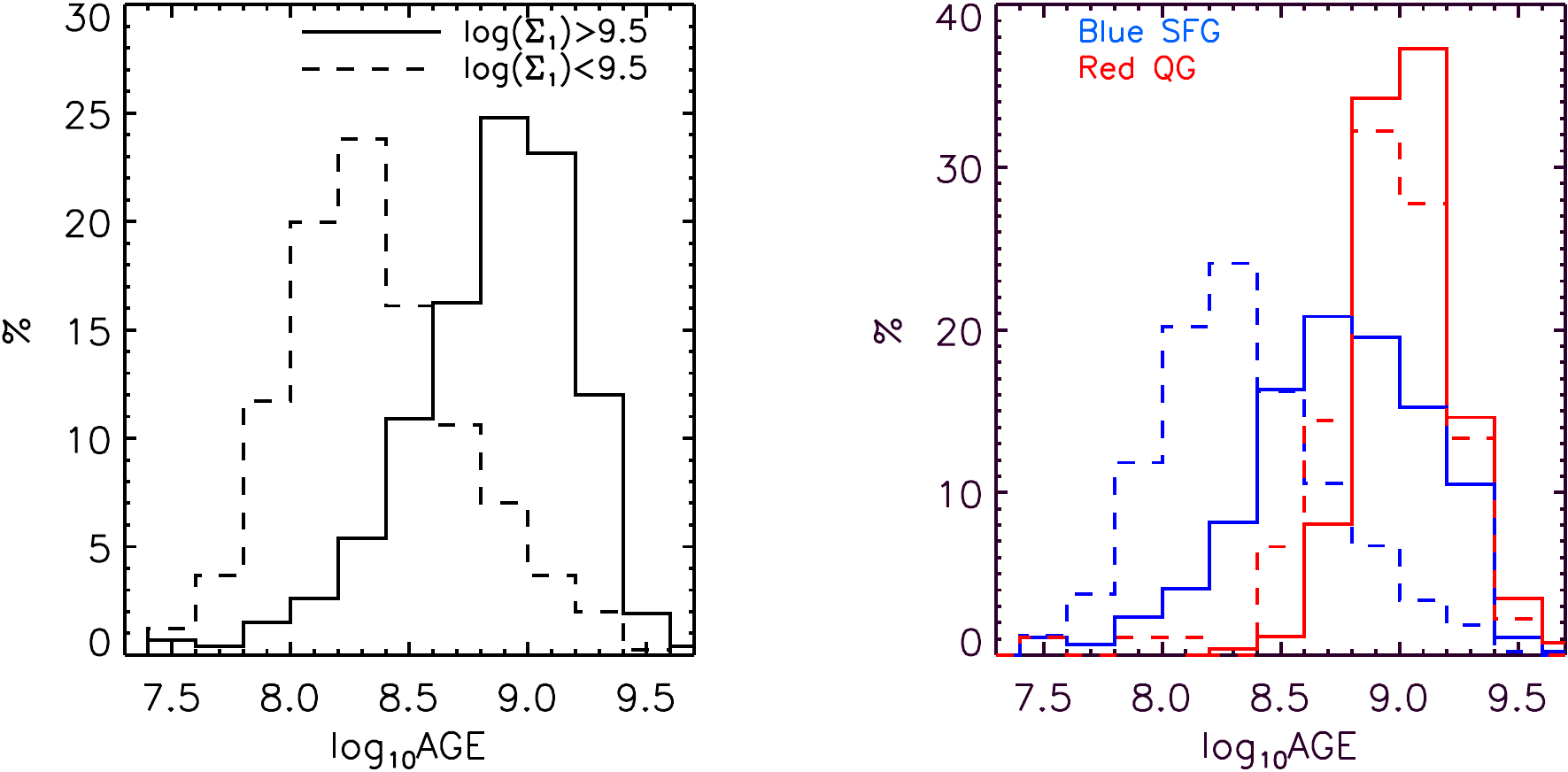}
\caption{Left: Age histograms of galaxies having high and low central density, determined
  by $\Sigma_{1}$. Solid and dashed lines represent galaxies having $log(\Sigma_{1})>9.5$
  and $log(\Sigma_{1})<9.5$, respectively. Right: Age
  histograms of UVJ selected blue SFGs and red QGs. Galaxies having high central 
  density ($log(\Sigma_{1})>9.5$) are relatively older than others for both blue SFGs and red QGs.}
\label{fig:age_hist}
\end{figure}

\begin{figure}
\epsscale{0.6}
\plotone{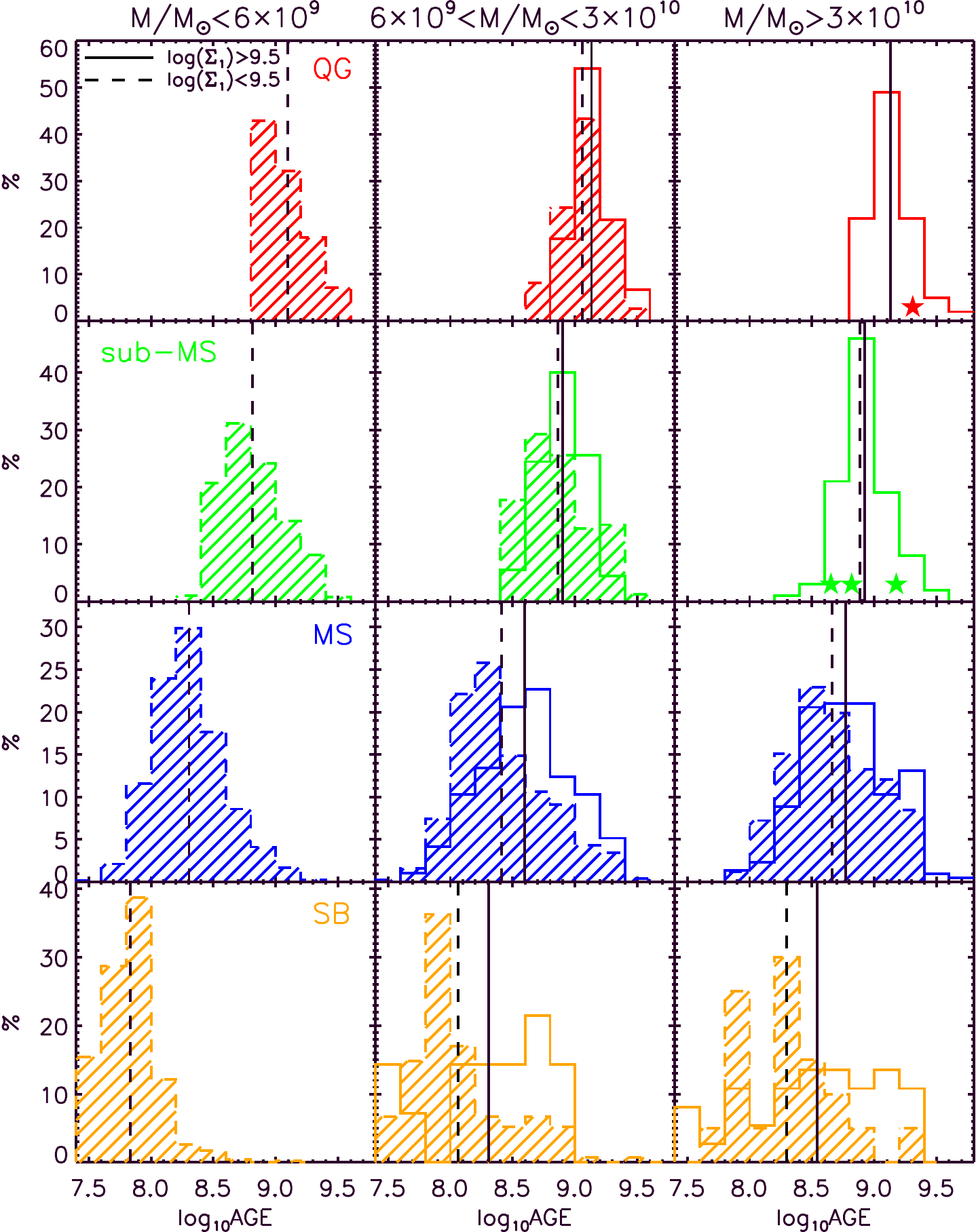}
\caption{The relative age distributions of QG, sub-MS, MS and SB galaxies in 
  three stellar mass bins. Plain and shaded histograms represent galaxies 
  having $\log(\Sigma_{1})>9.5$ and $log(\Sigma_{1})<9.5$, respectively. 
  Note that in the $M_{*}/M_{\odot}>3\times10^{10}$ mass bin, there is only one QG galaxy 
  and three sub-MS galaxies having $log(\Sigma_{1})<9.5$, which are marked 
  with red and green star symbols. Vertical lines stand for mean ages of high central 
  (solid line) and low central density (dashed line) galaxies for each population. 
  On average, the QGs are the oldest, while the SB galaxies are the youngest. 
  High central density massive sub-MS galaxies are older than the rest of the 
  star-forming galaxies. }
\label{fig:age_hist_ms}
\end{figure}

As in previous studies, we also find that the projected central mass density
$\Sigma_1$ strongly depends on $R_{SB}$, in the sense that nearly all
quiescent galaxies have high $\Sigma_1$, and, in the high mass bin, all of
them have log($\Sigma_{1}$)$>9.5$. There are, however, MS and sub-MS galaxies
that have similarly high values of $\Sigma_1$ as QGs.  The new aspect of our
study is to add the stellar age to the central density and the sSFR.  In
Figure~\ref{fig:den1_age_hm}, we further investigate the relationship between
morphology and age and find that as $R_{SB}$ decreases galaxies get
older. Galaxies on the MS, on the other hand, have $\Sigma_{1}$ that spans a
broad range of values. If fast quenching follows the formation of a compact
central core in most galaxies \citep{bar13, zol15}, then SFGs with high
$\Sigma_1$ are expected to be, on average, older than SFGs with lower central
densities. The left panel of Figure~\ref{fig:age_hist} shows the histograms of 
the median mass-weighted age for galaxies with normal and high central
densities (log($\Sigma_{1}) > 9.5$), regardless of their mass and
star--formation activity.  The right panel shows the histograms sub--divided
according to their status of either star--forming or passive in the UVJ
diagram.  Galaxies of all types and mass that have high central density are on
average older, by $\approx 700$ Myr, than the others. The right panel shows
that the difference still remains among SFGs, but that it tends to vanish
among passive ones. To explore this trend in more detail, in
Figure~\ref{fig:age_hist_ms} we further sub--divide the histograms into three mass bins
 and four $R_{SB}$ bins, respectively, for both
high-- and low--central density galaxies. First, there is a clear trend in age
from young to old, for all masses, in going from SB to QG galaxies with
peak--to--peak values $\approx 500$ Myr and $\approx 1$ Gyr. The
  similar age trend is also shown from the predictions using the SAMs that the
  ages increase from the MS galaxies to green valley to quiescent galaxies
  \citep{pan17}. We find that there is no galaxy having
  log($\Sigma_{1})>9.5$ at the smallest stellar mass bin,
  $1\times10^{9}<M_{*}/M_{\odot}<6\times10^{9}$.  In case of
  $M_{*}/M_{\odot}>6\times10^{9}$, galaxies with high central densities are
systematically older than those with low central densities (with the exception
of QG galaxies having $M_{*}/M_{\odot}>3\times10^{10}$, where there is only a 
single low--central density case). The mean age difference is very small for 
low--mass QG galaxies, but it becomes more pronounced, in both the mean values 
and the distributions, for MS and SB galaxies, of the order of $100$ Myr to 
a few $100$ Myr.  Even though the mean age difference between high-- and 
low--central density sub-MS galaxies is small, the distribution of ages suggests 
that the high central density sub-MS galaxies quench sooner and have more 
evolved stellar structures than the low central density ones, in agreement with 
the conclusions by \cite{wil15} based on the UV spectra of high-- and low--density 
galaxies at $z\sim 3$. This is also consistent with the idea that the progenitors of at
least some compact, high--central density galaxies are themselves similarly
compact before quenching.

These trends between age, $R_{SB}$ (sSFR) and $\Sigma_1$ are consistent with
the idea that galaxies develop a massive central structure as they evolve from
star--forming to quiescence and that the higher the density of this structure
the earlier and more rapid the quenching process is. They do not, however,
imply or even necessarily suggest a causal relationship between the central
density and the quenching process. 
The fact that central density of quenched 
galaxies is always observed to be in a narrow range at the
high end of the distribution could simply reflect that star formation
typically ends by the time the central parts of the galaxies have reached
those values of the projected stellar density. Finally, we observe that
massive SFGs are older than their low--mass counterparts, implying that the
former quench faster and earlier than the latter, presumably due to some
mechanism that depends on the stellar mass. 

\subsubsection{Is disk fading responsible for the compactification of galaxies
  as they quench?}

\begin{figure}
\epsscale{0.7}
\plotone{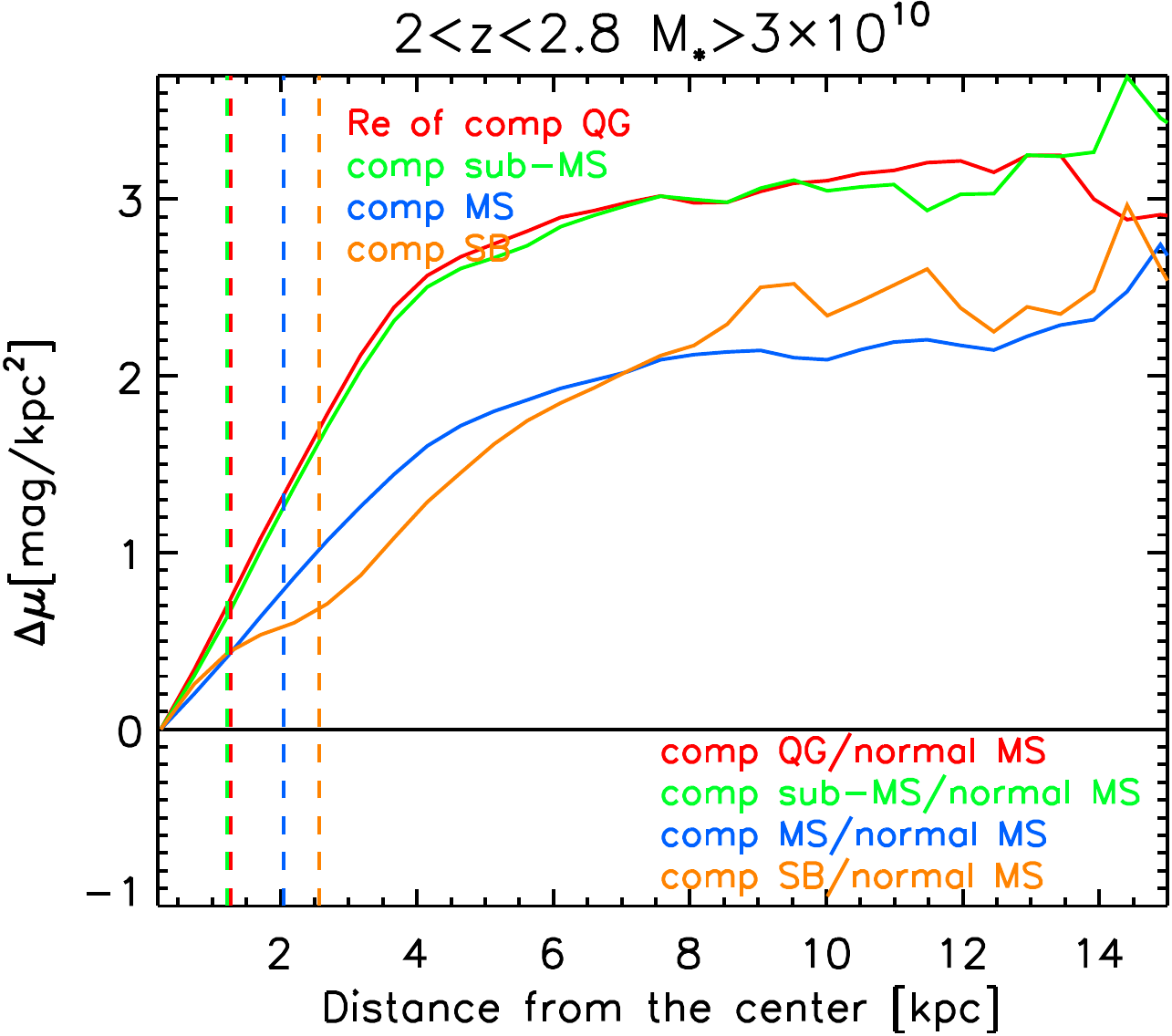}
\caption{Ratio of the average light profiles of high central density massive 
($log(\Sigma_{1})>9.5$ and $M_{*}>3\times10^{10}M_{\odot}$) galaxies 
  (compact SB, MS, sub-MS and QG) to that of the low central density
  massive MS galaxies ($\log(\Sigma_{1}<9.5)$; normal MS) at $2<z<2.8$. 
  27 QG, 27 sub-MS, 98 MS, 9 SB galaxies, and 55 normal MS are stacked 
  to increase SNR in the outer parts of the light profiles. Vertical lines with 
  different colors stand for the half-light radii ($R_{e}$) of stacks of 
  compact QG (red), sub-MS (green), MS (blue) and SB (orange). 
  The high central density massive QG/sub-MS galaxies
  have very similar light profiles and they lose their light faster than the
  normal MS, indicating that those QG/sub-MS galaxies are not faded MS
  galaxies. Note that there is about 7\% variation in the angular
    diameter distance between $z=2$ and $z=2.8$.}
\label{fig:disk}
\end{figure}

Slow-track quenching describes the passive fading of the disk
  galaxies. As they exhaust their gas, disks gradually fade, and bulges
become prominent. At low redshift, \cite{fan13} found that the bulges become
more pronounced and the disks fade away when galaxies evolve from blue to red
and suggested that slow track quenching is the effective quenching
process. Motivated from this, we test fading of disks at $2<z<2.8$. We target
galaxies having $M_{*} > 3\times10^{10}M_{\odot}$ at $2<z<2.8$ in particular 
because the number of massive, compact
galaxies drops at lower redshifts. The high central density ($\log(\Sigma_{1}) >9.5$) 
galaxies on, above, and below the MS (compact SB, MS, sub-MS, and QG) 
and the low central density ($\log(\Sigma_{1})<9.5$) MS galaxies (normal MS) 
are stacked to increase the SNR in the
outer parts of the light profiles. We, in turn, plot the differences of
surface brightness ($\Delta \mu$) between the high central density (SB, MS,
and sub-MS, QG) galaxies and normal MS galaxies in
Figure~\ref{fig:disk}. The surface brightness comes from the HST/F160W
  imaging, which corresponds to roughly the rest-frame B/V band at the
  redshifts examined. $\Delta \mu$ increases rapidly for the high central
density QG and sub-MS galaxies, indicating that they lose their light
faster than the normal MS galaxies. The stacks of the sub-MS galaxies have an
almost identical light profile ($n=4.03$, $R_{e}$=2.5 kpc) to those of the QGs
having $n=3.96$ and $R_{e}$=2.61 kpc.  If the disk fading is responsible for a
dense core of the high central density QG and sub-MS galaxies, we
should observe the remnants of disks around the core in stacks. However, we do
not find any evidence of the dead disks left in the outer regions of the high
central density QG/sub-MS galaxies (see also \cite{szo10,szo13}).  This
implies that the high central density QG/sub-MS galaxies are not faded
MS galaxies.  It might be the case that the disk fading (slow track quenching)
is just less efficient at high redshift because there is not enough time for
slow fading to occur or it might occur mainly in massive halos like
  clusters, which haven't formed yet. Alternatively, fast quenching may be
dominant at high redshift because of the higher gas fractions at early times
invoking more energetic events and causing a rapid quenching of SF
\citep{pan17}.

\subsubsection{Is the major merger-driven process a dominant quenching mechanism?} 
\label{sec:mergerscenario}

In early simulations, strong bursts of star formation by gas-rich mergers
leave a compact merger remnant behind \citep{spr05, hop08}.  Subsequently, the
compact remnant rapidly quenches star formation and if the gas fraction is
large (e.g. $f_{gas}>50$\%) its stellar mass is sufficiently large that
resulting morphology of the galaxy resembles that of the compact quiescent
galaxies observed at $z\sim2$. The simulated gas-rich merger remnants,
however, also have extended stellar halos that originate from the violent
relaxation of the pre-existing stellar components \citep{wuy10}. It has been
suggested that their morphology is not consistent with that of
the massive, compact quiescent galaxies described here, since no extended
stellar halos are observed around compact QGs and sub-MS galaxies, either
individually or from stacked images, like the one in Figure~\ref{fig:disk}
(e.g. see Figure 16 of \cite{wil14}, which compares the S\'{e}rsic light profile
parameters of the merger remnants to those of individual compact galaxies as
well as those of stacked images). The results presented here are in agreement 
with the conclusion of \cite{wil14}. For example, most starbursts found in
this study have larger sizes and more diffuse light profiles, as well
  as lower G and higher $M_{20}$, on average than other types of galaxies in
the log(SFR)-log(M$_{*}$) diagram, in particular including compact quiescent
galaxies of comparable stellar mass.  We visually inspected the starbursts
galaxies with $M_{*}>3\times10^{10}$ and $log(\Sigma_{1})>9.5$ whose
redshift is $z>2.8$ (about half of the starburst sample) and indeed confirm
the presence of compact compact structures embedded in a more diffuse light
distribution. Most of these sources are found to host heavily dust--obscured
star formation and, in approximately 50\% of the cases, an AGN (based in the
detection of X-ray flux in the Chandra images). If most starburst galaxies are
mergers observed during the burst of star formation, the compact source would
be consistent with the remnant of a highly dissipative gas process
\citep{hop08,wuy10}. The problem is how to get rid of the extended stellar
component during a purely passive transition into the quenching phase
\citep{van08, cas11, cas13}, since the light profile of most massive QGs is
that of a ``naked'' nugget and is quite different from that of the starburst
galaxies, as shown in Figure~\ref{fig:disk}. We caution, however, that it is
not clear if the simulations of wet mergers have sampled a large enough volume
of the parameter space of the progenitors such that compact passive galaxies
consistent with the observations can still be produced, and/or if the
observations have reached the sensitivity to rule out low surface brightness
extended halos.

\subsection{Evolution of galaxies in the log(SFR)-log(M$_{*}$) plane and morphological transformation}

Do galaxies transform their morphology as their stellar mass grows and their
star--formation activity evolves and eventually quenches? Or, stated
differently, is morphological transformation implied by the observed trends of
morphology, stellar age and position on and around the MS at $1.2<z<4$? As
others before us, we also find that the distribution of the projected central
density $\Sigma_1$ narrows toward the top end of the overall distribution as
galaxies move away and below from the MS. We did not find any evidence that
this is the result of a fading disk, which would leave the compact bulge as
visible evidence. In other words, it is unlikely that compact quiescent
galaxies form directly from normal SFGs on the MS through simple disk fading,
as the star formation fades. \cite{wel15} tracked the assembly histories of
compact quiescent galaxies at $z=2$ in the Illustris simulation, and found
that there are two dominant mechanisms of formation of compact quiescent
galaxies: 1) compact galaxies form via starbursts from major mergers between
$z\sim 2-4$, or 2) compact galaxies form at very early epochs and maintain
their morphologies until $z\sim2$. Based on our observational findings
(Section~\ref{sec:mergerscenario} and Figure~\ref{fig:disk}), the first
scenario seems unlikely, since starbursts, the most likely candidates for wet
mergers, are characterized by a significant amount of diffuse light around the
compact central sourtce, presumably tidal debris, stripped stars and in
general the violently relaxed stellar component, which seems not to be
observed in compact quenched galaxies, either individually or in very deep
stacked images. The alternative scenario suggested by our findings is that a
significant fraction of the compact massive galaxies that dominate the counts
at $z\sim 2$ assemble at earlier times when the universe was denser, and
subsequently evolve mostly through in--situ star formation keeping their
compact morphologies \citep{wil14, wil17, van15, lil16}.

\begin{figure}
\epsscale{1.0}
\plotone{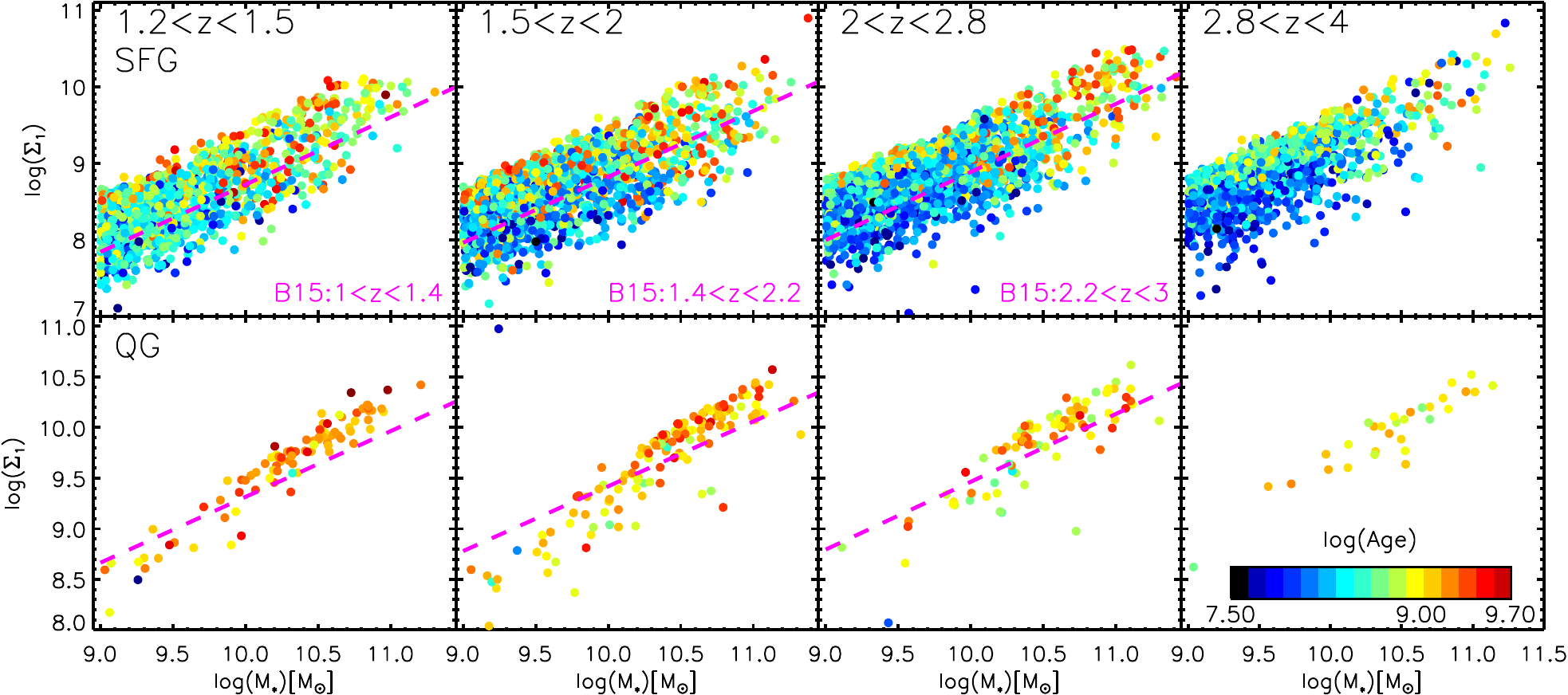}
\caption{Projected central mass density vs. stellar mass for UVJ selected 
star-forming galaxies (SFG: top) and quiescent galaxies (QG: bottom) at four redshift bins. 
The median mass-weighted stellar age is color-coded and the best-fit log($\Sigma_{1}$)--log(M$_{*}$) 
relations computed from \cite{bar17} for two populations are overplotted in the redshift bins in common. 
Our measures are in good agreement with \cite{bar17} and we show a clear correlation 
between $\Sigma_{1}$ and M$_{*}$ at all explored redshifts.}
\label{fig:sigma1_mass}
\end{figure}

\begin{figure}
\epsscale{0.9}
\plotone{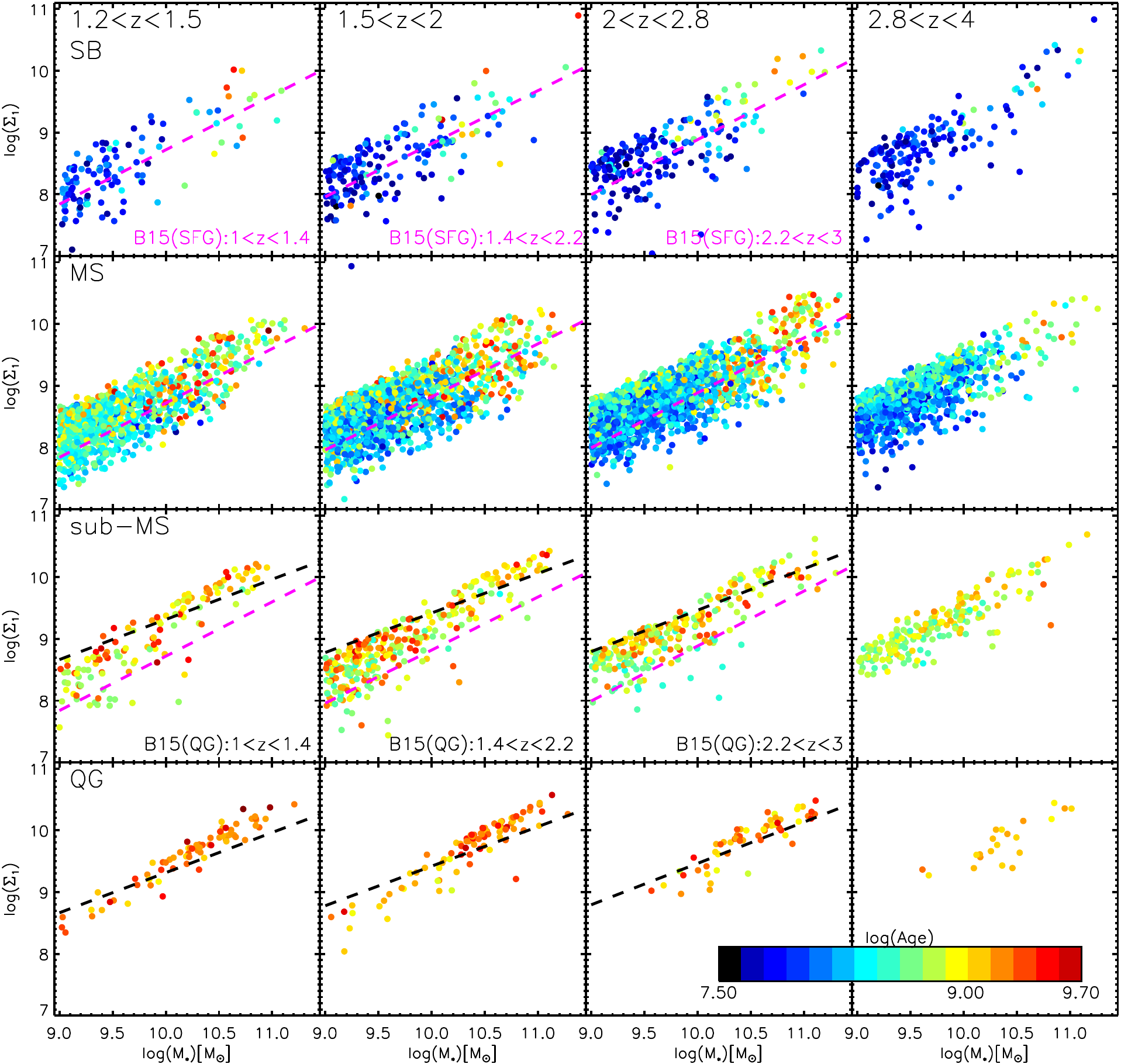}
\caption{Projected central mass density vs. stellar mass for four galaxy populations 
relative to the MS at four redshift bins. The median mass-weighted stellar age is 
color-coded. The best-fit relation for UVJ selected QGs (black dashed line) from \cite{bar17} are 
overplotted in the bottom panel (QG galaxies). For SB and MS galaxies, the best-fit relation 
for UVJ selected SFGs (magenta dashed line) from \cite{bar17} are used while best-fit relations 
for both SFG and QG are overplotted for sub-MS galaxies. Narrower dispersion of $\Sigma_{1}$ with 
$M_{*}$ represents that different quenching processes play a role at low and large masses (cf. sub-MS panels).}
\label{fig:sigma1_mass_ms}
\end{figure}

We investigate the relationship between $\Sigma_1$ and stellar mass as
a function of stellar age in our four redshift bins for star--forming and
quiescent galaxies selected in the UVJ plane and relative to their distance
from the MS, respectively, in Figures~\ref{fig:sigma1_mass}
and~\ref{fig:sigma1_mass_ms}. Our measures are in good agreement with the
analogous ones by \cite{bar17}, and, for comparison, in the figures we display
their best fit overplotted to our data in the redshift bins in common. There
is a clear correlation between central density and stellar mass, which becomes
tighter for galaxies with decreasing activity of star formation. Compared to
\cite{bar17}, our analysis adds the extra dimension of stellar age, which
shows that 1) at fixed stellar mass galaxies with higher central density tend
to be older, and 2) at larger stellar mass and hence larger central density
galaxies tend to be older. Since the distribution of $\Sigma_1$ is nearly
constant with redshift at all mass (Figure~\ref{fig:hist_sigma1}), the
implication is that, as galaxies form stars, both their stellar mass and central 
stellar density grow, and the growth continues until the central density
enters in a relatively narrow range of values at which point the growth stops
and galaxies quench. 

\begin{figure}
\epsscale{0.9}
\plotone{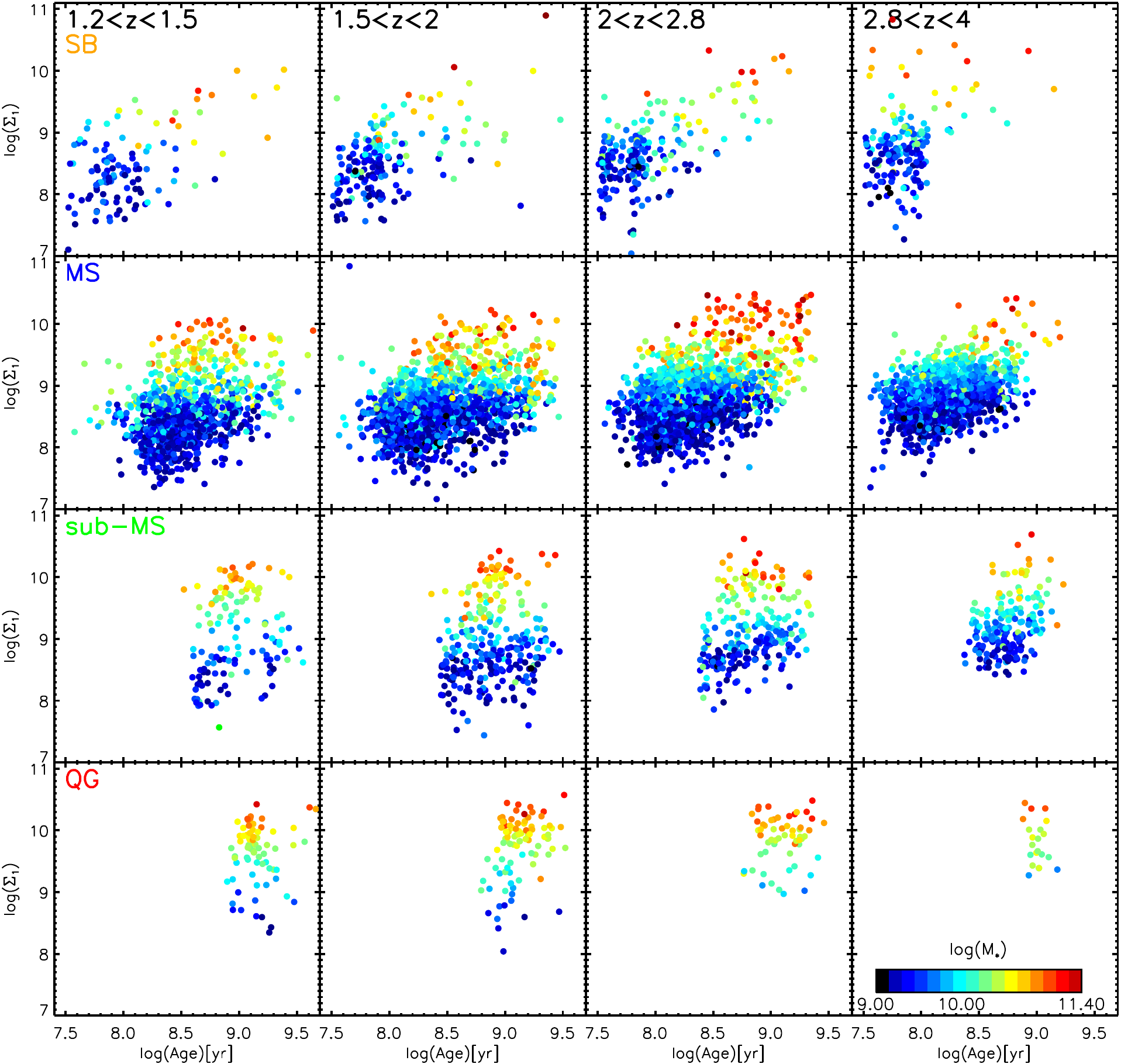}
\caption{Projected central mass density ($\Sigma_{1}$) vs. 
median mass-weighted stellar age (Age) for four galaxy populations 
relative to the MS at four redshift bins. The stellar mass is 
color-coded. When galaxies are in the MS and SB regions, both $\Sigma_{1}$ 
and $M_{*}$ increase with increasing stellar age. But, when they are quenched, the 
growth of both $\Sigma_{1}$ and $M_{*}$ stops and the correlation with ages ends. 
Thus, QGs show a ``vertical" shape in the bottom panel. }
\label{fig:sigma1_age}
\end{figure}

\begin{figure}
\epsscale{0.9}
\plotone{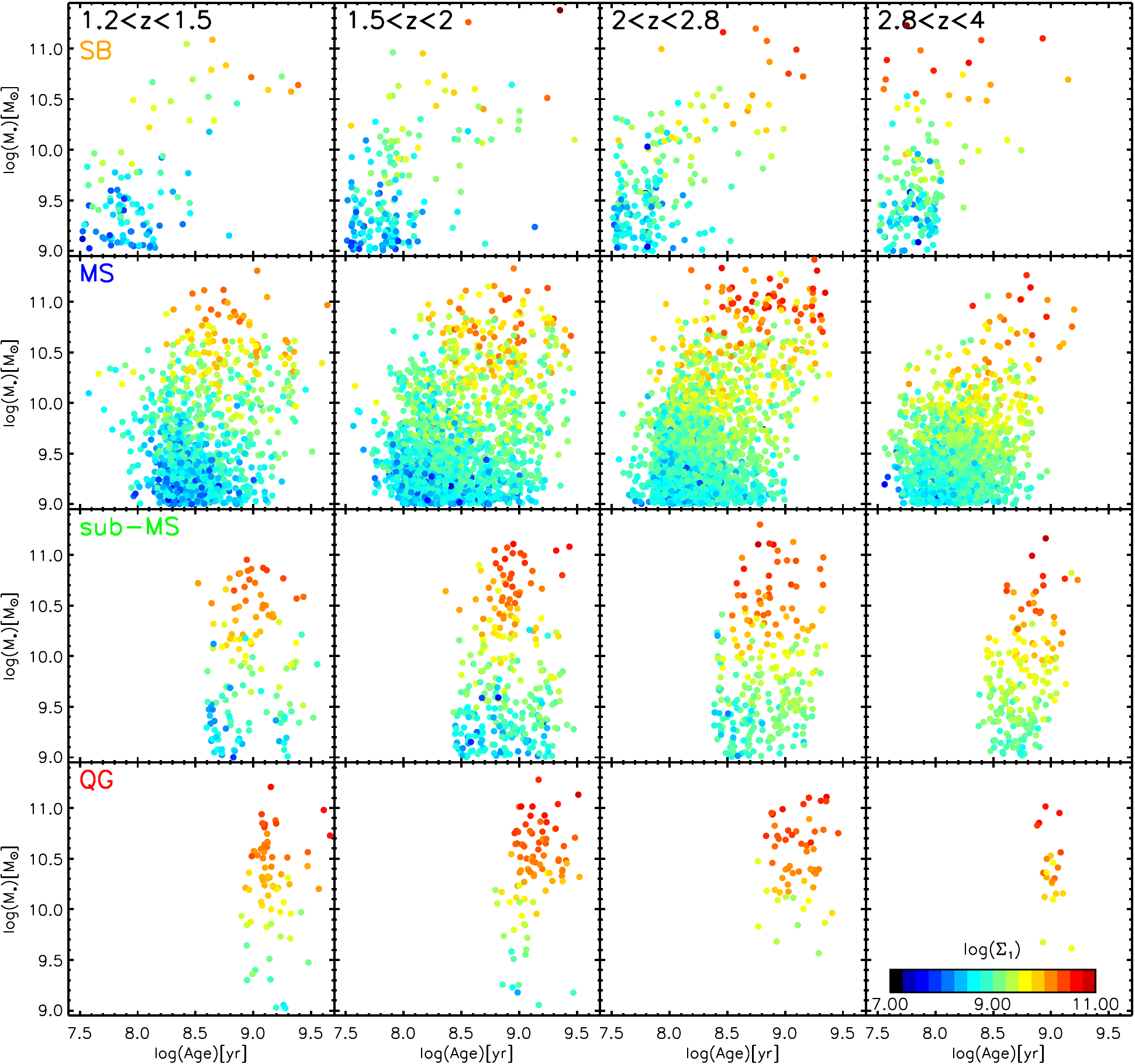}
\caption{Stellar mass ($M_{*}$ vs. 
median mass-weighted stellar age (Age) for four galaxy populations 
relative to the MS at four redshift bins. The projected central mass 
density is color-coded. As stellar age increases, galaxies become more massive 
and their central density increases. But, the correlation between $M_{*}$ and Age 
is no longer exist, once galaxies stop their star formation. }
\label{fig:mass_age}
\end{figure}

We additionally study the correlation between $\Sigma_{1}$ and
  age in Figure~\ref{fig:sigma1_age}, 
where the points representing the galaxies have been color-coded with 
their stellar masses and the 
correlation between stellar mass and age, with the points color-coded according 
to the value of $\Sigma_{1}$ in Figure~\ref{fig:mass_age}. Both figures show 
that, as galaxies form stars, i.e. when they are in the MS 
and SB regions, both $\Sigma_{1}$ and $M_{*}$ increase with increasing 
stellar age. Namely, during the star formation phase galaxies become 
both more massive and their central density increases as time increases. 
As galaxies start to quench, the correlations with ages becomes weaker. 
However, galaxies with older ages are more 
massive and also have larger central density $\Sigma_1$. 
Once galaxies have ceased star formation, the growth of both 
$\Sigma_1$ and $M_*$ stops and the correlations with ages ends as well. 
The plots for QGs become ``vertical'' (the bottom panels 
in Figure ~\ref{fig:sigma1_age} and \ref{fig:mass_age}) because 
stellar age differences are become smaller and harder to measure 
relative to the large absolute age of the galaxies.

Of the three variables, Age, $\Sigma_1$ and $M_*$, Age should
naturally be the independent one. Indeed, our results show that both stellar
mass and central density increase with age as galaxies evolve: stellar mass
reflects the history of star formation and $\Sigma_1$ reflects the history of
dissipative gas accretion in galaxies. However, more massive galaxies are also
more efficient at promoting dissipative gas accretion, which results in
the relatively tight correlation between $\Sigma_1$ and $M_*$. But while the 
measures of $\Sigma_1$ and $M_*$ are relatively accurate, the measures of 
stellar age are comparatively noisier, which washes out a bit the
correlation between age and the other two variables.

In conclusion, the scaling relationship of $\Sigma_1$ and stellar mass
  and their evolution with redshift for massive galaxies show that massive
  galaxies quench when their central density ($\Sigma_1$) how grown to value
  in a narrow range from the maximum observed one (which is mass dependent),
  and the narrowness, threshold (for quenching) and maximum value are
  essentially independent from redshift, as also argued by \cite{bar17}. Note
  that this {\it does not} imply any causal relationship between high core
  density and quenching: the two phenomena simply happen together at some
  point during the growth of a galaxy.  Qualitatively, the larger dispersion
  of $\Sigma_1$ values at fixed stellar mass observed in star--forming
  galaxies simply reflects the different points in times during their
  evolution when the galaxies are observed, as evidenced by the trend that at
  fixed stellar mass older galaxies have larger central density. If a galaxy
  can keep growing in mass, it eventually reaches a critical size,
  approximately log($M_*)\approx 10.5-11$ (see also \cite{lil13, pen10}), at
  which point the probability that it quenches rapidly increases with the
  mass, with the quenching caused by something that happens when the galaxy
  reaches that stellar mass. At large mass, the relatively narrow dispersion
  of $\Sigma_1$ suggests that the quenching is probably caused by processes
  internal to the galaxies and the quenching probability is a steep function
  of the mass (small dispersion).  This trend is particularly evident in the
  sub--MS panel of Figure~\ref{fig:sigma1_mass_ms}, \ref{fig:sigma1_age}, and
  \ref{fig:mass_age} (i.e. ``quenching'' galaxies). Quenching, however, could
  also happen for other reasons before the galaxy reaches the critical size,
  and this seems suggested by the larger scatter in the $\Sigma_1$ vs. $M_*$
  relationship (cf.  sub-MS panel) at lower masses. In this case quenching
  would be caused by something external to the galaxies (environmental
  quenching). Evidence that environmental quenching plays a role in shaping
  the red sequence has been found by \cite{qua12}. More recently,
  \cite{tal14}, \cite{guo17}, and our recently submitted work (Ji et al. 2017,
  ApJ) found direct evidence of environmental quenching in low-mass ($\approx
  10^9 M_{\odot}$) galaxies up to $z\sim2.5$. Ji et al. (2017) reach the same
  conclusions with \cite{qua12} that the environment plays a greater role in
  assembling the red sequence for low-mass galaxies than more massive ones
  (for which other mechanisms control quenching), based on both clustering
  arguments and on the quenched fraction as a function of stellar mass, which
  rapidly increases with stellar mass and peaks around the critical value 
  log($M_*)\approx 10.5-11$.  Taken
  together with the former argument about the internal process and the other
  evidence of environmental quenching, the thickening of the dispersion of the
  $\Sigma_1$ vs. stellar mass relationship at lower masses is therefore
  ``consistent'' with external quenching mechanisms that become more and more
  important at lower masses. Summarizing, we suggest the possibility that
  external mechanisms, most likely related to the environment, effectively
  quench the galaxies before internal ones, freezing the central density at
  the current value, which is lower than the one the galaxy would have had if
  it had the possibility to quench by internal mechanisms.

\subsection{Old galaxies in the Main Sequence}

\begin{figure}
\epsscale{1.0}
\plotone{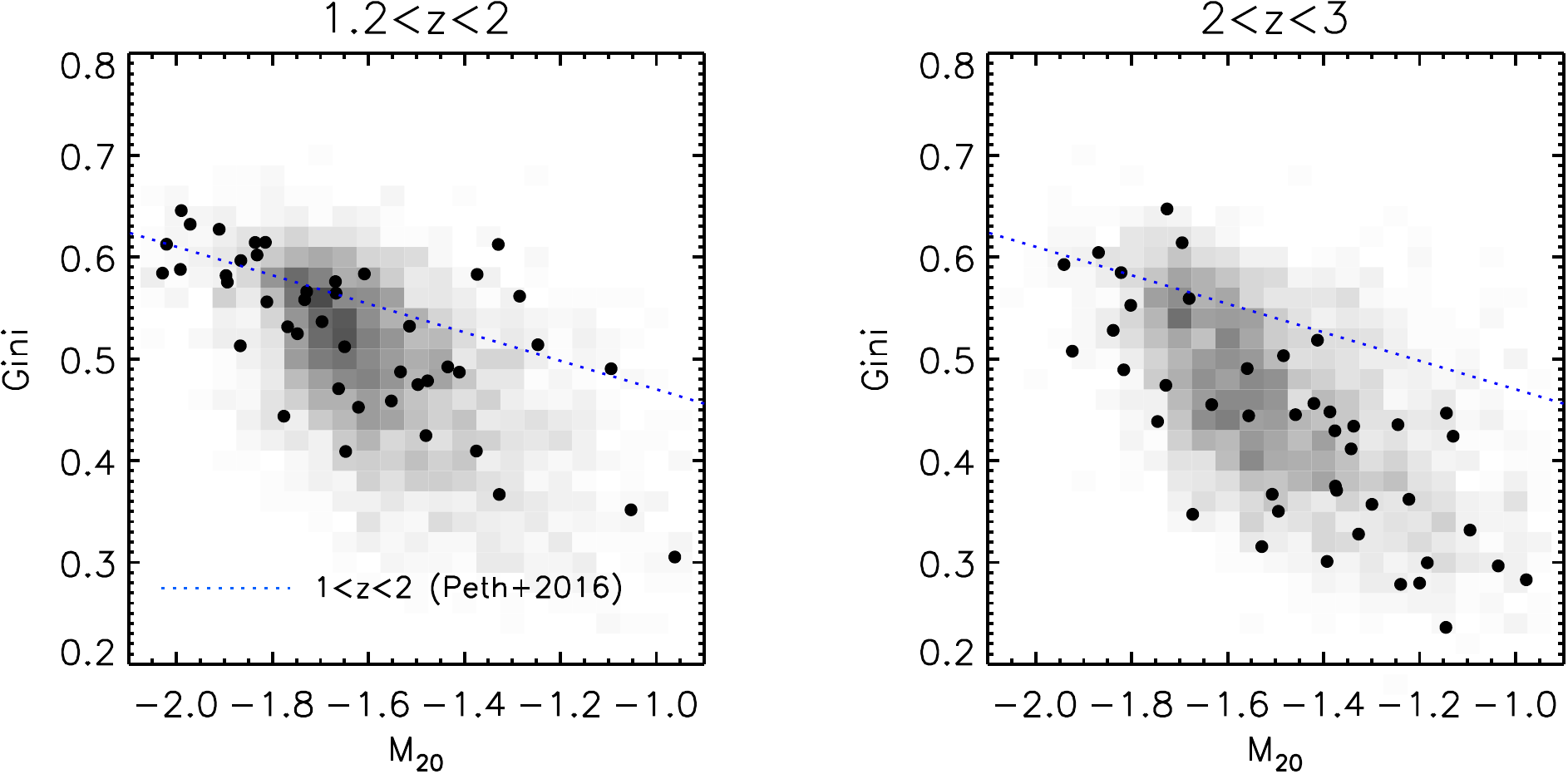}
\caption{Gini vs. $M_{20}$ for old Main Sequence galaxies (black points) at 
two redshift bins. We find no galaxy with Age$>1$ Gyr and
  $M_{*}>3\times10^{10}$ M$_{\sun}$ in the main Sequence at $z>3$. The
  distribution of all Main Sequence galaxies is colored according to the
  density of sources. The darkest color represents densest regions. We
  overplot the definition of mergers of $1<z<2$ galaxies from \cite{pet16} (blue dotted line). 
  We find that there is no difference in galaxy morphologies of old MS galaxies (black points) 
  and the rest of the galaxies in the MS. }
\label{fig:mergers_old}
\end{figure}

\begin{figure}
\epsscale{0.7}
\plotone{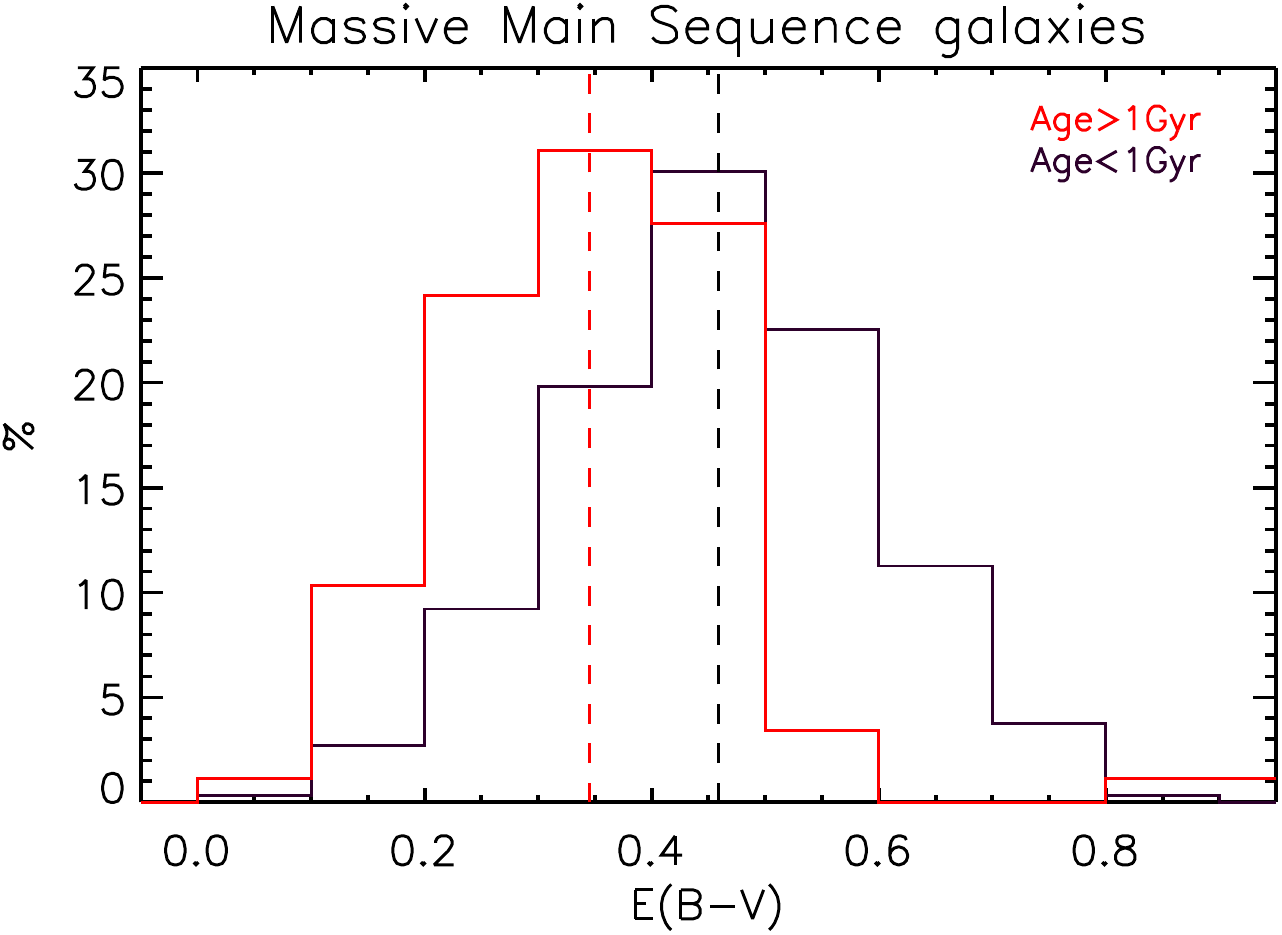}
\caption{The distribution of the values of the best--fit $E(B-V)$ parameter
  for massive ($M_{*}/M_{\odot}>3\times 10^{10}$) MS galaxies in the two age bins
  that define the old MS galaxies. The $E(B-V)$ distributions of the two
  groups of galaxies are similar in shape, but the peak of that of the old
  galaxies is $\approx 0.1$ smaller. The is an indication of the age--obscuration 
  degeneracy of SED fitting procedures. The older fitted age and the smaller $E(B-V)$ color 
  excess of old MS galaxies are within the covariance of the degeneracy between age and 
  obscuration of our SED fitting procedure.}
\label{fig:ebv_old}
\end{figure}

The addition of the ``third dimension'' of age (median mass--weighted stellar age) to
the log(SFR) vs. log(M$_{*})$ plane reveals the presence of galaxies with old
stellar age, i.e. $Age>10^9$ year, still on the Main Sequence. These old MS
galaxies are observed in both mass bins considered here, but seems to be more
common among the massive galaxies ($M_{*}/M_{\odot}>3\times 10^{10}$: 29.7\%) 
than in the low mass ones
($M_{*}/M_{\odot}<3\times10^{10}$: 3.7\%). In the high--mass bin,
they appear to span the full range of $\Sigma_1$ values, and seem to be most
abundant at $1.5<z<2.8$, while they are nearly absent in the highest redshift
bin. Among the low--mass galaxies they seem confined to high $\Sigma_1$ values
and their number progressively increases with decreasing redshift,
i.e. increasing cosmic time, perhaps signaling a genuine aging of the
star--forming galaxies as they approach the quenching phase.

We further investigate the nature of the massive old MS galaxies to
explore the possibility that they are somewhat different from the rest of the
MS galaxies of similar mass. As Figure~\ref{fig:den1_age_hm} shows, these
galaxies do not preferentially occupy a special region of the MS and, in
particular, they do not sit near the lower boundary of the MS, closer to the
quenching region. A look at the best-fit SFH of these galaxies shows the same
distribution of the five analytic functions that we have used as in the rest
of the MS.

We also look at their morphology and made a visual inspection to search
for evidence of an excess of mergers or interacting systems, finding that
these galaxies seem similar to the rest of those in the MS. More
quantitatively, a number of studies \citep{lot04, lot08, pet16} have shown
that the Gini and $M_{20}$ coefficients provide useful diagnostics of merging
and interaction at $z<2$. Following their results, we use the
$G$--$M_{20}$ diagram to inspect the incidence of mergers among massive old MS
galaxies at $1<z<2$ using the definition of from \cite{pet16}, who suggested
that most of the mergers at these redshifts are located above the blue dotted
line in the left panel of Figure ~\ref{fig:mergers_old}. As the figure shows,
there is no evidence that the relative distributions of the old galaxies and
of the rest of the MS differ, both at $1.2<z<2$, where the index was
  calibrated, and also at $2<z<3$. This confirms the visual analysis that the
frequency of merging events and/or of interactions among the old MS galaxies
is the same as that of the rest of the MS. This suggests that the old age
derived in our fitting procedure is unlikely to be the results of rejuvenation
effects of passive galaxies due merging events, although the accretion of gas
or of very faint gas-rich satellites cannot be excluded.

Additionally, we look at the fraction of AGN classified from the latest deep
Chandra images in GOODS-North by \cite{xue16} and by \cite{luo17} in
GOODS-South. AGN hosts are classified among X-ray sources if they
  satisfy one of the six criteria listed in \cite{luo17} (see section 4.7,
  paragraph 2). There are 97/380 AGN hosts among massive galaxies in
  the MS, namely 26\% with a poisson uncertainty, 0.03. Among the old
  galaxies, there are 16/87 AGN hosts, i.e. 18\% with a poisson uncertainty,
  0.04. These numbers show no evidence that AGN activity among the old MS
  galaxies is different from the rest of the MS.

We observe, however, a difference in the distribution of the values of
$E(B-V)$ from the best-fit SED procedure among the two groups of galaxies. As
Figure~\ref{fig:ebv_old} shows, while the shape of the two distributions is
rather similar, the peak of the old galaxies' one is $\approx 0.1$
smaller, very likely a manifestation of the age--obscuration degeneracy of SED
fitting procedures based on broad--band photometry, no matter how deep or
accurate. We conclude that most likely there is no intrinsic difference in the
physical properties of what we called ``old MS galaxies'', i.e. galaxies whose
best--fit stellar age is larger than 1 Gyr, and the rest of the galaxies in
the MS. The older fitted age and the smaller $E(B-V)$ color excess are within
the covariance of the degeneracy between age and obscuration of our SED fitting
procedure. 

\subsection{Does the MS slope depend on morphologies of galaxies?}

\begin{figure}
\epsscale{0.6}
\plotone{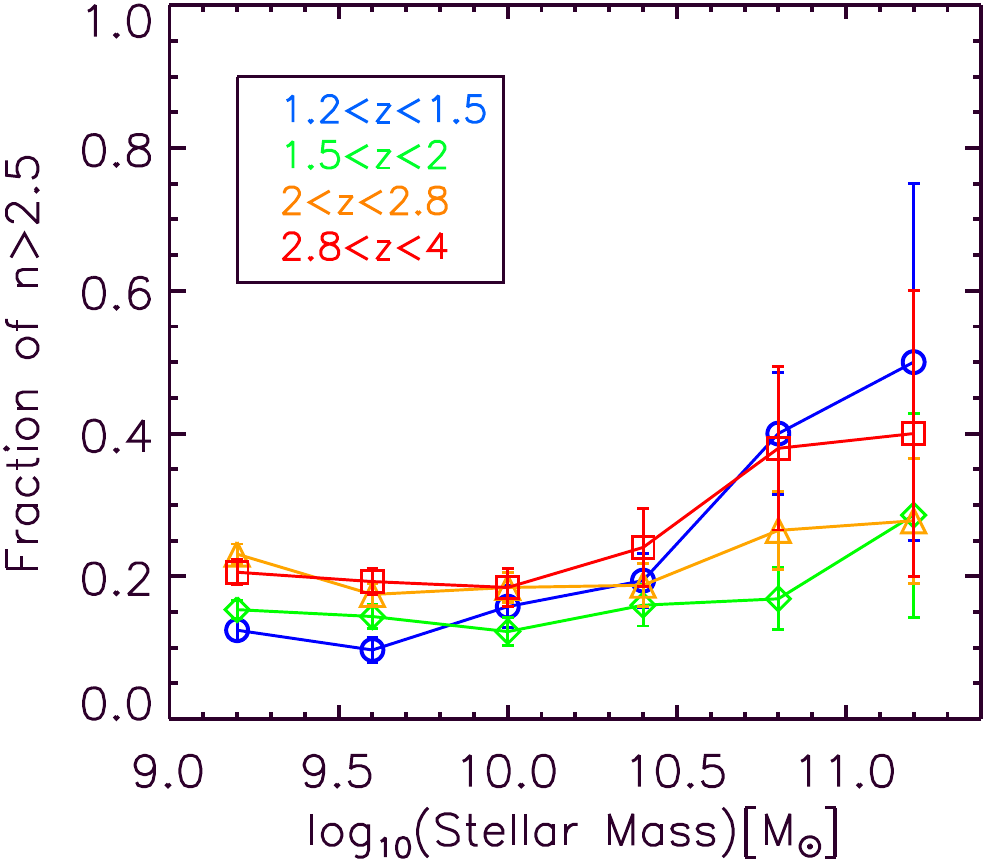}
\caption{Fraction of SFGs having S\'{e}rsic index (n) $>2.5$ as a function of
  the stellar mass (with $\Delta log(M_{*})=0.4$) for four redshift bins.  We find a weak increase
  of the fraction of bulge-dominant galaxies ($n>2.5$) at all explored redshifts. 
  The error-bar represented here is 1$\sigma$ uncertainties based on Poisson statistics.}
\label{fig:fbulge}
\end{figure}

\begin{figure}
\epsscale{0.7}
\plotone{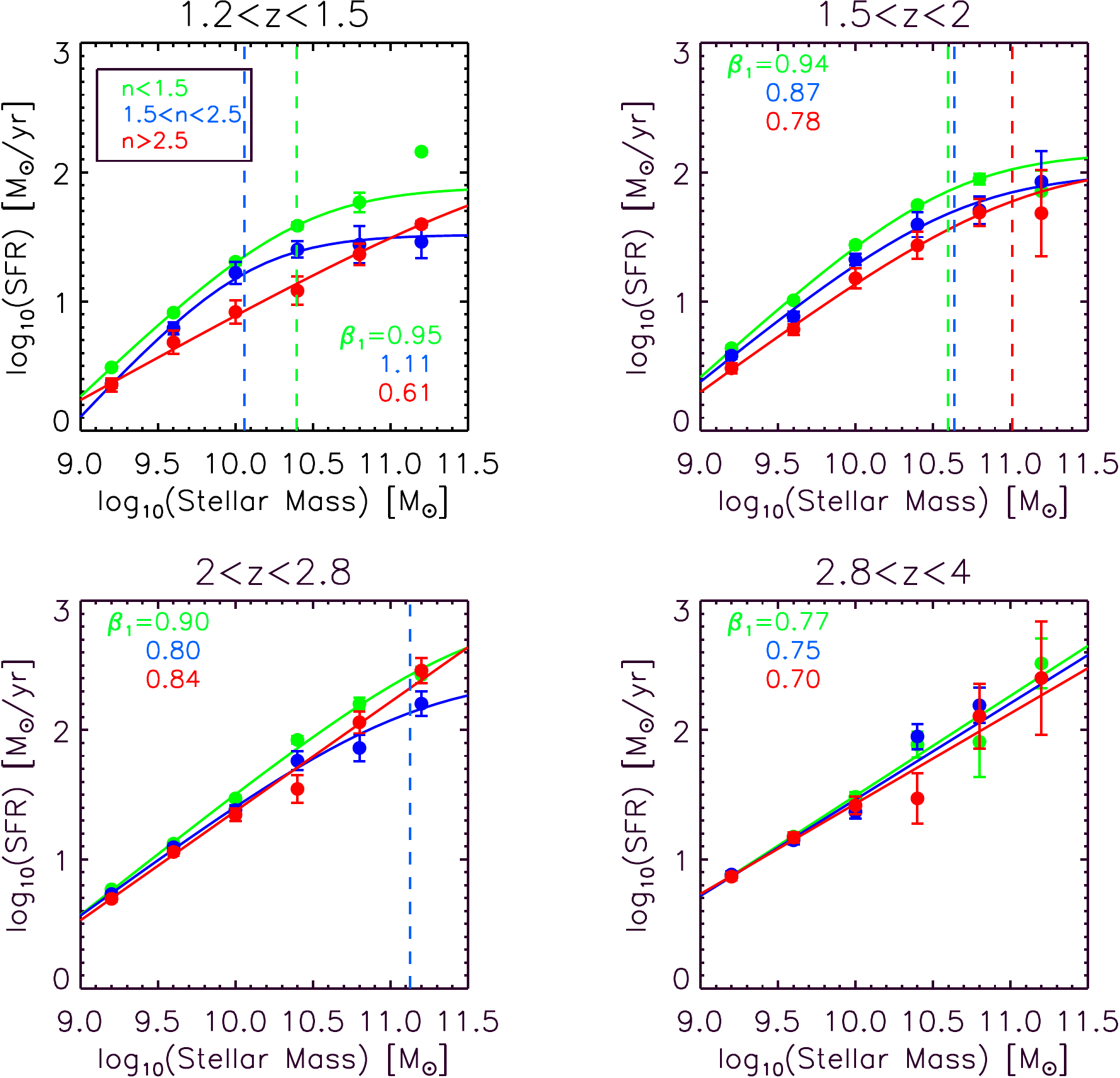}
\caption{MS of SFGs having different S\'{e}rsic indexes: we classify galaxies
  into three classes, $n<1.5$ (green), $1.5<n<2.5$ (blue) and
  $n>2.5$ (red), and then fit the slope of the log(SFR)-log(M$_{\star}$) relation separately with
  the standard error ($\sigma / \sqrt{N}$). The vertical line represents $M_{0}$ of each population, 
   and $\beta_{1}$ is the MS slope below
  $M_{0}$ or the slope in case of explained by a single power-law. Note that
  there is no galaxy having $M_{*} >10^{11}M_{\odot}$ and $1.5<n<2.5$ at
  $2.8<z<4$. Also, at $1.2<z<1.5$, the most massive bin of $n<1.5$ is excluded from a fit 
because of a poor statistic (only two galaxies 
having almost identical SFR values).}
\label{fig:ms_sersic}
\end{figure}

In Section~\ref{sec:sfr}, we show that the bending of the MS slope is a function of redshift and 
is prominent at $z<2$. By $z\sim2$, the slope is close to $\sim0.8$ at all stellar masses. 
The bending of the MS implies that massive galaxies are experiencing a
decrease of SF, i.e. massive galaxies have lower sSFR relative to less massive
galaxies. The reason for this bending of the MS is yet uncertain. \cite{abr14} 
showed that MS slope is almost unity without a bending of the MS in the
local universe when they use $M_{disk}$ instead of $M_{*}$. This suggests that
including bulge components might cause the bending of the MS at high masses
since the bulge does not contribute significantly to SF. \cite{lan14} also concluded that 
the bulge growth at high masses leads to the departure
of the MS slope from unity using the CANDELS/3D-HST. Recently, \cite{sch16} 
extended this study out to $z=1$ by measuring the slope of the
SFR-M$_{disk}$ relation. In disagreement with previous studies, they found a
flattening of the MS at high masses using M$_{disk}$ and concluded that the
secular growth of quiescent bulges in SFGs is not the main driver for the change
of the MS, at least at $z=1$.

To test for a possible relationship between the bulge growth and the bending
of the MS at high masses, we use the S\'{e}rsic index to broadly classify
disk--like and spheroid--like galaxies. S\'{e}rsic index has been widely used
to distinguish early-type quiescent galaxies with $n>2.5$ \citep{kaj15}
 and star-forming galaxies with $n<1.5$ \citep{shi15}. We
compute the fraction of bulge-dominant galaxies, simply identified as SFGs
having $n>2.5$, as a function of stellar masses in
Figure~\ref{fig:fbulge}. There is a weak increment of the fraction of $n>2.5$
over all stellar masses at all explored redshifts. However, it is insufficient to support the
idea that higher bulge fractions at high masses cause the bending of the MS
due to small number statistics (large error-bars at the massive end). 

We further study whether the break of the MS depends on morphologies in
Figure~\ref{fig:ms_sersic}. Here, SFGs are separated into galaxies having
disky ($n<1.5$), intermediate ($1.5<n<2.5$) and bulge-dominant structures
($n>2.5$). Because we do not see a depression in the slope 
at the massive end at $z>2$, we do not expect to see a trend with morphology at 
these redshifts. If the bulge growth in a massive galaxy at $z<2$ is a key driver of the
decrease of the MS slope at high masses, the MS slope measured only using
galaxies having $n>2.5$ is expected to show the bending of the MS and it
should be different from one measured using galaxies having $n<1.5$. 
However, we find that the bending of the MS is evident for galaxies having $n<1.5$ and
$1.5<n<2.5$ out to $z=2$, while the MS of galaxies having $n>2.5$ is rather
explained by a single power-law, except for galaxies at $1.5<z<2$. At $z<2$, $n<1.5$ have relatively 
steeper slopes (see $\beta_{1}$ in the
Figure~\ref{fig:ms_sersic}) and higher SFRs at a given stellar masses on
average than $n>2.5$. But, the average SFRs at the highest mass bin weakly 
depends on $n$, indicating that massive galaxies have lower SFRs irrespective of
their morphologies. Apparently, the MS does not depend on galaxy morphologies
based on S\'{e}rsic index. \cite{whi15} also reported the weak
dependence on $n$ at $z<1$ amongst star-forming galaxies, but do not see strong evidence at $z>1$.
They suggested that bulges in massive $z\sim2$ galaxies
are actively building up and the stars in the bulges are relatively younger
than old bulges within SFGs at $z<1$. However, using gas masses estimated by stacking
\emph{Herschel} data, \cite{sch16} provided evidence that the low sSFRs in
massive galaxies on the MS are caused by a slow downfall of the star formation
efficiency (SFE = SFR/M$_{\mathrm{gas}}$) at $z<2$, and not by the growth of bulges. 

In sum, we do not find evidence that the inclusion of galaxies having bulge-dominant
structures causes the bending of the MS at $1.2<z<2$. Rather, massive galaxies have
relatively lower SFRs on average regardless of their morphologies. This is
consistent with the notion that the decrease of SF, and ultimately quenching,
is driven by internal processes, e.g. AGN and/or stellar feedback, which
depend on the mass of the galaxy (mass quenching: \cite{pen10, pen12, lil13, lil16}).

\section{Summary} 

We have studied the morphology and stellar ages of galaxies located on and
around the Main Sequence of star formation in the log(SFR)-log(M$_{*}$) plane
at $1.2<z<4$.  In order to constrain the MS more accurately, we have
re--measured the physical properties of galaxies (stellar mass, age, SFR)
using new SED fitting procedures.  In our adopted methodology, we explore 
five analytical star formation history models (constant SFR; linearly
increasing; exponentially increasing; delayed; exponentially decreasing) and
find the Best Fit SFH for individual galaxies. We test this ``Best Fit SFH''
procedure using mock galaxies from high--resolution N--body simulations
coupled to the semi--analytical models. We find that the Best Fit SFH
procedure recovers the intrinsic properties of the galaxies more accurately
than using one fixed SFH for all galaxies. Our key results are summarized as
follows:

\begin{enumerate}
\item At $z<2.8$, the main sequence (MS) observed in this study is tight with
  a constant dispersion, $\sigma\sim0.37$dex, and the slope is curved at the
  turnover mass, $M_{0}$, which mildly increases from $10^{10.6}$ to
  $10^{11.5}M_{\odot}$ with redshifts. At $z<2$, below $M_{0}$, the MS slope
  is about 0.85--1.0, and the slope becomes flattened above the turnover
  mass. The MS of galaxies at $2<z<4$ is rather explained by a single
  power-law, with the slope $\sim 0.8$, and the dispersion of the MS is higher
  than one for lower redshifts. With the carefully measured MS, we classify
  galaxies into four populations based on their positions in the
  log(SFR)-log(M$_{*}$) plane, using starburstiness, $R_{SB}$: starbursts
  1-$\sigma$ above the MS, star-forming galaxies on the MS, sub-MS galaxies
  located 1-$\sigma$ below the MS, and the quiescent galaxies.
  
  \item We identify a significant number of galaxies located below the MS
  (sub-MS) having intermediate rest-frame colors and morphologies between the
  quiescent galaxies and star-forming galaxies on the MS. These galaxies have
  lower E(B-V) than MS galaxies on average, supporting that the sub-MS
  galaxies may be under transition from normal star-forming to the quiescent
  population. In particular, among the sub-MS galaxies with $M_{*}>3\times
  10^{10}M_{\odot}$, most of them are compact
  ($\Sigma_{1}>10^{9.5}M_{*}/kpc^{2}$). These galaxies are systematically
  older than normal star-forming galaxies at same stellar masses.

\item Using both traditional diagnostics of morphology (Sersic index $n$,
  $R_e$ and $\Sigma_1$), as well non--parametric (Gini and $M_{20}$
  coefficients), we observe clear morphological differences among galaxies
  located in different locations in the log(SFR)-log(M$_{*}$) plane, which
  also correlate with stellar age.  In particular, we reproduce the trends
  between the projected central mass density ($\Sigma_{1}$) and $R_{SB}$: 
  as the star formation activity decreases
  (i.e., $R_{SB}$ decreases) and galaxies become older, the spread of
  $\Sigma_1$ becomes narrower and confined to the top end of the global
  distribution at all explored redshifts.  The projected central density gets
  compressed toward an upper limit at low $R_{SB}$. This upper limit does not
  change with redshift but it depends on the stellar mass of the galaxies.

\item We find a general trend between the galaxies' median mass--weighted
  stellar age and their position relative to the MS, where the age steadily
  increases for galaxies located at increasingly lower (s)SFRs below the MS.
  Galaxies on the MS, however, have $\Sigma_{1}$ that spans a broad range of
  values, i.e. SFGs with high $\Sigma_{1}$ are not older than SFGs with lower
  central densities. Thus, the central stellar density of galaxies spans a
  relatively large dynamic range of values while they are on the MS, i.e. they 
  are during the star-forming phase. The
  dynamic range becomes restricted (to about 1/4 in Log space) toward the high
  end of the distribution as galaxies quench and become passive. We stress
  that this growth of the stellar mass density of the central regions, which
  reflects of history of dissipation that took place in each galaxy, is not
  necessarily causally connected with the quenching process.

\item Stacks of light profiles of massive, compact quenched/quenching galaxies
  at $z\sim2.5$ do not show evidence for faded disks in their outskirts.  The
  lack of the extended stellar halos, namely the violently relaxed stellar
  component of the merging galaxies, around the compact quiescent galaxies
  suggests that these galaxies are unlikely the remnants of highly dissipative
  wet mergers, unless the stellar contents of the merging galaxies was so
  small to remain undetected in existing images. This interpretation is
  consistent with the presence of extended light in starburst galaxies, which
  probably formed via gas-rich major merging. We find that in general the
  light distribution of starburst galaxies is significantly more diffuse and
  their size larger than compact passive galaxies. Massive, compact starbursts
  ($M_{*}/M_{\odot}>3\times10^{10}$ and $log(\Sigma_{1})>9.5$) are rare and
  {\it essentially} do not exist at lower redshifts.  Therefore, it is
  unlikely that starbursts can shrink their size to match that of compact
  quiescent galaxies within a short timescale.

\item By adding the stellar age as a third dimension in the $\Sigma_{1}$ and
  $M_{*}$ plane, we show that older galaxies have a larger central density 
  at fixed stellar mass and the dispersion of $\Sigma_{1}$ observed in 
  the star--forming galaxies is relatively larger than one for galaxies with 
  decreasing activity of star formation. As galaxies evolve in 
  size and mass by forming stars,
  their central density also increases with age, reflecting the integrated
  history of dissipation that they underwent. The shrinking of the dispersion
  of $\Sigma_{1}$ after quenching depends on the stellar mass, with more
  massive galaxies showing a smaller dispersion. Together with independent
  evidence of environmental quenching of smaller mass satellites around more
  massive centrals (e.g. Guo et al. 2017; Ji et al. 2017) this can be
  interpreted as evidence that different quenching processes are at work at
  different mass regimes.  At large mass, the quenching is caused by
  processing internal to the galaxies that depends on the galaxy's reaching a
  particular value of the stellar mass, while at lower mass is related to the
  environment namely when the quenching takes place is independent on
    the stellar mass that the galaxy has grown up to that moment.
    
\item The flattening of the MS at high masses persists even when we examine
  only the disk-dominated galaxies (having $n<1.5$). At $z<2$, we find no
  significant difference in the MS slopes between the spheroid-dominated
  ($n>2.5$) and disk-dominated ($n<1.5$) galaxies. Furthermore, we find that
  the average SFR for the highest mass bin rarely depends on S\'{e}risc index,
  indicating that massive galaxies have low SFRs regardless of their
  morphologies. We suggest that the decrease in SFR at high masses, e.g.  star
  formation quenching, is not driven by the bulge growth, but it is the result
  of internal processes, which is dependent on the stellar masses.
  
\end{enumerate}

Based on our empirical study of the morphologies of galaxies and their stellar
age at $1.2<z<4$, we suggest that the monotonic increase of projected central 
mass density (i.e. growth of the central parts of the galaxies) as galaxies grow is
an indication of a general phenomenon of structural transformation of
galaxies. Massive, compact galaxies observed at $z\sim2$ are simply assembled
at very early times and evolve through in situ star formation to form compact
quiescent galaxies.  Instead of major wet merging events, a direct accretion
of cold gas can drive the formation of massive, compact galaxies, either via
violent disk instabilities in a compact disk or direct cold mode accretion of
the gas traveling into the galaxy center and forming stars in-situ.

In this paper, we focus on morphological analysis mostly using the
projected central density in the log(SFR)-log(M$_{*}$) plane with an
additional dimension, stellar age, to quantify the process of
``compactification'' of galaxies as they quench their star formation.
However, it is insufficient to explain the overall features of the complex
morphological evolution and transformation of a galaxy. In the second paper,
we will complete the discussion of characteristics of galaxies in the
log(SFR)-log(M$_{*}$) plane and their morphological evolution by comparing
$\Sigma_1$ with the non--parametric morphological indicators, $G$ and
$M_{20}$.

\acknowledgments

The authors thank the anonymous referee for very useful comments helping to improve the presentation of the paper. This work is based on observations taken by the CANDELS Multi-Cycle Treasury Program with the NASA/ESA HST, which is operated by the Association of Universities for Research in Astronomy, Inc., under NASA contract NAS5-26555.
KEW gratefully acknowledge support by NASA through Hubble Fellowship grant$\sharp$HF2-51368 awarded by the Space Telescope Science Institute, which is operated by the Association of Universities for Research in Astronomy, Inc., for NASA. 
RSS acknowledges the generous support of the Downsbrough family.

\appendix

\section{A weak dependence of magnitude limits on the analysis of morphologies}

Morphological parameters used in this study are sensitive to the magnitude, hence signal-to-noise 
ratio \citep{lot04,vdw12}. Using CANDELS 4-epoch data, \cite{vdw12} found that H-band magnitude limit for GALFIT is
$H<23.5$ for n and $H<24.5$ for $R_{e}$, with galaxies fainter than these magnitude limits
are expected to produce biased results. Non-parametric measures can also be unreliable for faint sources, and 
they are generally robust for bright sources \citep{lot04}. Using CANDELS/Deep 
fields (10-epoch), \cite{gro11} suggested 
that morphologies of galaxies having $H<24.7$ can be effectively identified using non-parametric measures. 

\begin{figure*}
\epsscale{0.8}
\plotone{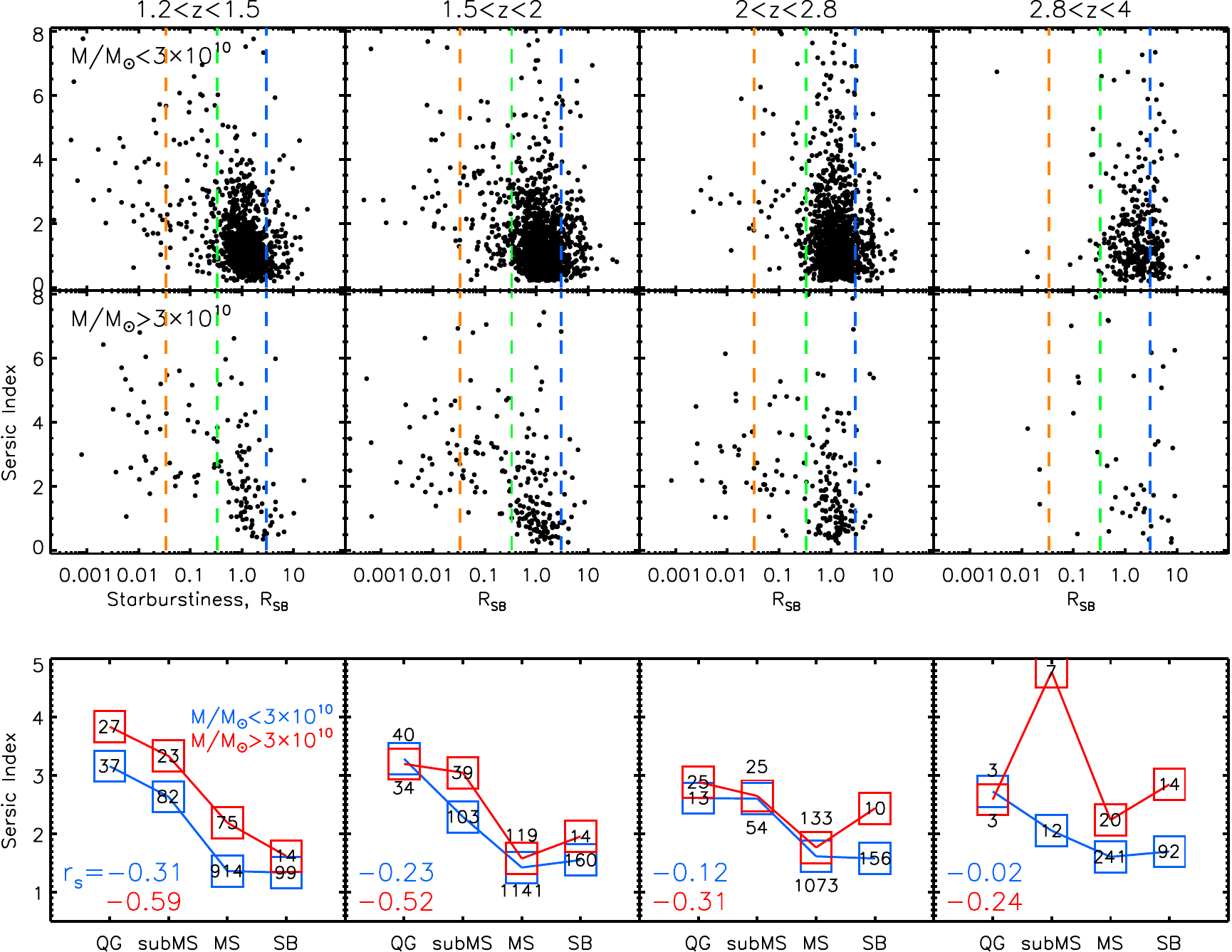}
\plotone{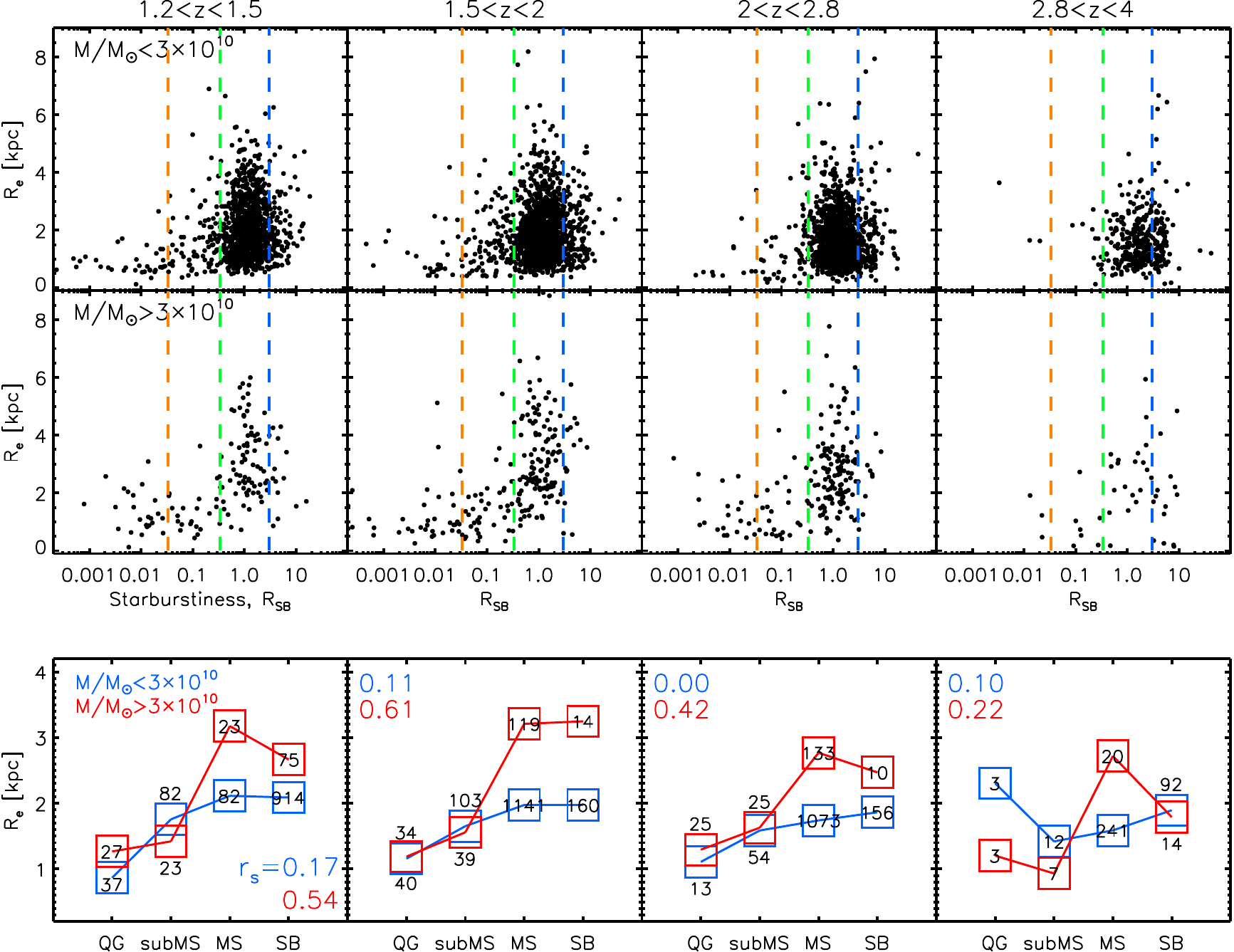}
\caption{We study the correlation between $R_{SB}$ and S\'{e}rsic 
index (top)/ $R_{e}$ (bottom) for four different galaxy populations on the log(SFR)-log(M$_{*}$) relation
across four redshift bins using galaxies having $H<24.4$. All lines, colors and 
symbols correspond to the definitions in Figure~\ref{fig:mean_sersic}. $r_{s}$ is the Spearman's rank 
correlation coefficient between n/$R_{e}$ and $R_{SB}$ for each redshift and mass bin. $r_{s}$ is almost 
identical to the one using all sample without limiting magnitude, indicating that restricting galaxy 
sample with its magnitude would not significantly change our morphological analysis.}
\label{fig:mean_sersic_ex}
\end{figure*}

\begin{figure*}
\epsscale{0.8}
\plotone{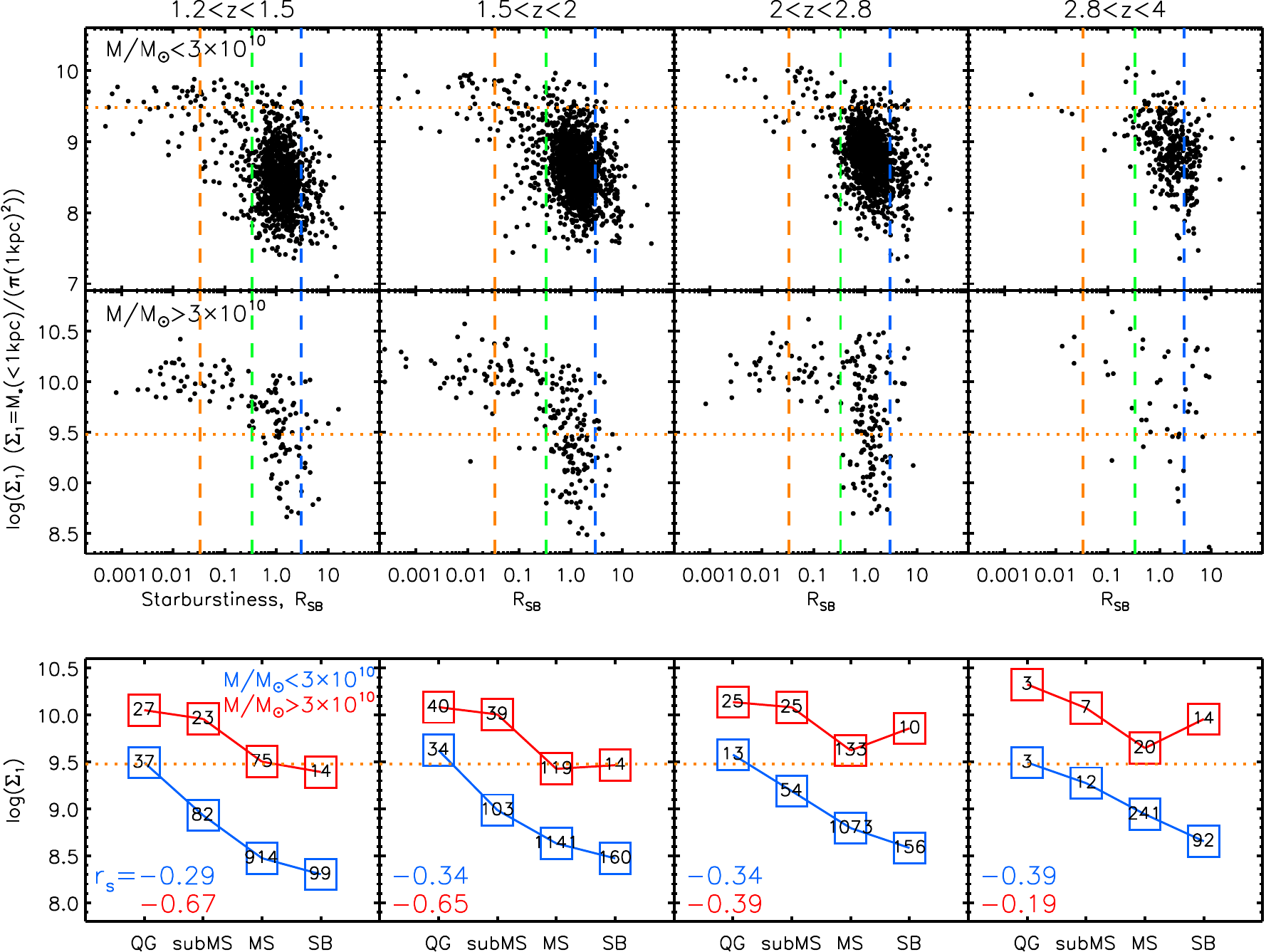}
\caption{The extrapolated projected central density, $\Sigma_{1}$,
  vs. $R_{SB}$ (two top panels) and the average of $\Sigma_{1}$ for the four galaxy
  populations (bottom panel) using galaxies having $H<24.4$. The horizontal line 
is for the classification of high 
 central density galaxy, $\log(\Sigma_{1})=9.5$, from Barro et al. (2015).}
\label{fig:mean_den1_ex}
\end{figure*}

\begin{figure*}
\epsscale{0.8}
\plotone{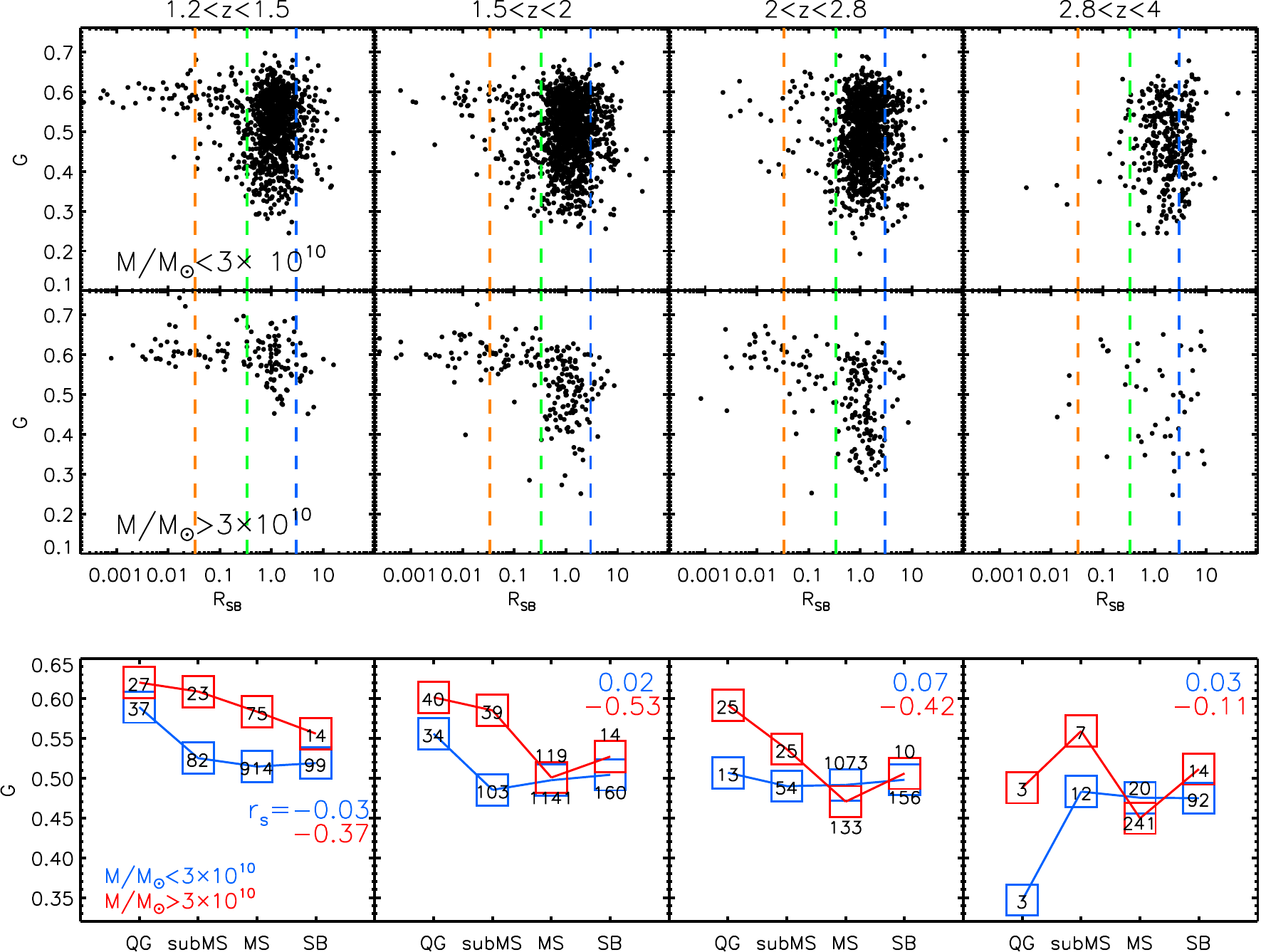}
\plotone{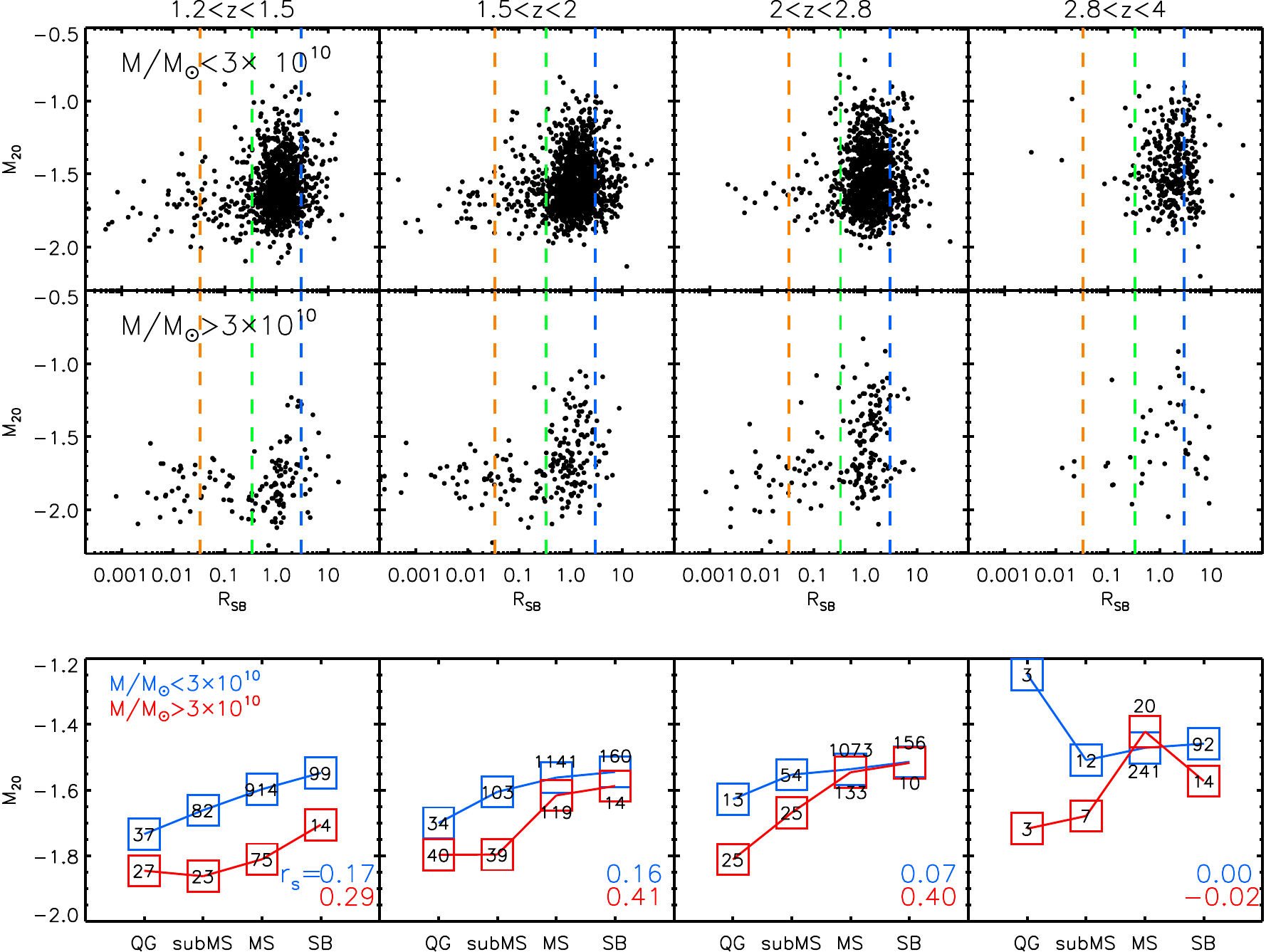}
\caption{Top: Gini vs. $R_{SB}$ using galaxies having $H<24.4$. 
Bottom: $M_{20}$ vs. $R_{SB}$ using galaxies having $H<24.4$. All lines, colors and
  symbols correspond to Figure~\ref{fig:mean_sersic_ex}.}
\label{fig:m20_ex}
\end{figure*}

Because our galaxy sample goes to lower stellar mass limits ($\sim 10^{9}M_{\odot}$) and most of them have $H<26$, 
it is important to understand how the limiting magnitude 
affects our morphological analysis. We use deeper images (10-epoch) than \cite{vdw12}, so 
the limiting magnitude for GALFIT is about 0.9 magnitudes deeper, 
which is $H=25.4$ ($H=24.4$) for $R_{e}$ (n). About 92\% (60\%)
of our sample have $H<25.4$ ($H<24.4$). In this section, we repeat the same analysis for 
all galaxy morphology measures used in this study, S\'{e}rsic index, $R_{e}$, 
$\Sigma_{1}$, Gini, and $M_{20}$, using only galaxies having $H<24.4$ in Figure~\ref{fig:mean_sersic_ex},
~\ref{fig:mean_den1_ex},~\ref{fig:m20_ex}. The overall distribution of all morphological parameters 
looks less scattered than ones using entire sample without limiting magnitudes (Figure~\ref{fig:mean_sersic}, 
\ref{fig:mean_den1}, \ref{fig:mean_gini}). However, 
the correlations between $R_{SB}$ and morphological parameters 
(based on the Spearman's rank correlation coefficient, $r_{s}$) are rarely changed at 
all redshifts and stellar masses. This suggests that our result does not suffer from this systematic
bias. Therefore, throughout this paper, we do not limit the sample based on the galaxy's brightness. 

\section{$\Sigma_{1}$ vs. $\Sigma_{50}$}

\begin{figure*}[ht]
\centering
\includegraphics[width=6.in]{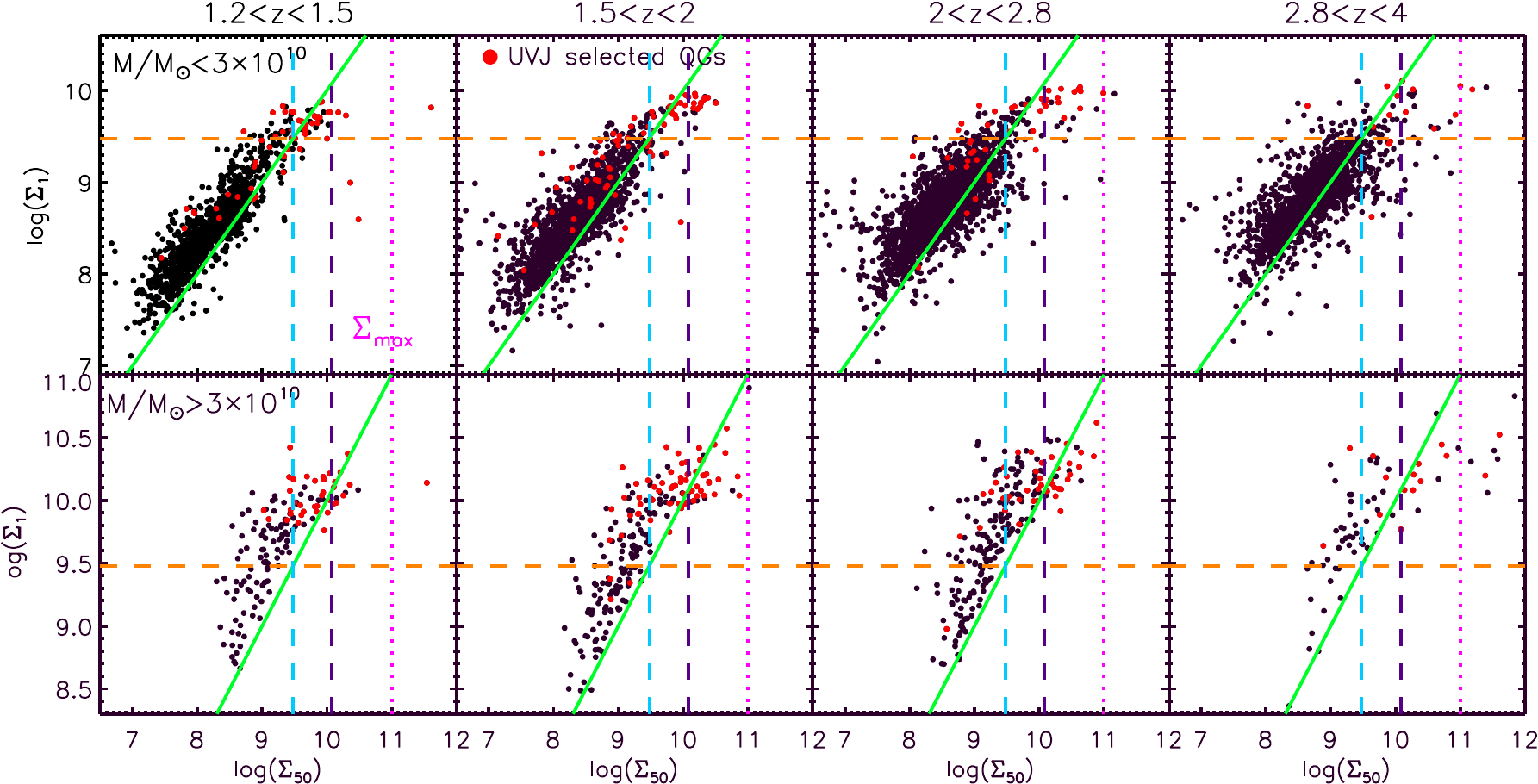}
\caption{$\Sigma_{1}$ vs. $\Sigma_{50}$. The green line is the one-to-one correlation 
between $\Sigma_{1}$ and $\Sigma_{50}$. High density and ultra-high density threshold 
based on the $\Sigma_{50}$ are sky blue and violet vertical lines. The orange horizontal 
line indicates log$(\Sigma_{1})=9.5$, and galaxies above this line are considered as high 
central density galaxies. $\Sigma_{max}$ is equal to $10^{11} M_{\odot}/kpc^{2}$, 
which is the maximum stellar density can be observed in any systems \citep{hop09}.}
\label{fig:comp_den}
\end{figure*}

We directly compare two projected mass densities in Figure~\ref{fig:comp_den}. 
The majority of galaxies have higher $\Sigma_{1}$ than $\Sigma_{50}$, and all the galaxies with 
log$(\Sigma_{1})>9.5$ have log$(\Sigma_{50})>9.5$. There are some galaxies with $\Sigma_{50}>\Sigma_{1}$, 
most of which have $R_{e} <1kpc$. \cite{hop09} found that the maximum stellar surface 
density of any galactic systems is close to log$(\Sigma_{max})\sim 11$. But we find 
15 galaxies with log$(\Sigma_{50}) > 11.0$, whereas $\Sigma_{1}$ never reaches $\Sigma_{max}$. 
Although these 15 galaxies are bright (all of them have $H<26$ and 80\% of them have H$<24.4$), 
they are extraordinarily small, $<0.3 kpc~(<0.1$ pixel). They might be unresolved relative to the point spread function 
(PSF). Or, GALFIT S{\'e}rsic profile of these galaxies might be wrong because GALFIT cannot 
converge for galaxies with $R_{e} <0.5$ pixel \citep{pen10}. 

\begin{figure}
\centering
\includegraphics[width=6.in]{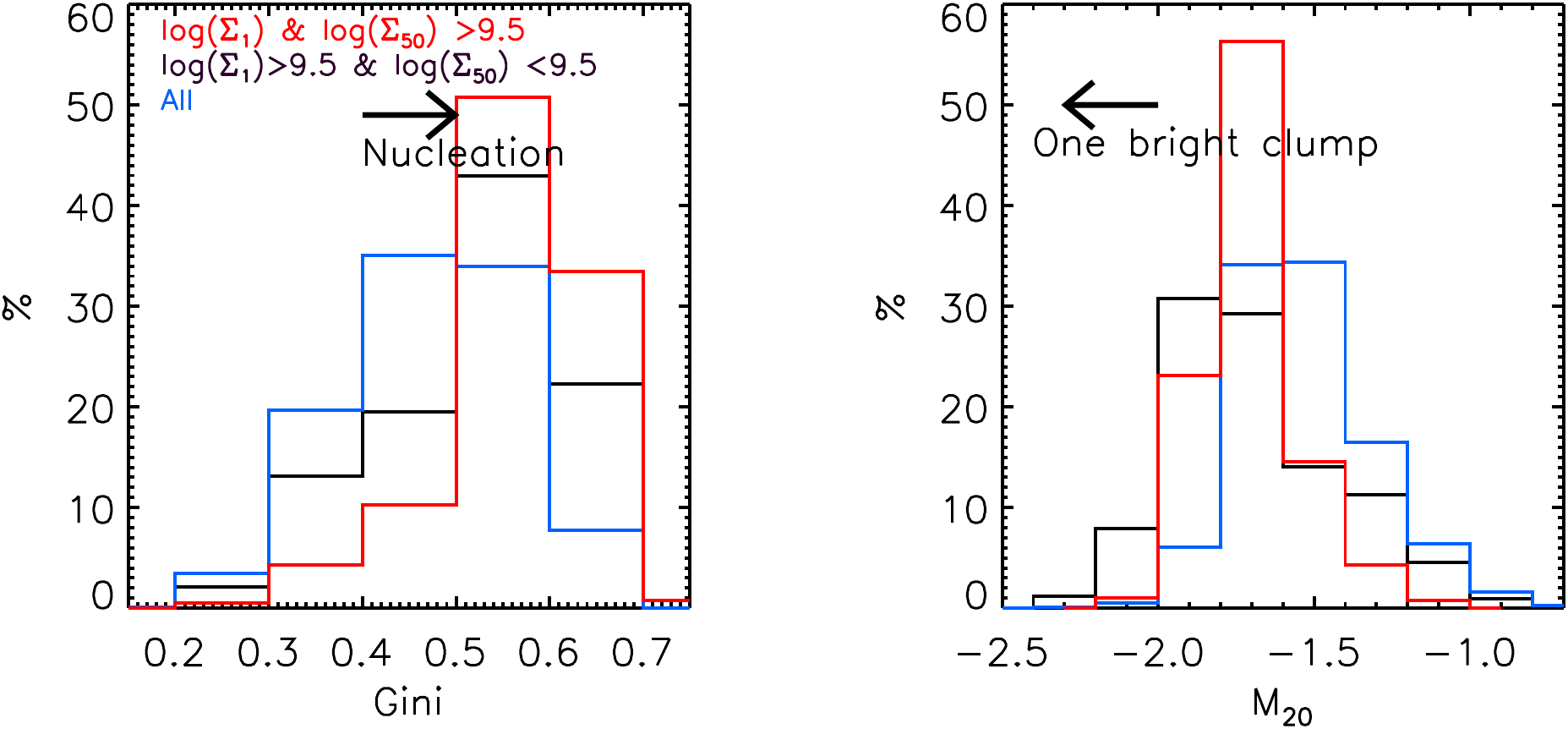}
\caption{Relative distribution of G and $M_{20}$ for all galaxies (blue), galaxies 
having log$(\Sigma_{1})$ \& log$(\Sigma_{50})>9.5$ (red), galaxies having 
log$(\Sigma_{1})>9.5$ \& log$(\Sigma{50})<9.5$ (black). The mean values of G ($M_{20}$) are 0.47 
(-1.56), 0.52 (-1.69) and 0.57 (-1.70) for blue, black and red histograms. The black 
histogram has a similar distribution with the red one, indicating that these galaxies might be nucleated 
although their log$(\Sigma_{50}) < 9.5$.  }
\label{fig:hist_morph}
\end{figure}

To reveal how well the projected mass densities represent the compactness of 
galaxies, we study the correlation between projected mass densities and 
non-parametric measures ($G$, $M_{20}$) by computing the Spearman's rank correlation 
coefficient, $r_{s}$. Because the parametric measures obtained from S\'{e}rsic profile 
fitting are already correlated with $\Sigma_{50}$ and $\Sigma_{1}$, it is not worthy to 
study correlations between S\'{e}rsic index (or $R_{e}$) and densities.  
Statistically, $\Sigma_{1}$ is correlated with $G$ and $M_{20}$ better than 
$\Sigma_{50}$ based on $r_{s}$ (the $r_{s}$ between $G$ and $\Sigma_{50}$ ($\Sigma_{1}$) 
are 0.45 (0.49) and the $r_{s}$ between $M_{20}$ and $\Sigma_{50}$ ($\Sigma_{1}$) are 
-0.25 (-0.43)). This correlation is also observed in the histogram of $G$ and $M_{20}$ 
in Figure~\ref{fig:hist_morph}. Galaxies having log$(\Sigma_{50}) <9.5$ \& log$(\Sigma_{1})>9.5$ 
(black histogram) show similar $G$ and $M_{20}$ distributions with those having 
log$(\Sigma_{50})>9.5$ \& log$(\Sigma_{1})>9.5$ (red histogram). The average values of G ($M_{20}$) 
are 0.47 (-1.56), 0.52 (-1.69), and 0.57 (-1.70) for blue (all galaxies), black, and 
red histograms, respectively. Black and red histograms have very similar average values 
of $G$ and $M_{20}$, indicating that these galaxies having log$(\Sigma_{1})>9.5$ and log$(\Sigma_{50})<9.5$ 
can be nucleated. It is possible to lose compact galaxies to some 
extent, only when $\Sigma_{50}$ is used. Our results show that the compactness of a 
galaxy might be better explained by $\Sigma_{1}$ than $\Sigma_{50}$.

\end{document}